\documentclass{aa}

\usepackage{amsmath,amssymb,graphicx,txfonts,epsf,natbib,psfrag}
\usepackage{sidecap}

\begin{document}

\title{Chaos in the inert Oort cloud}
\titlerunning{Chaos in the inert Oort cloud}
\author{Melaine Saillenfest\inst{1} \and Marc Fouchard\inst{1} \and Takashi Ito\inst{2} \and Arika Higuchi\inst{2}}
\authorrunning{M. Saillenfest et al.}
\institute{IMCCE, Observatoire de Paris, PSL Research University, CNRS, Sorbonne Universit\'e, Universit\'e de Lille, 75014 Paris, France. \and National Astronomical Observatory of Japan, 2-21-1 Osawa, Mitaka, Tokyo 181-8588, Japan.\\\email{melaine.saillenfest@obspm.fr}}
\date{Received 12 July 2019 / Accepted 9 August 2019}

\abstract
{Distant trans-Neptunian objects are subject to planetary perturbations and galactic tides. The former decrease with the distance, while the latter increase. In the intermediate regime where they have the same order of magnitude (the `inert Oort cloud'), both are weak, resulting in very long evolution timescales. To date, three observed objects can be considered to belong to this category.}
{We aim to provide a clear understanding of where this transition occurs, and to characterise the long-term dynamics of small bodies in the intermediate regime: relevant resonances, chaotic zones (if any), and timescales at play.}
{The different regimes are explored analytically and numerically. We also monitored the behaviour of swarms of particles during 4.5 Gyrs in order to identify which of the dynamical features are discernible in a realistic amount of time.}
{There exists a tilted equilibrium plane (Laplace plane) about which orbits precess. The dynamics is integrable in the low and high semi-major axis regimes, but mostly chaotic in between. From about 800 to 1100 astronomical units (au), the chaos covers almost all the eccentricity range. The diffusion timescales are large, but not to the point of being indiscernible in a 4.5 Gyrs duration: the perihelion distance can actually vary from tens to hundreds of au. Orbital variations are damped near the ecliptic (where previous studies focussed), but favoured in specific ranges of inclination corresponding to well-defined resonances. Moreover, starting from uniform distributions, the orbital angles cluster after 4.5 Gyrs for semi-major axes larger than 500 au, because of a very slow differential precession.}
{Even if it is characterised by very long timescales, the inert Oort cloud mostly features chaotic regions; it is therefore much less inert than it appears. Orbits can be considered inert over 4.5 Gyrs only in small portions of the space of orbital elements, which include (90377) Sedna and 2012VP113. Effects of the galactic tides are discernible down to semi-major axes of about 500 au. We advocate including the galactic tides in simulations of distant trans-Neptunian objects, especially when studying the formation of detached bodies or the clustering of orbital elements.}


\maketitle

\section{Introduction}
   Beyond Neptune, the orbits of distant small bodies around the barycentre of the solar system are subject to two kinds of perturbations: an internal perturbation from the planets (mainly the giant ones), and an external perturbation from the galactic tides, passing stars, and molecular clouds. Historically, a distinction is made between the trans-Neptunian or Kuiper belt objects (with all their subclasses) and the Oort cloud. This distinction was made because the trans-Neptunian population is indeed observed on orbits lying beyond or close to Neptune, whereas the long-period comets coming from the Oort cloud are only observed when they are injected into the inner solar system, making them observable from Earth. These different classes of objects are thought to have been initially populated through distinct mechanisms (see e.g. the recent review by \citealp{MOR.NES:19}). However, there is no dynamical boundary between the trans-Neptunian and the Oort cloud populations, and numerical simulations show a continuous transfer of objects in both directions (\citealp{FOUetal:17a}, \citealp{KAIetal:19}). This means that objects that were initially dominantly perturbed by the planets are driven into a region where the galactic tides dominate, and vice versa.
   
   However, the external perturbations are often neglected in simulations of trans-Neptunian objects, even when they feature very distant orbits (\citealp{GALetal:12}, \citealp{SAIetal:17a,SAIetal:17b}, \citealp{BAT.BRO:16}, \citealp{BECetal:17}), whereas the internal perturbations are usually neglected in simulations of the Oort cloud, at least beyond a distance threshold (see e.g. \citealp{HIGetal:07}, \citealp{FOUetal:18}). These simplifications are not necessarily wrong, but a clear understanding of where the transition occurs is still missing, as well as the behaviour of small bodies when they cross the limit.
   
   In reality, there necessarily exists an intermediate region where perturbations from the planets and from the galactic tides have the same order of magnitude. This region, which is itself a continuous transition rather than a clear boundary, can be thought of as the dynamical frontier between the trans-Neptunian and the Oort cloud populations. Since both types of perturbations are expected to be small in this region, we call it the `inert Oort cloud' throughout the article. Authors generally consider that nothing has happened in this region since the formation of the solar system, excluding a very unlikely star passage going completely through, or an even more unlikely close encounter with a giant molecular cloud. Strong orbital perturbations could only have occurred there in the very early evolutionary stages of the solar system, when it was still in a dense stellar cluster. For this reason, the inert Oort cloud is sometimes called `fossilised', or `detached', meaning that the objects it contains could have been placed very early on their current orbits through the interaction with neighbour solar siblings \citep[see e.g.][]{BRAetal:12,JILetal:2015}. For such a frozen configuration to be achieved, the objects within this region should have a perihelion far away from the giant planets, and a semi-major axis small enough for the Galactic tides not to be able to significantly change the perihelion distance over long timescales.
   
   A rough idea of the location of the inert Oort cloud can be obtained from previous works. \citet{GLAetal:02} showed that the scattering effect by Neptune is significant over long timescales only for perihelion distances below about $45$ astronomical units (au). The precise limit actually increases with the semi-major axis \citep{GALetal:12}, because energy kicks result in larger variations of semi-major axis if the semi-major axis is large. In fact, some observed objects with perihelion beyond $45$~au are known to experience scattering \citep{BANetal:17}. In any case, we look here for a rough limit only. The scattering process mostly affects the semi-major axis of small bodies, which diffuses chaotically, while the perihelion distance does not vary much. Later on, \citet{GOMetal:05}, \citet{GALetal:12}, and \citet{SAIetal:16,SAIetal:17a}, showed that the Lidov-Kozai mechanism raised by the giant planets inside a mean-motion resonance with Neptune is able to raise the perihelion of small bodies beyond $60$~au in a few thousands of million years. Contrary to scattering effects, this mechanism induces a variation of perihelion distance and inclination, while the semi-major axis remains at the resonance location. This mechanism, however, is only efficient for semi-major axes smaller than about $500$~au. From these studies, one can deduce that the action of the planets is limited to orbits with perihelion distances smaller than about 80~au, and that for perihelion beyond $45$~au, the semi-major axis should be smaller than $500$~au for the planets to possibly have a substantial effect through mean-motion resonances. As regards the effects of the galactic tides, \citet{FOUetal:17a} showed that an object with perihelion in the Jupiter-Saturn region, that is, below $15$~au from the Sun, should have a semi-major axis larger than $1600$~au for the tides to be able to raise its perihelion beyond $45$~au in less than the age of the solar system. In other words, the tides can move its perihelion out of reach of any significant planetary scattering.
   
\begin{figure}
      \includegraphics[width=\columnwidth]{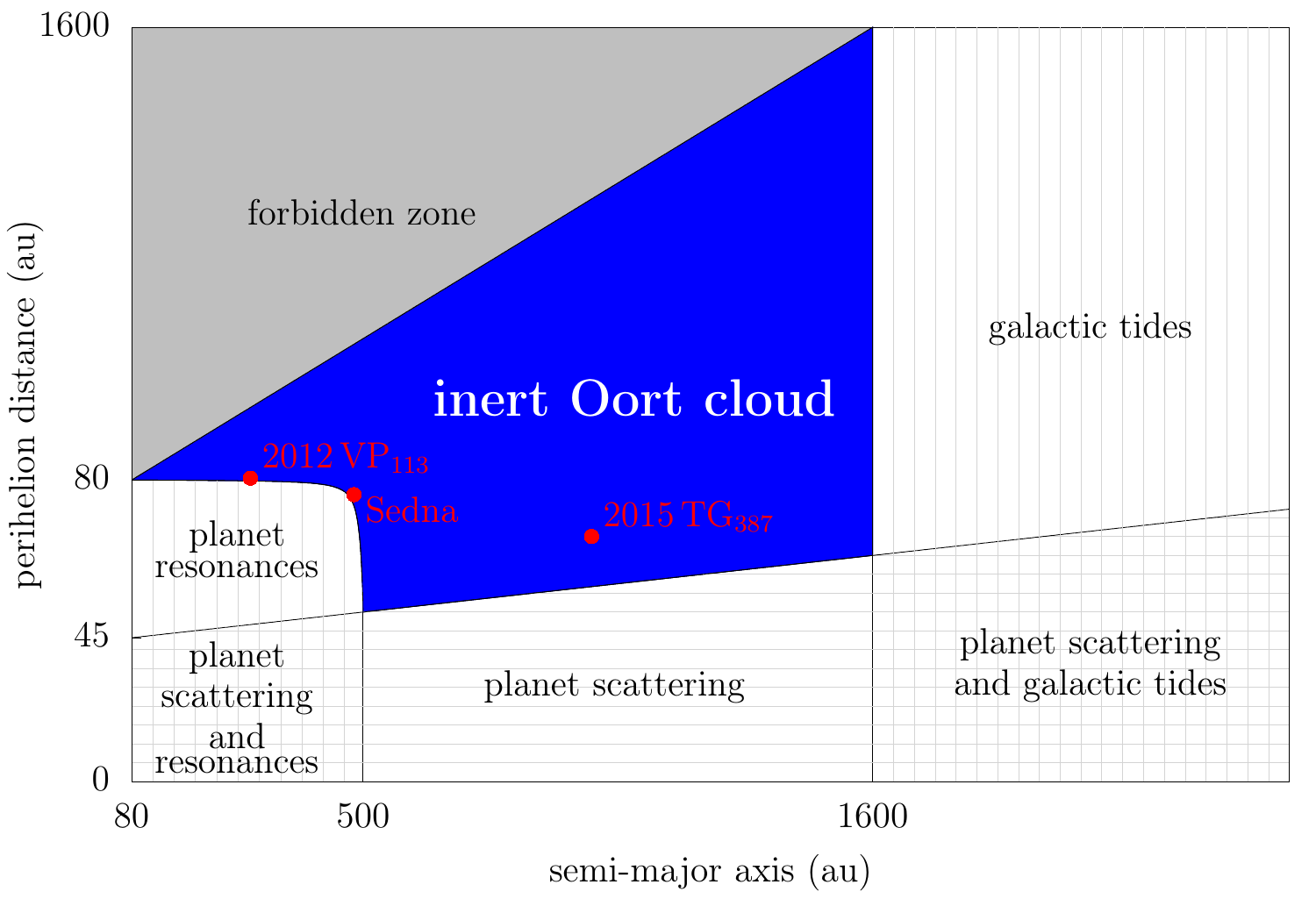}
      \caption{Naive view of the inert Oort cloud. It is defined as the region where neither the planets nor the galactic tides have substantial effects on the orbit of small bodies. In this schematic view, the planet scattering makes small bodies move horizontally, whereas the galactic tides and the isolated mean-motion resonances with planets (labelled `planet resonances' on the graph) make them move vertically. The orbital inclination of small bodies is not considered in this picture, though it is known to play a role as well \citep{SAIetal:17a}.}
      \label{fig:inertOC}
   \end{figure}

   Consequently, the inert Oort cloud can be considered as the region where the semi-major axis is smaller than $1600$~au and the perihelion distance is larger than $45$~au, but the semi-major axis should be larger than $500$~au if the perihelion distance is smaller than $80$~au. The resulting zone is schematised in Fig.~\ref{fig:inertOC}.
   
   At the early stages of the solar system history, most of the small bodies had nearly circular and coplanar orbits that were close to, or even intersecting, the trajectories of the planets (see e.g. \citealp{TSIetal:05}). By scattering, their semi-major axes then spanned a large range of values, populating the bottom part of Fig.~\ref{fig:inertOC}. It is therefore common in simulations to see small bodies wander around the inert Oort cloud, roughly following the straight lines of Fig.~\ref{fig:inertOC} (see \citealp{DONetal:04b} or \citealp{GOMetal:15}). However, a few bodies have been discovered within this region: (90377) Sedna \citep{BROetal:04}, 2012\,VP$_{113}$~\citep{TRU.SHE:14}, and 2015\,TG$_{387}$~\citep{SHEetal:19}. After the discovery of 2012\,VP$_{113}$, \citet{TRU.SHE:14} conjectured the existence of a massive stable population lying in this region. However, one can already notice that these bodies all have very eccentric orbits, implying that either dramatic events occurred during the early evolutionary stages of the solar system (e.g. as a result of a dense stellar environment, see \citealp{BRAetal:12}), or this inert Oort cloud may not be as inert as it appears at first glance. The Planet~9 hypothesis could point in this direction \citep{TRU.SHE:14,BAT.BRO:16}, but strong external perturbations would still be required to emplace the planet itself on its distant orbit. Anyway, it should not distract us from studying the complex interplay between planets and galactic tides in this transitional regime. Even though planets and galactic tides have almost negligible effects on the inert Oort cloud over long timescales, their combined effects could pile up and still induce substantial orbital changes over a 4.5-Gyr evolution.
   
   The aim of the present paper is to characterise and explore the long-term dynamics of the inert Oort cloud, driven by the perturbations from both the galactic tides and the giant planets. We will investigate the dynamical mechanisms at play in this region (resonances, chaos) and draw a quantitative picture of the relevant timescales.
   
   Section~\ref{sec:model} is devoted to the dynamical model used and its underlying simplifications. General considerations about the long-term dynamics are exposed in Sect.~\ref{sec:preliminary}. They are followed in Sect.~\ref{sec:explor} by a detailed exploration of the trajectories allowed through Poincar{\'e} surfaces of section. In Sect.~\ref{sec:app}, we discuss the implications of this mixed-type dynamics for real objects, and we map the inert region in the space of orbital elements. We finally conclude in Sect.~\ref{sec:disconc}.
   
\section{Unified model of planets and galactic tides}\label{sec:model}
   We consider a small body of negligible mass with respect to the giant planets of the solar system. The Hamiltonian function governing its orbital motion can be decomposed into the Sun-body Keplerian part\footnote{Even though the orbits of distant bodies are essentially barycentric, and not heliocentric, the Hamiltonian function is simpler when expressed in heliocentric coordinates. This is not a problem because we then use average coordinates, in which barycentric and heliocentric elements are equivalent (the wobbles of the Sun are averaged out).}, a perturbation due to the planets, and a perturbation due to the galactic tides:
   \begin{equation}\label{eq:Hbase}
      \mathcal{H} = \mathcal{H}_0 + \varepsilon_\mathrm{P}\mathcal{H}_\mathrm{P} + \varepsilon_\mathrm{G}\mathcal{H}_\mathrm{G} \,.
   \end{equation}
   Expressed using Keplerian elements, the two-body part is
   \begin{equation}
      \mathcal{H}_0 = -\frac{\mu}{2a} \,,
   \end{equation}
   where $a$ is the semi-major axis of the small body and $\mu$ is the gravitational parameter of the Sun.
   
   We assume that the small body never goes inside the orbits of the planets. The Hamiltonian $\varepsilon_\mathrm{P}\mathcal{H}_\mathrm{P}$ can therefore be expanded in Legendre polynomials. As explained in the Introduction, mean-motion resonances are inefficient in the inert Oort cloud, such that we are allowed to use the averaged perturbation from the planets (whose orbital periods are much smaller than the one of the small body). At this level of approximation, effects coming from the small eccentricities and mutual inclinations of the planets are perfectly negligible for such distant small bodies. Consequently, using circular and coplanar orbits for the planets, we obtain
   \begin{equation}
      \varepsilon_\mathrm{P}\mathcal{H}_\mathrm{P} = \varepsilon_{\mathrm{P}_0}\mathcal{H}_{\mathrm{P}_0} + \varepsilon_{\mathrm{P}_2}\mathcal{H}_{\mathrm{P}_2} + \varepsilon_{\mathrm{P}_4}\mathcal{H}_{\mathrm{P}_4} + \mathcal{O}(\varepsilon_{\mathrm{P}_6}) \,.
   \end{equation}
   These terms correspond to the monopole (index $0$), quadrupole (index $2$), and hexadecapole (index $4$), respectively. Their expressions can be taken from \cite{SAIetal:16}, or \cite{LAS.BOU:10} in a general context:
   \begin{equation}
      \left\{
      \begin{aligned}
         \varepsilon_{\mathrm{P}_0}\mathcal{H}_{\mathrm{P}_0} &= -\frac{1}{r}\sum_{i=1}^N\mu_i \,,\\
         \varepsilon_{\mathrm{P}_2}\mathcal{H}_{\mathrm{P}_2} &= \frac{1}{2}\left(\frac{1}{r^3}-\frac{3}{2}\frac{x^2+y^2}{r^5}\right)\sum_{i=1}^N\mu_ia_i^2 \,,\\
         \varepsilon_{\mathrm{P}_4}\mathcal{H}_{\mathrm{P}_4} &=  -\frac{3}{64}\left(\frac{8}{r^5} - 40\frac{x^2+y^2}{r^7} + 35\frac{(x^2+y^2)^2}{r^9}\right)\sum_{i=1}^N\mu_ia_i^4\,.
      \end{aligned}
      \right.
   \end{equation}
   In these expressions, $(x,y,z)$ are the coordinates of the small body in a reference frame centred on the Sun, where the $(x,y)$ is the orbital plane of the planets, and $r\equiv\sqrt{x^2+y^2+z^2}$. We call this frame the `ecliptic' reference frame. The quantities $\mu_i$ and $a_i$ are the gravitational parameter and the semi-major axis of the planet $i$, for a total of $N$ planets. Because of their small semi-major axes, the planets are supposed to be unaffected by the galactic tides, such that this reference frame is inertial (we consider no precession of the ecliptic pole around the galactic pole).
   
   We now consider the coordinates $(X,Y,Z)$ of the small body in a fixed reference frame centred on the Sun, where the $(X,Y)$ plane is the galactic plane. We call it the `galactic' reference frame. We note $(X',Y',Z')$ the coordinates of the small body in an analogous reference frame, but for which at any time the $X'$ axis points towards the galactic centre. Because of the motion of the Sun in the Galaxy, the latter reference frame is rotating. At lowest-order of approximation, the Sun describes a circular orbit with constant velocity lying in the galactic plane \citep[e.g.][]{FOU:04}. We have in this case the relation
   \begin{equation}
      \begin{pmatrix}
         X' \\
         Y' \\
         Z'
      \end{pmatrix}
      =
      \begin{pmatrix}
          \cos\theta &  \sin\theta & 0 \\
         -\sin\theta &  \cos\theta & 0 \\
                   0 &           0 & 1
      \end{pmatrix}
      \begin{pmatrix}
         X \\
         Y \\
         Z
      \end{pmatrix}
      \,,
   \end{equation}
   where the time derivative of $\theta$ is a constant and corresponds to the angular velocity of the galactic centre seen from the Sun. In the following, we write it $\nu_\mathrm{G}$. In the quadrupolar approximation, the Hamiltonian function describing the orbital perturbations of the small body from the galactic tides can be written
   \begin{equation}
      \varepsilon_\mathrm{G}\mathcal{H}_\mathrm{G} = \nu_\mathrm{G}P_\theta + \mathcal{G}_1\frac{X'^2}{2} + \mathcal{G}_2\frac{Y'^2}{2} + \mathcal{G}_3\frac{Z'^2}{2} \,,
   \end{equation}
   where $\mathcal{G}_1$, $\mathcal{G}_2$ and $\mathcal{G}_3$ are constants encompassing the shape of the galaxy, its mass density, and the inertial forces due to the rotation of the frame \citep{FOU:04}. The momentum $P_\theta$ is conjugate to the angle $\theta$; it has been introduced such that the Hamiltonian function is autonomous. Using the usual approximation $\mathcal{G}_2 = -\mathcal{G}_1$, we get
   \begin{equation}
      \varepsilon_\mathrm{G}\mathcal{H}_\mathrm{G} = \nu_\mathrm{G}P_\theta + \varepsilon_{\mathrm{G}_\mathrm{V}}\mathcal{H}_{\mathrm{G}_\mathrm{V}} + \varepsilon_{\mathrm{G}_\mathrm{R}}\mathcal{H}_{\mathrm{G}_\mathrm{R}} \,,
   \end{equation}
   where
   \begin{equation}
      \left\{
      \begin{aligned}
         \varepsilon_{\mathrm{G}_\mathrm{V}}\mathcal{H}_{\mathrm{G}_\mathrm{V}} &= \mathcal{G}_3\frac{Z^2}{2} \\
         \varepsilon_{\mathrm{G}_\mathrm{R}}\mathcal{H}_{\mathrm{G}_\mathrm{R}} &= \mathcal{G}_2\left(\frac{Y^2-X^2}{2}\cos(2\theta)-XY\sin(2\theta)\right) \,.
      \end{aligned}
      \right.
   \end{equation}
   The symbols V and R are used here in reference to the vertical and radial components of the galactic tides, respectively.
   
   The perturbations due to the planets and due to the galactic tides being both very small with respect to the Keplerian part, they act on a much longer timescale. Therefore, we use a perturbative approach to order one. The resulting Hamiltonian function is obtained by averaging $\mathcal{H}$ (Eq.~\ref{eq:Hbase}) over an orbital period. The momentum conjugate to the mean anomaly of the small body becomes a constant of motion, which implies the conservation of the secular semi-major axis (that we still denote $a$). Dropping the constant parts, the secular Hamiltonian is
   \begin{equation}\label{eq:Hsecbase}
      \bar{\mathcal{H}} = \nu_\mathrm{G}P_\theta + \varepsilon_{\mathrm{P}_2}\bar{\mathcal{H}}_{\mathrm{P}_2} +
      \varepsilon_{\mathrm{P}_4}\bar{\mathcal{H}}_{\mathrm{P}_4} + \varepsilon_{\mathrm{G}_\mathrm{V}}\bar{\mathcal{H}}_{\mathrm{G}_\mathrm{V}} + \varepsilon_{\mathrm{G}_\mathrm{R}}\bar{\mathcal{H}}_{\mathrm{G}_\mathrm{R}} \,.
   \end{equation}
   We will now introduce explicit expressions for the small parameters:
   \begin{equation}\label{eq:eps}
      \begin{aligned}
         \varepsilon_{\mathrm{P}_2} &= \frac{1}{a^3}\sum_{i=1}^N\mu_ia_i^2 \hspace{0.3cm}&,\\
         \varepsilon_{\mathrm{G}_\mathrm{V}} &= a^2\mathcal{G}_3 \hspace{0.3cm}&,
      \end{aligned}
      \hspace{0.3cm}
      \begin{aligned}
         \varepsilon_{\mathrm{P}_4} &= \frac{9}{16}\frac{1}{a^5}\sum_{i=1}^N\mu_ia_i^4 \,,\\
         \varepsilon_{\mathrm{G}_\mathrm{R}} &= a^2\mathcal{G}_2 \,.
      \end{aligned}
   \end{equation}
   We note $(e,I,\omega,\Omega)$ the Keplerian elements  of the small body in the ecliptic reference frame, with $e$ its eccentricity, $I$ its inclination, $\omega$ its argument of perihelion, and $\Omega$ its longitude of ascending node. We will use the subscript $\mathrm{G}$ for the same quantities measured in the galactic reference frame (excepting $e$ that does not change). Performing the required averages, the different components of Eq.~\eqref{eq:Hsecbase} can be written
   \begin{equation}\label{eq:Hp}
      \left\{
      \begin{aligned}
         \bar{\mathcal{H}}_{\mathrm{P}_2} &= \frac{1-3\cos^2I}{8(1-e^2)^{3/2}} \,,\\
         \bar{\mathcal{H}}_{\mathrm{P}_4} &= \frac{1}{64(1-e^2)^{7/2}}\Bigg((2+3e^2)(-3+30\cos^2I-35\cos^4I) \\
         &\hspace{2.3cm}
         + 10e^2(1-7\cos^2I)\sin^2I\cos(2\omega)\Bigg) \,,
      \end{aligned}
      \right.
   \end{equation}
   and
   \begin{equation}\label{eq:Hg}
      \left\{
      \begin{aligned}
         \bar{\mathcal{H}}_{\mathrm{G}_\mathrm{V}} &= \frac{\sin^2I_\mathrm{G}}{4}\left(1+\frac{3}{2}e^2-\frac{5}{2}e^2\cos(2\omega_\mathrm{G})\right) \,,\\
         \bar{\mathcal{H}}_{\mathrm{G}_\mathrm{R}} &= -\frac{1}{4}\left(1+\frac{3}{2}e^2\right)\cos(2\Omega_\mathrm{G}-2\theta)\sin^2I_\mathrm{G} \\
         &+ \frac{5}{4}e^2\Bigg(\sin(2\omega_\mathrm{G})\sin(2\Omega_\mathrm{G}-2\theta)\cos I_\mathrm{G} \\
         &\hspace{1cm}
         - \cos(2\omega_\mathrm{G})\cos(2\Omega_\mathrm{G}-2\theta)\frac{1+\cos^2I_\mathrm{G}}{2}\Bigg) \,.
      \end{aligned}
      \right.
   \end{equation}   
   We write $\psi$ the inclination of the ecliptic plane in the galactic reference frame, and $\alpha$ its ascending node. Since we have neglected the precession of the ecliptic pole, the angles $\psi$ and $\alpha$ are constant. The ascending node of the ecliptic can therefore be used as the origin of longitudes in the galactic frame, meaning that $\alpha \equiv 0$. The corresponding conversion formulas between the two reference frames are given in Appendix~\ref{asec:conv}, and the values for the physical constants of the problem are gathered in Table~\ref{tab:physconst}. We note that the galactic and ecliptic reference frames used here are a natural choice considering the dynamics under study, but they are not the usual IAU ones (this is only a matter of origin of the longitudes).
   
   \begin{table}
      \caption{Values used for the physical constants of the problem.}
      \label{tab:physconst}
      \vspace{-0.7cm}
      \begin{equation*}
         \begin{array}{rr}
            \hline
            \text{quantity} & \text{value} \\
            \hline
            \hline
            \sum_{i=1}^N\mu_ia_i^2 & 4.5413 \ \mathrm{au}^5\mathrm{yr}^{-2} \\
            \sum_{i=1}^N\mu_ia_i^4 & 2037.2597 \ \mathrm{au}^7\mathrm{yr}^{-2} \\
            \mathcal{G}_2          & 7.0706\times 10^{-16}\ \mathrm{yr}^{-2} \\
            \mathcal{G}_3          & 5.6530\times 10^{-15}\ \mathrm{yr}^{-2} \\
            \psi                   & 1.05048854 \ \mathrm{rad}\\
            \hline
         \end{array}
      \end{equation*}
      \vspace{-0.3cm}
      \tablefoot{The planetary elements come from the theory of \cite{BRE:82}, the galactic constants $\mathcal{G}_2$ and $\mathcal{G}_3$ are taken from \cite{FOU:04}, and the inclination of the ecliptic is obtained from \cite{MUR:89}. Even though it is quite old, the theory of \cite{BRE:82} has the advantage of directly giving the secular component of the planetary dynamics. Since it is semi-analytical, this theory is also expected to be more robust than numerical ephemerides when considering very long timescales.}
   \end{table}

\section{The galactic Laplace plane}\label{sec:preliminary}
\begin{figure}
      \includegraphics[width=\columnwidth]{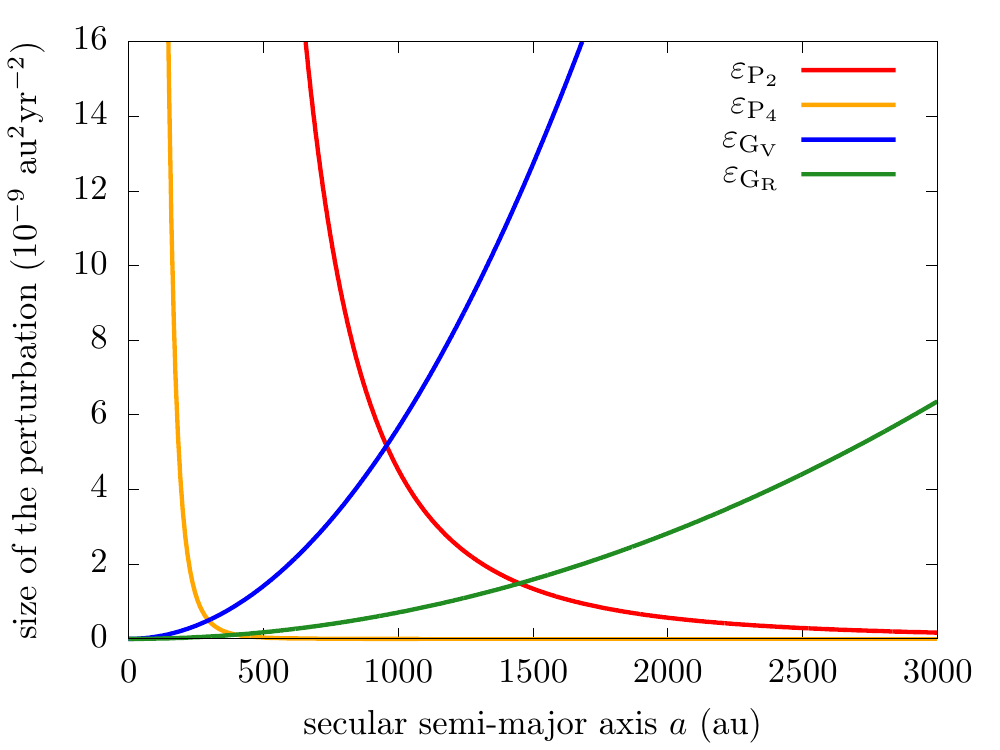}
      \caption{Size of the small parameters listed in Eq.~\eqref{eq:eps} with respect to the secular semi-major axis of the small body. The value of the physical parameters used are given in Table~\ref{tab:physconst}.}
      \label{fig:eps}
   \end{figure}

   The explicit expressions of the small parameters (Eq.~\ref{eq:eps}) have been chosen such that the Hamiltonian functions $\bar{\mathcal{H}}_{\mathrm{P}_2}$, $\bar{\mathcal{H}}_{\mathrm{P}_4}$, $\bar{\mathcal{H}}_{\mathrm{G}_\mathrm{V}}$, and $\bar{\mathcal{H}}_{\mathrm{G}_\mathrm{R}}$ have the same order of magnitude for $e=0$. The secular semi-major axis rules the relative importance of the different perturbation terms. Figure~\ref{fig:eps} shows that below $a\sim 600$~au, the planetary perturbations dominate over the galactic tides by more than a factor 10. The situation is reversed beyond $a\sim 1500$~au. In between, both kinds of perturbations have the same order of magnitude ($\varepsilon_{\mathrm{P}_2}$ and $\varepsilon_{\mathrm{G}_\mathrm{V}}$ cross at $a\sim 950$~au). However, since the eccentricity appears at the denominator in $\bar{\mathcal{H}}_{\mathrm{P}_2}$ (see Eq.~\ref{eq:Hp}), we expect that the planetary perturbations always have a substantial effect in the high-eccentricity regime.
   
   From Fig.~\ref{fig:eps}, it is also clear that in the weakly perturbed intermediate regime, the planetary perturbations are dominated by the quadrupolar term, whereas the galactic tides mostly consist in their vertical component (the radial component is always smaller by one order of magnitude, see Table.~\ref{tab:physconst}). In the remaining parts of the article, we will therefore limit the study to the simplified Hamiltonian function
   \begin{equation}\label{eq:F}
      \mathcal{F} = \varepsilon_{\mathrm{P}_2}\bar{\mathcal{H}}_{\mathrm{P}_2} + \varepsilon_{\mathrm{G}_\mathrm{V}}\bar{\mathcal{H}}_{\mathrm{G}_\mathrm{V}} \,,
   \end{equation}
   in order to draw a qualitative picture of the dynamics in the intermediate regime. This Hamiltonian has two degrees of freedom, and we will use the canonical Delaunay elements:
   \begin{equation}\label{eq:Delem}
      \left\{
      \begin{aligned}
         g &= \omega_\mathrm{G} \\
         h &= \Omega_\mathrm{G}
      \end{aligned}
      \right.
      \hspace{0.5cm}\text{conjugate to}\hspace{0.5cm}
      \left\{
      \begin{aligned}
         G &= L\sqrt{1-e^2} \\
         H &= L\sqrt{1-e^2}\cos I_\mathrm{G}
      \end{aligned}
      \right. \,,
   \end{equation}
   where $L=\sqrt{\mu a}$ is a constant. We note that $g$ only appears in $\bar{\mathcal{H}}_{\mathrm{G}_\mathrm{V}}$, whereas $h$ only appears in $\bar{\mathcal{H}}_{\mathrm{P}_2}$ (through $\cos I$). In this context, the galactic coordinates are therefore the most natural coordinates to use.
   
\begin{figure}
      \includegraphics[width=0.85\columnwidth]{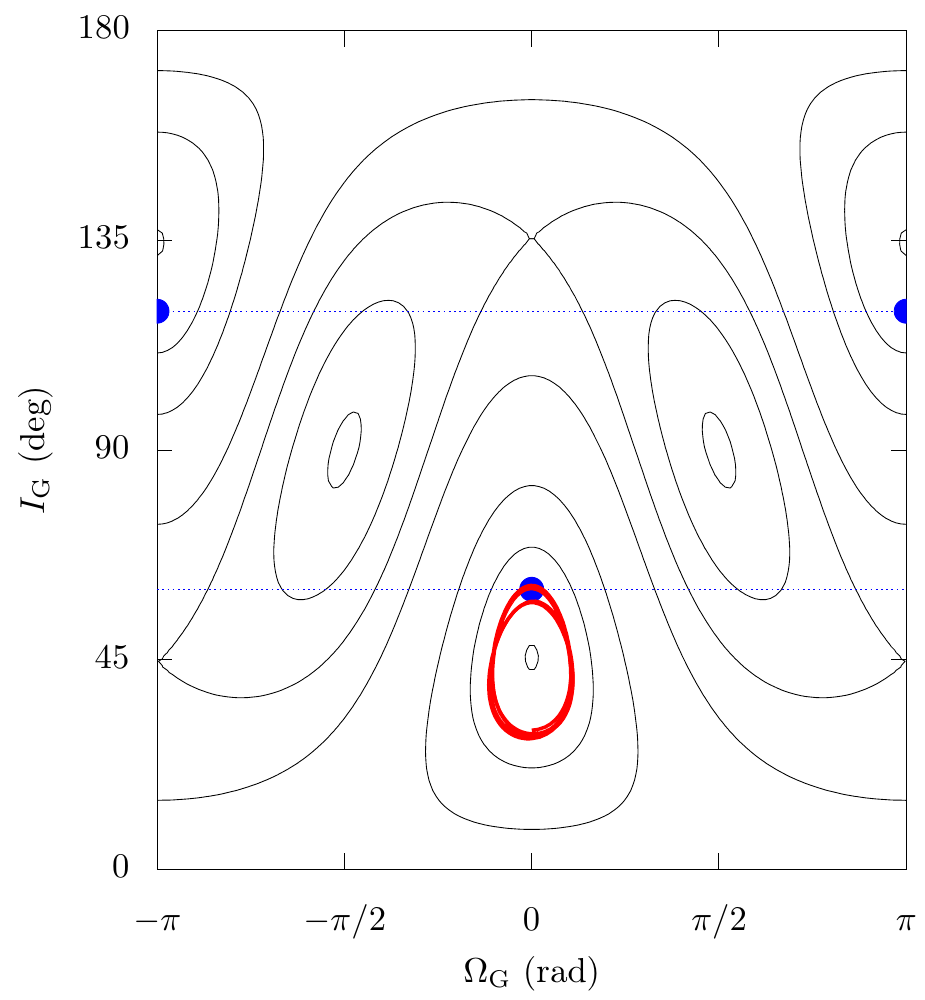}\\
      \vspace{0.1cm}\\
      \includegraphics[width=0.85\columnwidth]{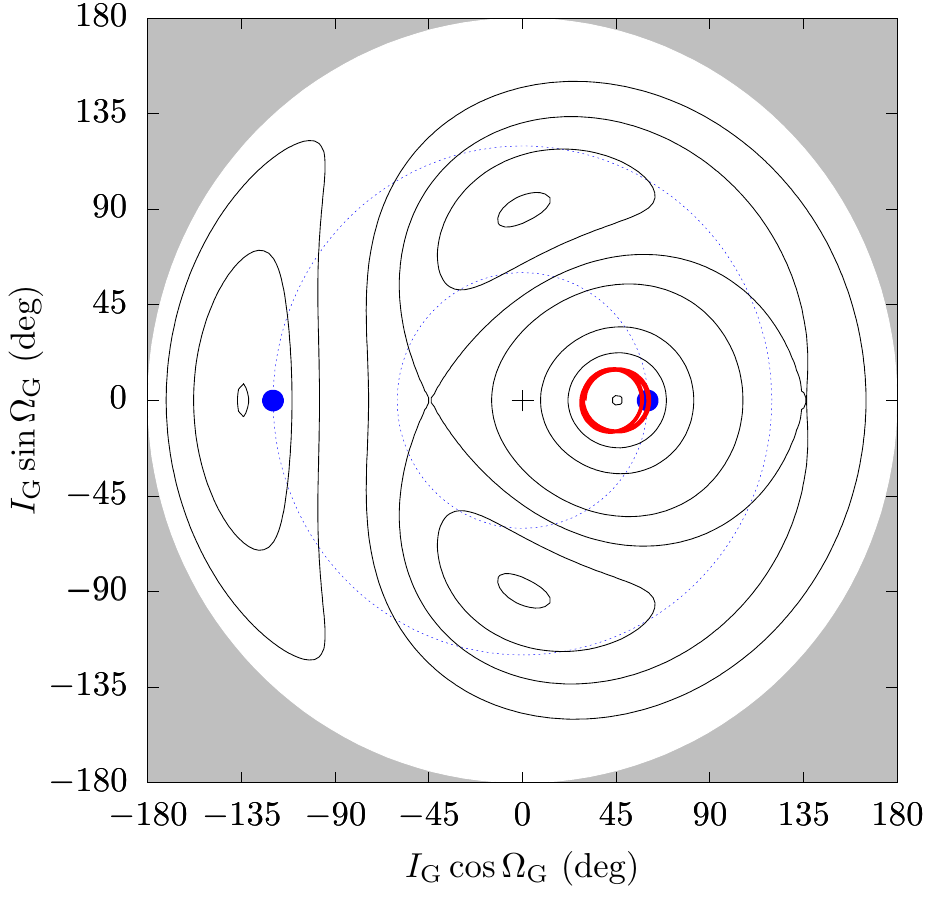}
      \caption{Level curves of the Hamiltonian function $\mathcal{F}$ (Eq.~\ref{eq:F}) for a circular orbit. The semi-major axis taken as parameter is $a=900$~au, and the level curves are shown in black. The two graphs show the same level curves for two sets of variables (in order to avoid being misled by coordinate singularities). The dotted curves represent the inclination $\psi$ of the ecliptic, and the blue spots show the location of the ecliptic plane. The red curve shows an example of nearly circular trajectory precessing around the normal to its local Laplace plane, obtained by numerical integration. The eccentricity slightly varies creating a deviation from the initial level curve. The oscillation period around the fixed centre is about $200$~Gyrs, in accordance with Fig~\ref{fig:LaplaceFreq}.}
      \label{fig:exLap}
   \end{figure}

   First of all, we note that the Hamiltonian $\mathcal{F}$ (see Eq.~\ref{eq:F}) is very similar to the Hamiltonian governing the secular orbital motion of a satellite perturbed by the Sun and by the $J_2$ flattening of its host planet. In the quadrupolar approximation used here, the two Hamiltonian functions are even identical for small eccentricities, as shown in Appendix~\ref{asec:satellite}. This means that the concept of `Laplace plane' introduced in the satellite case (see e.g. \citealp{TREetal:09}) has its equivalent for distant trans-Neptunian objects in the galactic potential. The Laplace plane is normal to the axis around which the orbital angular momentum precesses. In other words, the orbital inclination measured with respect to the Laplace plane is almost constant, while the corresponding longitude of ascending node circulates. More specifically, a Laplace plane corresponds to a fixed points of the dynamics. The results obtained by \cite{TREetal:09} in the satellite case remain valid here for circular orbits ($e=0$ is a fixed point for the eccentricity). We have the same geometry of phase space, as shown in Fig.~\ref{fig:exLap}: we recognise the `circular coplanar' equilibria at $\Omega_\mathrm{G}=0$ and $\pi$, among which the stable ones correspond to the classical Laplace plane (located equivalently at $\Omega=\pi$ and $0$). We also recognise the `circular orthogonal' equilibrium for $I_\mathrm{G}=I=90^\text{o}$ and $\Omega_\mathrm{G}=\Omega=\pm\pi/2$. Since the phase space is a sphere, we stress that all trajectories oscillate around one of the stable Laplace equilibria. The stability of the equilibria against eccentricity growth is different from the satellite case, but this has no consequence here considering the timescales involved (see Appendix~\ref{asec:satellite} for details).
   
\begin{figure}
      \includegraphics[width=\columnwidth]{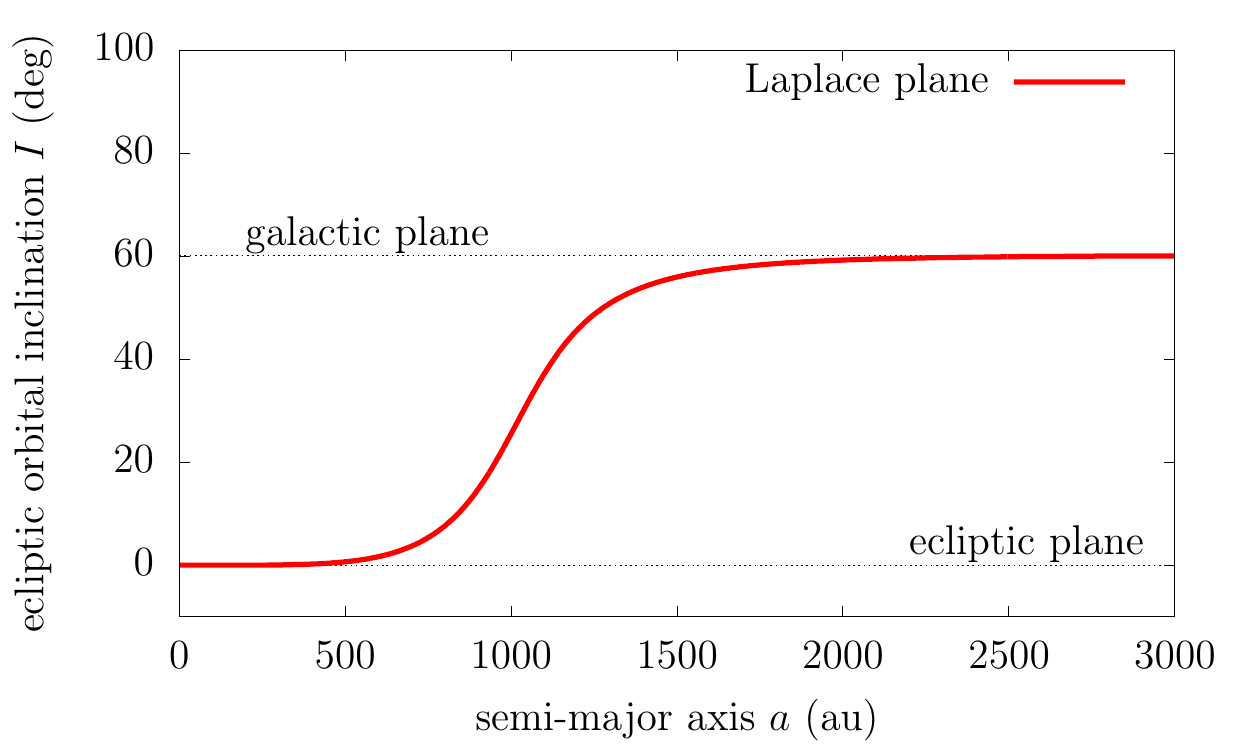}
      \caption{Inclination of the classical Laplace plane with respect to the planetary plane.}
      \label{fig:Laplace}
   \end{figure}

   Low-eccentricity orbits initially lying close to the ecliptic ($I_\mathrm{G}\approx\psi$ and $\Omega_\mathrm{G}\approx 0$) should all precess around the classical Laplace plane. Using the expression of the Hamiltonian $\mathcal{F}$ (see Eq.~\ref{eq:F}), the inclination of the classical Laplace plane is a root of a second order polynomial in $\tan I_\mathrm{G}$. Figure~\ref{fig:Laplace} shows the inclination of the classical Laplace plane according to the value of the secular semi-major axis $a$. For small values of $a$, this plane is very close to the ecliptic plane, whereas for large values of $a$, it is very close to the galactic plane. In between, the orbits of small bodies precess about an intermediary plane. This transition occurs in the region where $\varepsilon_{\mathrm{P}_2}$ and $\varepsilon_{\mathrm{G}_\mathrm{V}}$ have the same order of magnitude (compare Figs.~\ref{fig:eps} and \ref{fig:Laplace}).
   
\begin{figure}
      \includegraphics[width=\columnwidth]{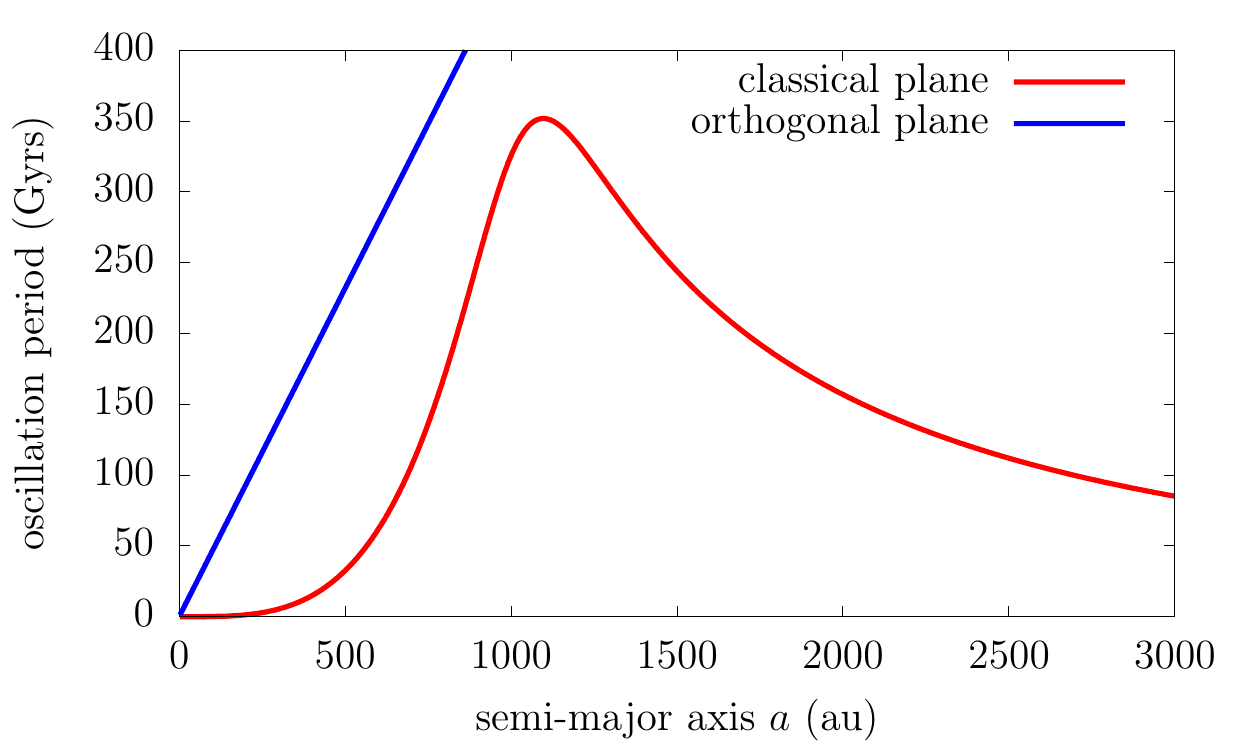}
      \caption{Period of small oscillations about the two kinds of circular Laplace equilibrium. The period of oscillations around the classical equilibrium exceeds $9$~Gyrs for semi-major axes between $350$~au and $13400$~au (out of the graph). The period of oscillations around the orthogonal equilibrium is linear with $a$, and it is everywhere dramatically long with respect to the age of the solar system (apart from very small values of $a$ where the overall model is questionable).}
      \label{fig:LaplaceFreq}
   \end{figure}

   However, for nearly circular orbits, the oscillations about the classical Laplace plane in the transition regime are extremely slow compared to the age of the solar system (see Fig.~\ref{fig:LaplaceFreq}), meaning that in practice these orbits hardly change at all and are indeed `inert'. More precisely, the half precession period of the orbit pole exceeds the age of the solar system for semi-major axes between $350$~au and $13400$~au. Below $350$~au, the precession is about the ecliptic pole ($I\approx const$), so that a set of orbits initially lying close to the ecliptic plane never departs from it. Beyond $13400$~au, the precession is about the galactic pole ($I_\mathrm{G}\approx const$), such that orbits explore all the values of $I$ between $\psi-I_\mathrm{G}$ and $\psi+I_\mathrm{G}$ in less than the age of the solar system. Considering a swarm of particles with initially small inclinations~$I$, this upper limit corresponds to the transition between a disc-like and an isotropic region, even though previous authors based their criterion on a full period for very eccentric orbits (see e.g. \citealp{HIGetal:07,FOUetal:18,VOKetal:19}). For such large values of $a$, the revolution period of $\Omega_\mathrm{G}$ tends to the value obtained when neglecting the planets (noted $P_{\Omega^*}$ by \citealp{HIGetal:07}, see their Fig.~2).
      
   We note that contrary to regular satellites slowly migrating from their formation region, that are expected to remain very close to their local Laplace plane (see e.g. the discussion of \citealp{POLetal:2018} about Iapetus), distant trans-Neptunian objects can be subject to `fast' changes of orbit: either a diffusion of $a$ by planetary scattering, or an overall randomisation due to passing stars. This last mechanism is thought to have been quite efficient in the inert Oort cloud during the early stages of the solar system. This means that the orthogonal equilibrium could be populated as well, or at least, it could a priori play a role in the dynamics of distant trans-Neptunian objects. In the circular case, the orthogonal orbits are however completely frozen, as shown in Fig.~\ref{fig:LaplaceFreq}.
   
   As can be guessed from the expression of the Hamiltonian function, the situation is different for the eccentricity degree of freedom. Indeed, if the orbit is very eccentric (and this is the case for all known distant trans-Neptunian objects) the planetary part of the Hamiltonian is large, bending the Laplace plane towards the ecliptic (Appendix~\ref{asec:eccLap}), and resulting in much shorter timescales than shown in Fig.~\ref{fig:LaplaceFreq}. Eccentric orbits have a much shorter revolution period of $\Omega_\mathrm{G}$ also when neglecting the planetary perturbations (function $P_{\Omega^*}$ of \citealp{HIGetal:07}). The behaviour of eccentric orbits is the subject of the next section.

\section{Exploration of the dynamics}\label{sec:explor}

   Since the Hamiltonian $\mathcal{F}$ (Eq.~\ref{eq:F}) is composed of two parts that dominate respectively in the low- and high-semi-major axis regimes (see Fig.~\ref{fig:eps}), the first step is to understand the two kinds of dynamics taken separately. Both $\varepsilon_{\mathrm{P}_2}\bar{\mathcal{H}}_{\mathrm{P}_2}$ and $\varepsilon_{\mathrm{G}_\mathrm{V}}\bar{\mathcal{H}}_{\mathrm{G}_\mathrm{V}}$ are integrable, and their dynamics are well known. In this section, we recall briefly their main aspects and study the interplay between the two kinds of perturbations.

   \subsection{Planetary regime}\label{sec:pladyn}
\begin{figure}
      \centering
      \includegraphics[width=0.8\columnwidth]{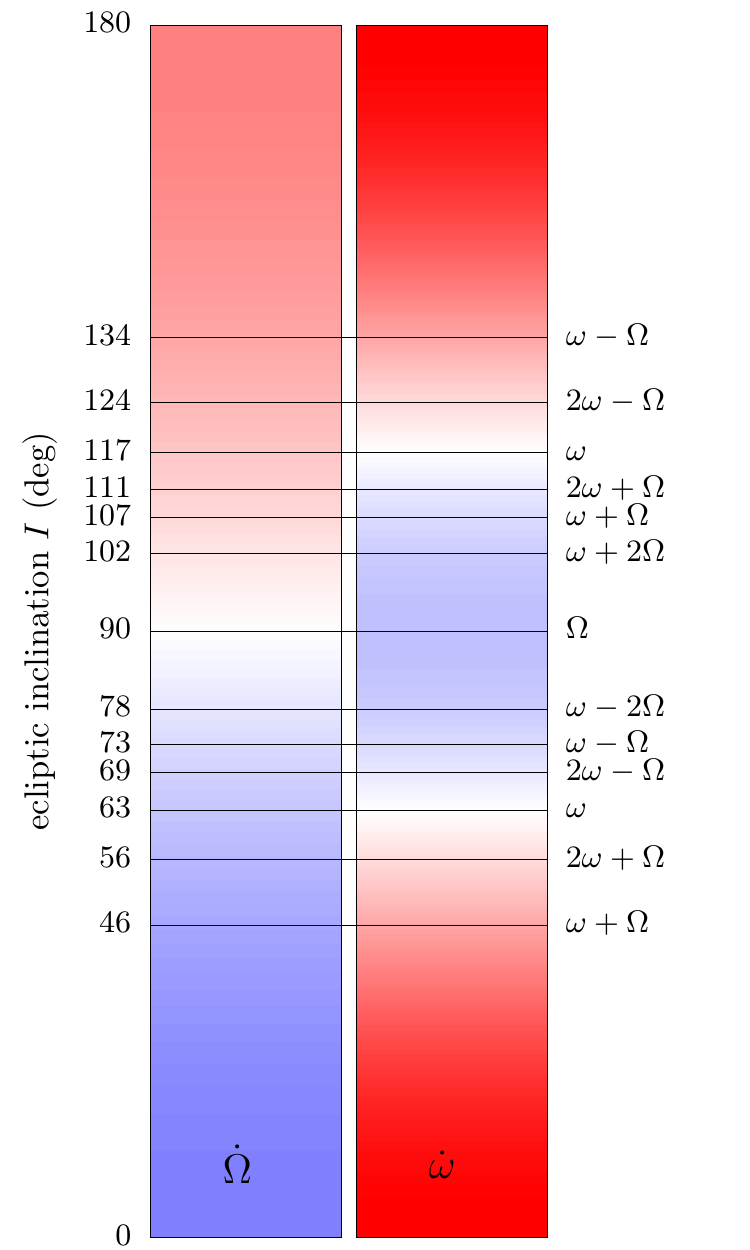}
      \caption{Precession velocity of $\Omega$ and $\omega$ in the planetary regime. The colour represents the velocity scale from negative values in blue to positive values in red, with the same colour scale for $\Omega$ and $\omega$. Both $\dot{\Omega}$ and $\dot{\omega}$ attain their maximum absolute value at $I=0^\text{o}$ and $180^\text{o}$. The locations where the integer combinations $k\dot{\omega}+j\dot{\Omega}$ vanish (limited to $\|k\|,\|j\|<3$) are shown by horizontal lines. The inclination values on the left are obtained from Eq.~\ref{eq:plregime}, and the corresponding constant angles are written on the right.}
      \label{fig:res}
   \end{figure}

   If $\varepsilon_{\mathrm{G}_\mathrm{V}}\ll\varepsilon_{\mathrm{P}_2}$, the dynamics is largely dominated by the planetary perturbations. Expressed in ecliptic coordinates, the Hamiltonian $\varepsilon_{\mathrm{P}_2}\bar{\mathcal{H}}_{\mathrm{P}_2}$ taken alone is trivially integrable  (see Eq.~\ref{eq:Hp}): the momenta are conserved, and the angles circulate with constant velocities. More specifically, $e$ and $I$ are constant, and the precession velocities are
   \begin{equation}\label{eq:plregime}
      \dot{\omega} = \varepsilon_{\mathrm{P}_2}\frac{3(5\cos^2I-1)}{8\sqrt{\mu a}(1-e^2)^2}
      \hspace{0.3cm};\hspace{0.3cm}
      \dot{\Omega} = \varepsilon_{\mathrm{P}_2}\frac{-3\cos I}{4\sqrt{\mu a}(1-e^2)^2} \,.
   \end{equation}
   The angle $\omega$ increases for $I<63^\text{o}$, decreases for $63^\text{o}<I<117^\text{o}$, and increases again for $I>117^\text{o}$. In contrast, $\Omega$ decreases for $I<90^\text{o}$ and increases for $I>90^\text{o}$. Figure~\ref{fig:res} shows a map of these precession velocities with respect to the ecliptic inclination, as well as the places where their main integer combinations vanish. These combinations cannot be called `resonances' at this stage, because the two degrees of freedom are strictly decoupled when considering the planetary perturbations alone, but we can expect that they will have a dynamical importance in the perturbed problem (see below). Some of these combinations are mentioned by \cite{GALetal:12} as affecting observed trans-Neptunian objects. As shown by \cite{SAIetal:16}, the hexadecapole (see Eq.~\ref{eq:Hp}) and successive planetary terms only make small libration islands of $\omega$ appear at $I\approx 63^\text{o}$ and $117^\text{o}$. These islands have a maximum width of $16.4$~au for the perihelion distance, which, for large semi-major axes, represents a very small variation of eccentricity. When expressed in galactic coordinates, the evolution of $I_\mathrm{G}$, $\omega_\mathrm{G}$ and $\Omega_\mathrm{G}$ driven by $\varepsilon_{\mathrm{P}_2}\bar{\mathcal{H}}_{\mathrm{P}_2}$ are combinations of sinusoids (see Appendix~\ref{asec:conv} for the conversion formulas).
   
   \subsection{Planetary regime weakly perturbed by galactic tides}\label{sec:res}
   When the planetary perturbations dominate over the galactic tides (i.e. for small semi-major axes and/or high eccentricities), the effects of galactic tides can be studied in a perturbative approach: the planetary component $\varepsilon_{\mathrm{P}_2}\bar{\mathcal{H}}_{\mathrm{P}_2}$ (see Eq.~\ref{eq:Hp}) acts as the integrable dominant part of $\mathcal{F}$, while the galactic component $\varepsilon_{\mathrm{G}_\mathrm{V}}\bar{\mathcal{H}}_{\mathrm{G}_\mathrm{V}}$ (see Eq.~\ref{eq:Hg}) acts as a small perturbation.
   
   Since our dominant part $\varepsilon_{\mathrm{P}_2}\bar{\mathcal{H}}_{\mathrm{P}_2}$ is already expressed in action-angle coordinates (i.e. it does not depend on the angles), the perturbative approach is straightforward. Expressed in ecliptic coordinates, our perturbing part $\varepsilon_{\mathrm{G}_\mathrm{V}}\bar{\mathcal{H}}_{\mathrm{G}_\mathrm{V}}$ is composed of several terms featuring various combinations of $\omega$ and $\Omega$ (see the complete list in Appendix~\ref{asec:conv}). Therefore, because of the galactic tides, such combinations become genuine resonances, whose characteristics can be obtained analytically. The procedure is detailed in Appendix~\ref{asec:pend}.
   
   Figures~\ref{fig:secres500} and \ref{fig:secres700} show the location and widths of all the strongest resonances (the ones that appear at first order in $\varepsilon_{\mathrm{G}_\mathrm{V}}$), obtained analytically. These figures are restricted to small perihelion distances, for the planets to remain by far the dominant term of the dynamics. We focus on prograde orbits, since the resonances for $I>90^\text{o}$ are obtained by replacing $\cos I$ by $-\cos I$ and $\Omega$ by $-\Omega$. As shown in the top panel of Figs.~\ref{fig:secres500}-\ref{fig:secres700}, the libration zone of $\Omega$ has by far the largest width in inclination (it actually corresponds to the emergence of the orthogonal Laplace equilibrium, see Sect.~\ref{sec:preliminary}). We note that the resonances $\omega-\Omega$ and $2\omega-\Omega$ are the first ones to overlap when the galactic tides increase (yellow and grey areas). As shown in the bottom panel of Figs.~\ref{fig:secres500}-\ref{fig:secres700}, the resonances $\omega+\Omega$ and $2\omega+\Omega$ are by far the largest ones in perihelion distance. The other resonances are quite small in comparison, and the libration zone of $\Omega$ even has a null width in~$q$. The resonances $\omega\pm 2\Omega$, visible in Fig.~\ref{fig:res}, do not even appear in Figs.~\ref{fig:secres500}-\ref{fig:secres700}: this means that they only exist at second order of $\varepsilon_{\mathrm{G}_\mathrm{V}}$, and have virtually no effect in the weakly perturbed planetary regime. As explained in Appendix~\ref{asec:pend}, in addition to resonances, the Hamiltonian function features a term that is responsible for the emergence of the classic Laplace plane: low-inclination orbits do not precess about the ecliptic pole, as in Sect.~\ref{sec:pladyn}, but about an inclined axis.
   
   As shown by Figs.~\ref{fig:secres500}-\ref{fig:secres700}, when we increase the semi-major axis or the perihelion distance, the resonances become very large and overlap massively. For overly large resonances, the whole dynamical structure outlined in this section is actually destroyed: the galactic tides cannot be treated as a small perturbation anymore.
   
   \begin{figure}
      \includegraphics[width=\columnwidth]{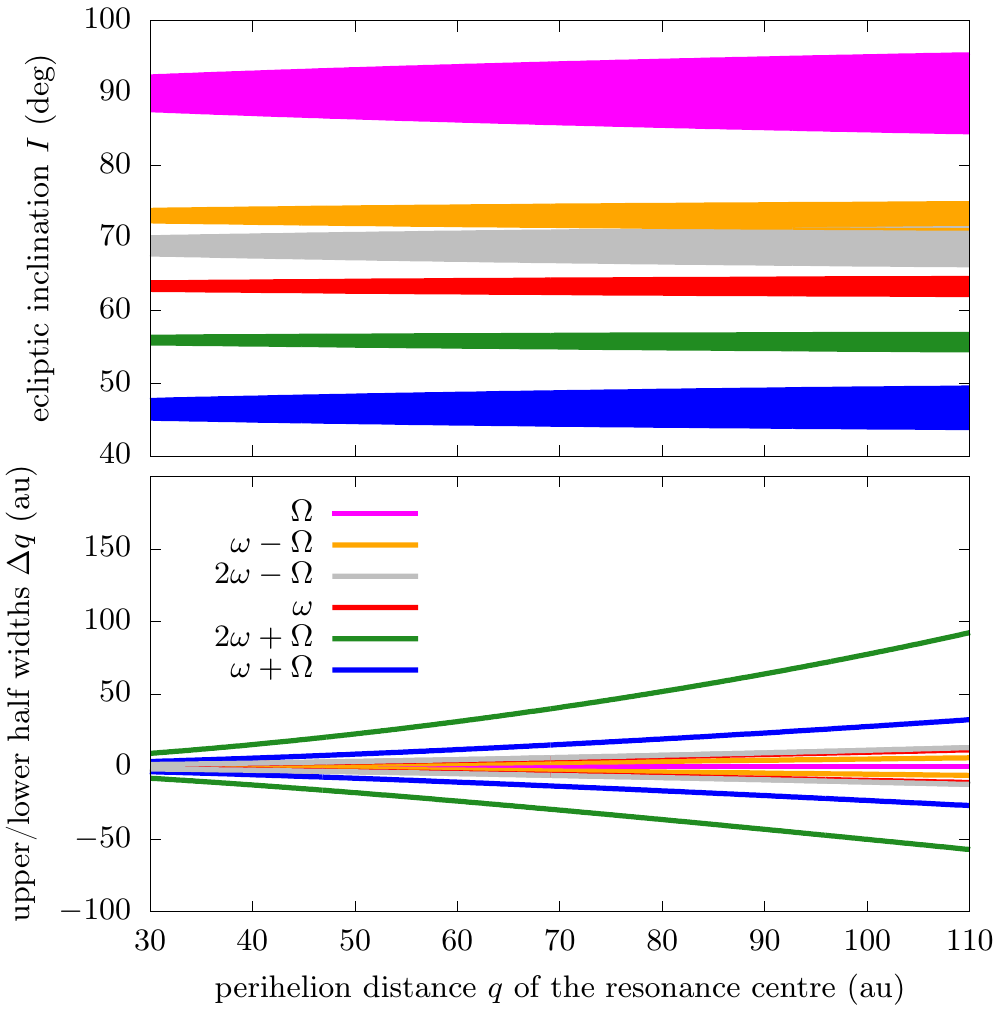}
      \caption{Location and widths of the strongest resonances in the planetary regime weakly perturbed by the galactic tides. The semi-major axis taken as parameter is $a=500$~au. For better visibility, the perihelion distance of the resonance centre is directly used as horizontal axis. \emph{Top:} location and width in inclination (filled areas). \emph{Bottom:} upper and lower half widths in perihelion distance, for a centre given by the horizontal axis.}
      \label{fig:secres500}
   \end{figure}
   
   \subsection{Galactic regime}\label{sec:galdyn}
   If $\varepsilon_{\mathrm{P}_2}\ll\varepsilon_{\mathrm{G}_\mathrm{V}}$, the dynamics is dominated by the galactic tides. The dynamics driven by $\varepsilon_{\mathrm{G}_\mathrm{V}}\bar{\mathcal{H}}_{\mathrm{G}_\mathrm{V}}$ taken alone has been studied by many authors. The solutions can actually be expressed analytically in terms of elliptic integrals (see \citealp{BRE.RAT:05}, \citealp{HIGetal:07}, \citealp{HIG.KOK:15} and references therein). The quantity
   \begin{equation}
      K = \sqrt{1-e^2}\cos I_\mathrm{G}
   \end{equation}
   is conserved and can be used as parameter in the Hamiltonian, which, in turns, has only one degree of freedom. Figure~\ref{fig:Hgalniv} shows the level curves of $\bar{\mathcal{H}}_{\mathrm{G}_\mathrm{V}}$ for different values of $K$. The limit $I_\mathrm{G}=0$ or $180^\text{o}$ is a stable fixed point whatever the eccentricity; it coincides with the classic Laplace plane in the large-$a$ regime (see Sect.~\ref{sec:preliminary}), and results in a frozen orbit. Using $K$ as parameter, this is equivalent to $e^2 = 1-K^2$ (border of the forbidden regions in Fig.~\ref{fig:Hgalniv}). The limit $e=0$ is a fixed point with circulating $\Omega_\mathrm{G}$, but it is unstable for $K^2<4/5$, that is, for $27^\text{o}<I_\mathrm{G}<153^\text{o}$. For $K^2<4/5$, there are two additional fixed points located at $\omega_\mathrm{G} = \pi/2$ or $3\pi/2$ and
   \begin{equation}
      e^2 = 1 - \frac{\sqrt{5K^2}}{2} \,,
   \end{equation}
  which is equivalent to the condition
   \begin{equation}
      e^2 = 1 - \frac{5}{4}\cos^2I_\mathrm{G} \,.
   \end{equation}
   These ones are stable, still with circulating $\Omega_\mathrm{G}$. Finally, as already noted by \cite{HIGetal:07}, the conservation of $K$ implies that the orbit cannot become retrograde if it is prograde, and vice versa. This is the same for the planetary perturbations, even if we go beyond the quadrupolar approximation \citep{SAIetal:16}, but this time this concerns the galactic inclination $I_\mathrm{G}$, not the ecliptic one $I$. Moreover, $\Omega_\mathrm{G}$ is always decreasing if $I_\mathrm{G}<90^\text{o}$ and always increasing if $I_\mathrm{G}>90^\text{o}$ (the period of its linear part, already mentioned in Sect.~\ref{sec:preliminary}, is noted $P_{\Omega^*}$ by \citealp{HIGetal:07}).
   
   \begin{figure}
      \includegraphics[width=\columnwidth]{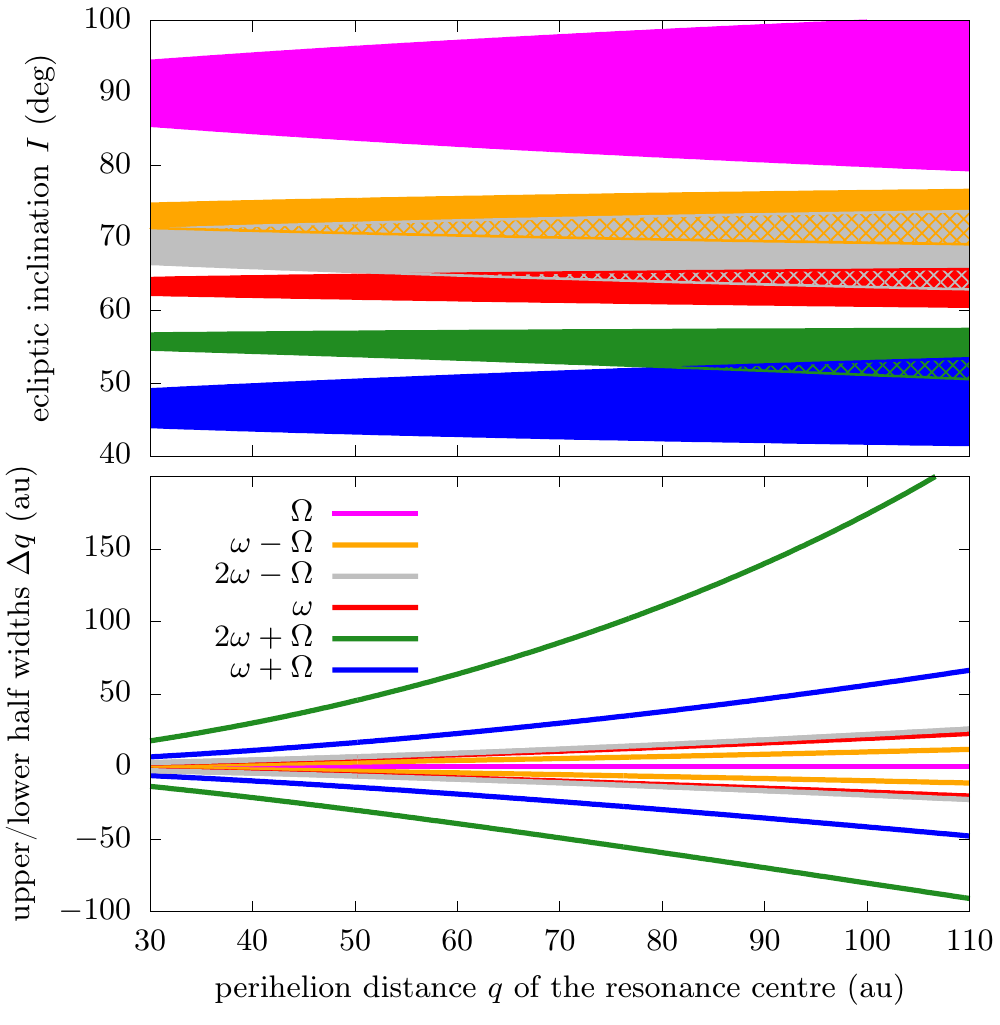}
      \caption{Same as Fig.~\ref{fig:secres500} for $a=700$~au. The hatched regions mean overlap.}
      \label{fig:secres700}
   \end{figure}
   
   \begin{figure}
      \centering
      \large
      \hspace{0.6cm}$K^2 = 0.5$
      \hspace{3.0cm}$K^2 = 0.9$\\
      \includegraphics[width=0.49\columnwidth]{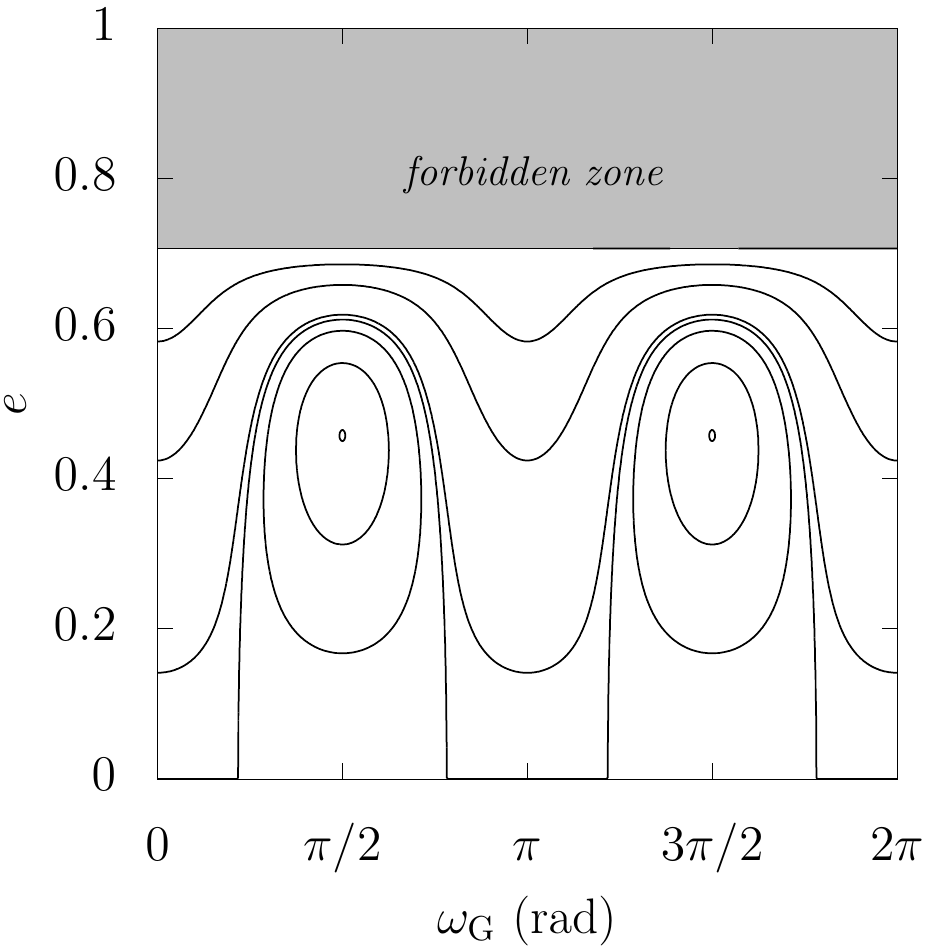}
      \includegraphics[width=0.49\columnwidth]{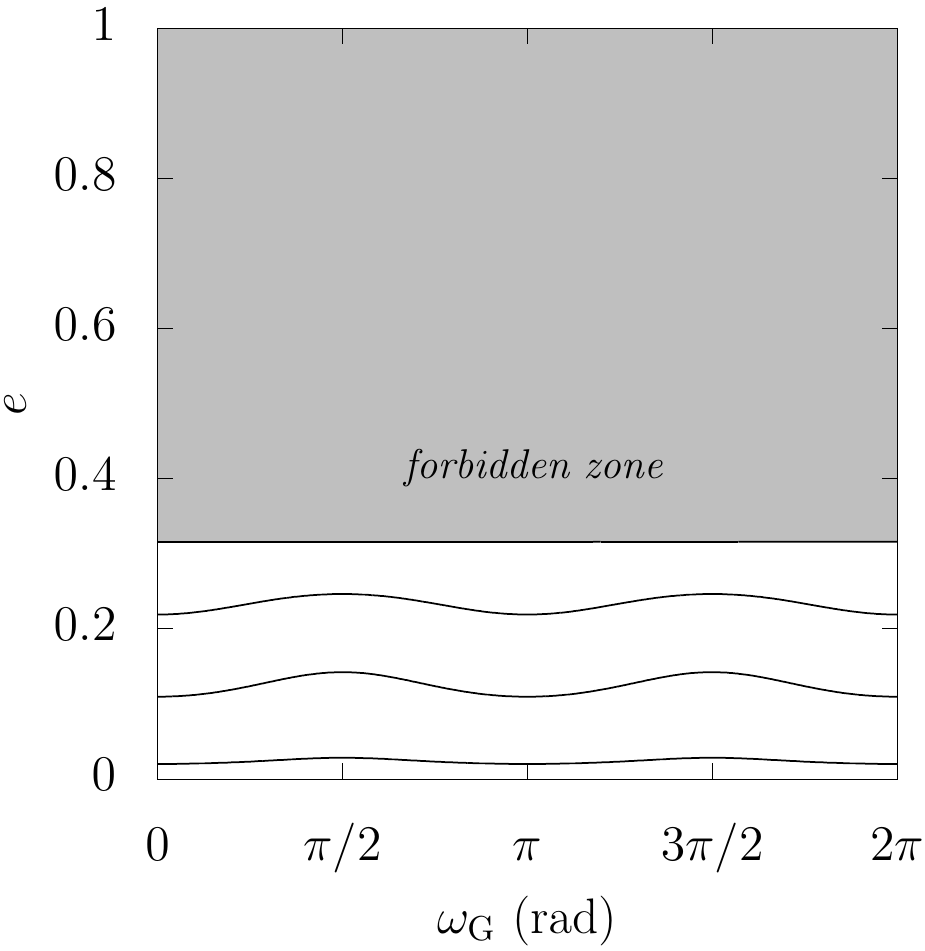} \\
      \includegraphics[width=0.49\columnwidth]{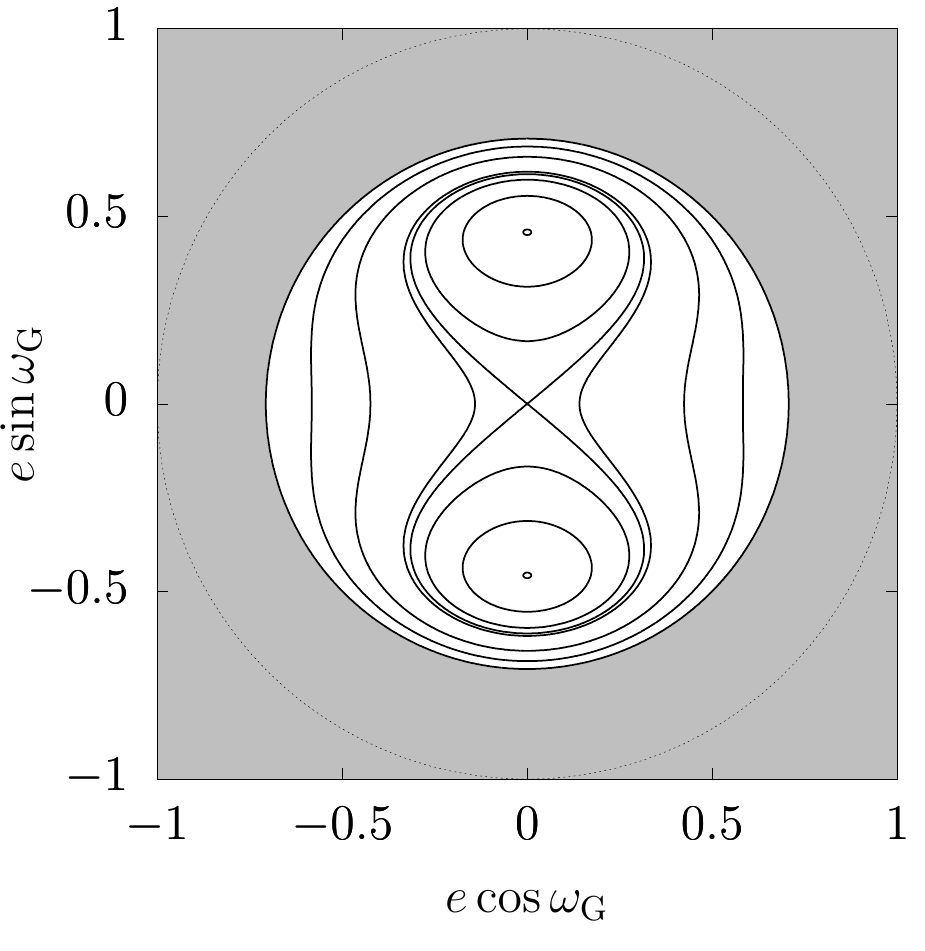}
      \includegraphics[width=0.49\columnwidth]{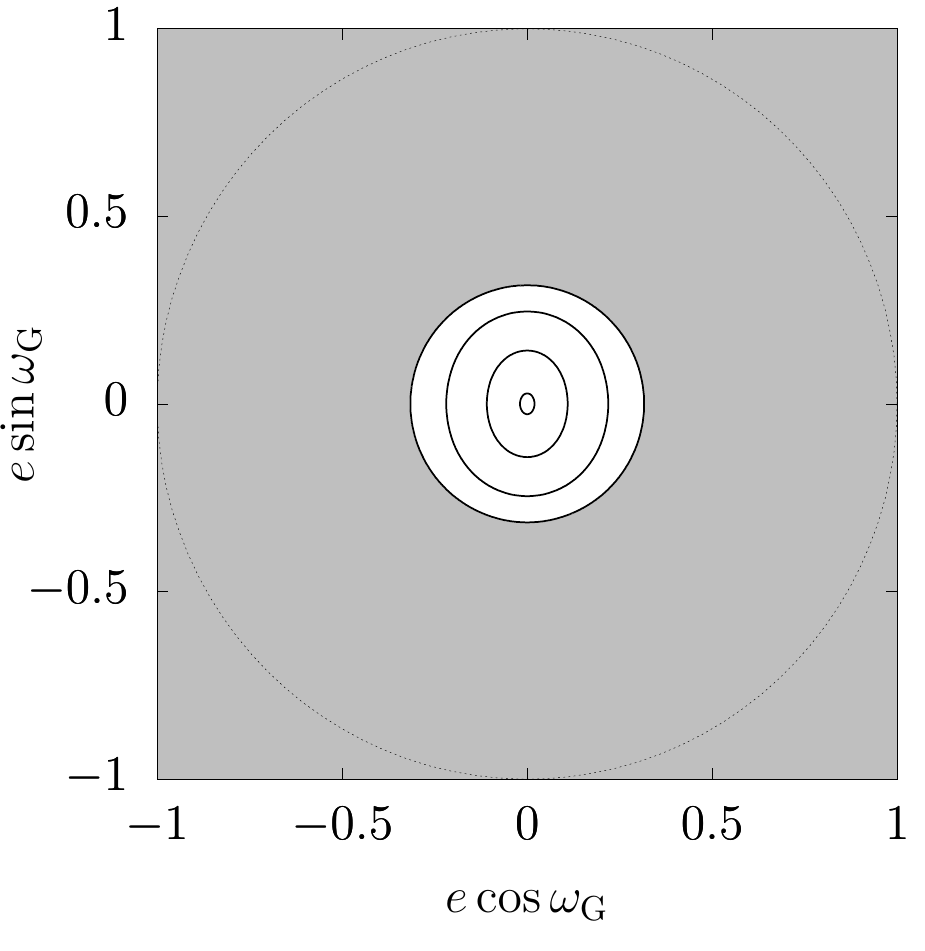}
      \caption{Level curves of the Hamiltonian function $\bar{\mathcal{H}}_{\mathrm{G}_\mathrm{V}}$ (Eq.~\ref{eq:Hg}). The parameters chosen are $K^2 < 4/5$ (left) and $K^2>4/5$ (right). The top and bottom rows show the same level curves for two sets of variables.}
      \label{fig:Hgalniv}
   \end{figure}
   
   \subsection{Intermediate non-integrable regime}\label{sec:intermed}
   In the intermediate regime (say, from $a\sim 500$ to $2000$~au, see Fig.~\ref{fig:eps}), the dynamics features two fully interacting degrees of freedom, represented by the two pairs of conjugate coordinates $(g,G)$ and $(h,H)$. The dynamics is chaotic in general, but can be explored through Poincar{\'e} sections, in the spirit of \cite{LIetal:14} and \cite{SAIetal:17b}. This method allows one to locate the regular trajectories and to determine the size of the chaotic zones. For $10$ values of $a$, and about $15$ values of the Hamiltonian spanning the different dynamical regimes for each value of $a$, we computed simultaneously four Poincar{\'e} sections (for increasing and decreasing $g$ and $h$). We give our conclusions below and show the most representative figures of our sample.
   
   \begin{figure}
      \centering
      \includegraphics[width=0.85\columnwidth]{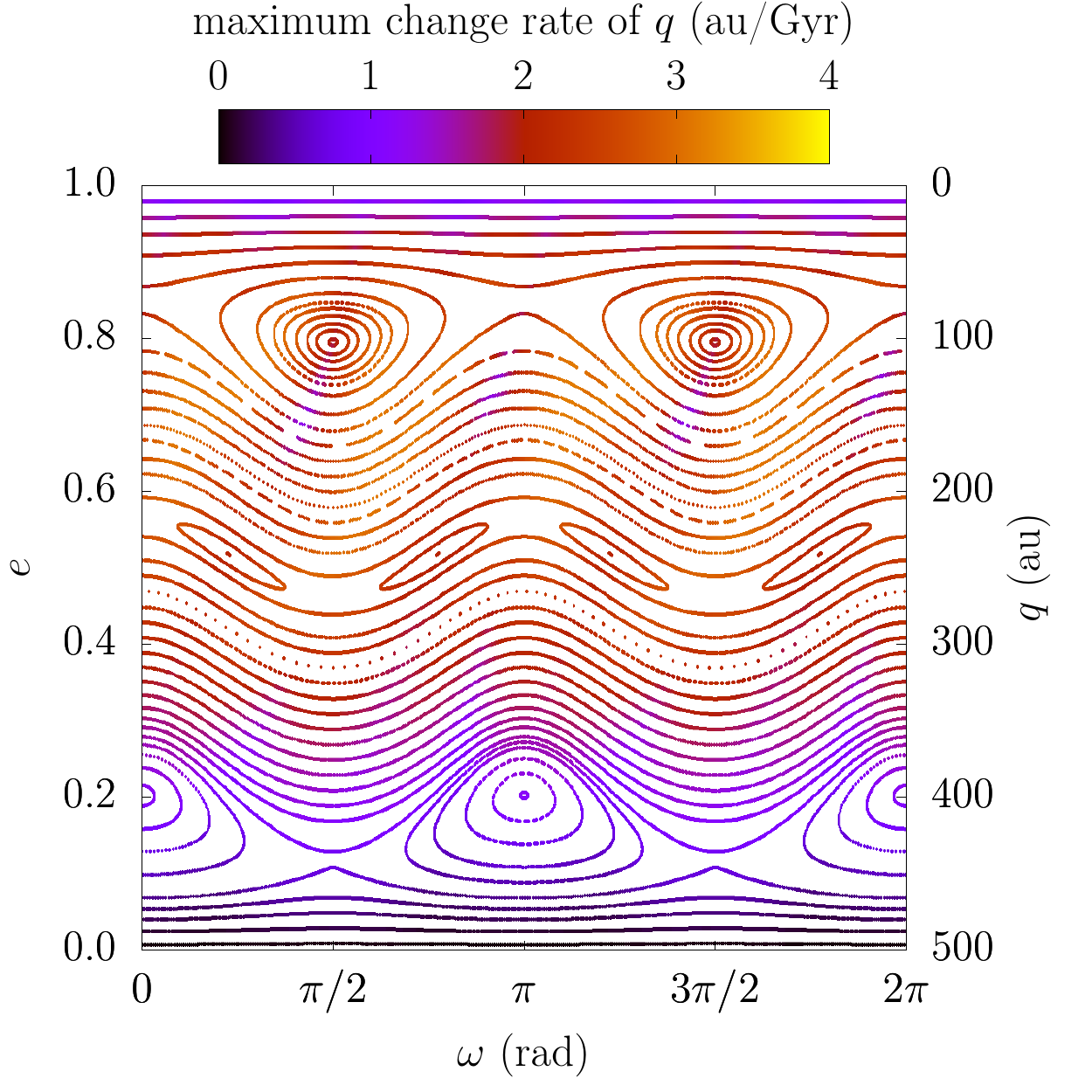}\\
      \vspace{0.2cm}
      \includegraphics[width=0.85\columnwidth]{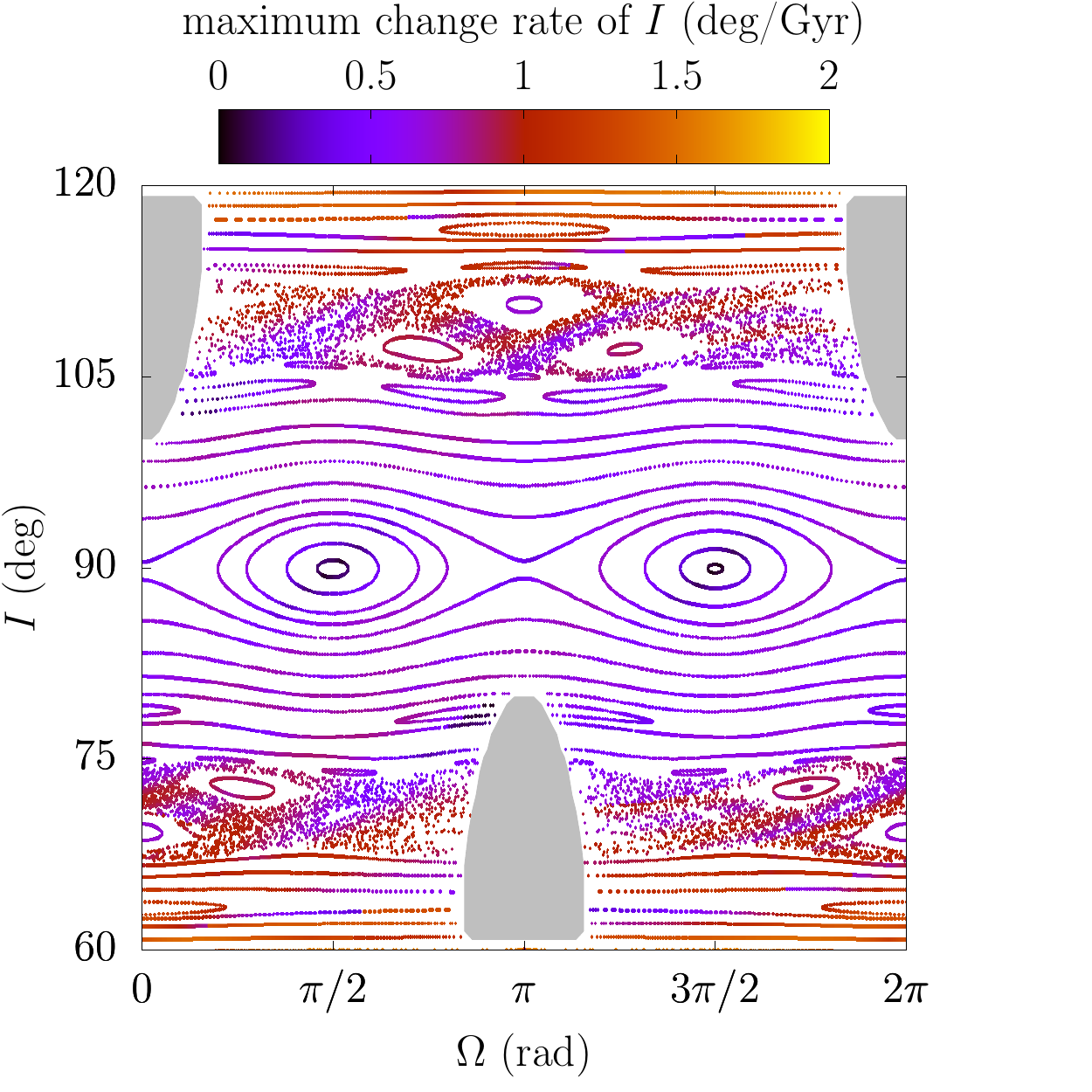}
      \caption{Poincar{\'e} sections of the dynamics driven by both the planetary and galactic perturbations (Hamiltonian from Eq.~\ref{eq:F}). The semi-major axis taken as parameter is $a=500$~au. The colour scale shows the maximum orbital change rate between two successive points (see titles), and the grey zones are forbidden. \emph{Top:} section for $\mathcal{F}=2\times 10^{-9}$~au$^2$yr$^{-2}$ at $\Omega_\mathrm{G}=0$ decreasing. The islands located at $e\approx 0.8$ are due to the resonance $2\omega+\Omega$, and the islands located at $e\approx 0.2$ are due to librations of $\omega$ while $I\approx 63^\text{o}$ (Lidov-Kozai mechanism). \emph{Bottom:} section for $\mathcal{F}=2\times 10^{-8}$~au$^2$yr$^{-2}$ at $\omega_\mathrm{G}=0$ decreasing. The chaotic bands are due to the overlap of the resonances $\omega-\Omega$ and $2\omega-\Omega$ (for $I<90^\text{o}$), or $\omega+\Omega$ and $2\omega+\Omega$ (for $I>90^\text{o}$), and the islands located at $I\approx 90^\text{o}$ are due to librations of $\Omega$ (orthogonal Laplace plane).}
      \label{fig:a500}
   \end{figure}
   
   \begin{figure*}
      \centering
      \includegraphics[width=0.85\columnwidth]{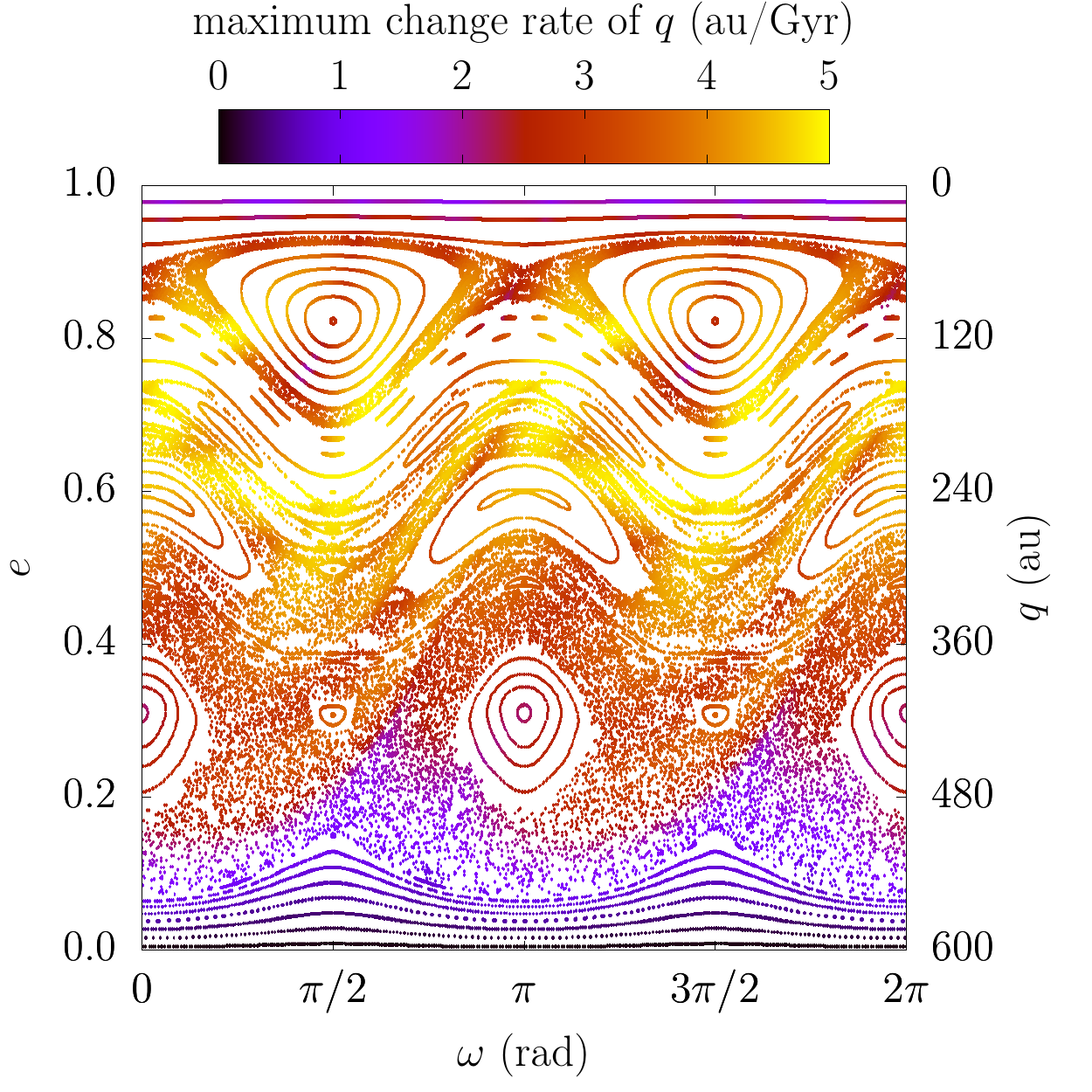}
      \hspace{1cm}
      \includegraphics[width=0.85\columnwidth]{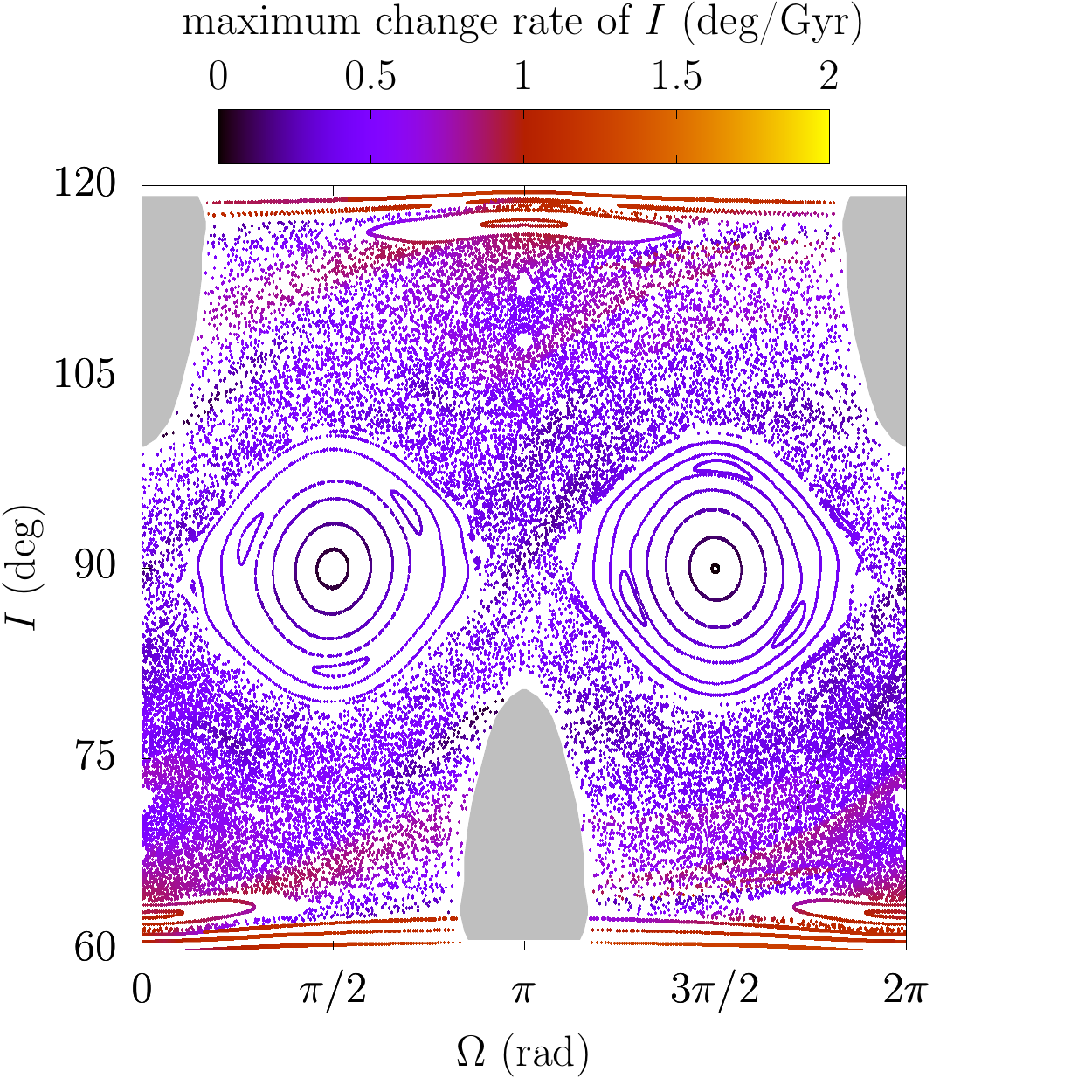}
      \caption{Same as Fig.~\ref{fig:a500} for $a=600$~au. \emph{Left:} section for $\mathcal{F}=2\times 10^{-9}$~au$^2$yr$^{-2}$ at $\Omega_\mathrm{G}=0$ decreasing. From top to bottom, the largest islands are due to the resonance $2\omega+\Omega$, the libration of $\omega$ at $I\approx 63^\text{o}$, and the resonance $2\omega-\Omega$. \emph{Right:} section for $\mathcal{F}=5\times 10^{-9}$~au$^2$yr$^{-2}$ at $\omega_\mathrm{G}=0$ decreasing. The islands located at $I\approx 90^\text{o}$ are due to librations of $\Omega$.}
      \label{fig:a600}
   \end{figure*}
   
   For semi-major axes smaller than $500$~au, the dynamics is dominated by the planetary perturbations, meaning that the eccentricity and ecliptic inclination do not vary much, while the angles $\omega$ and $\Omega$ circulate (see Sect.~\ref{sec:pladyn}). However, the two degrees of freedom now interact, meaning that genuine resonances appear between $\omega$ and $\Omega$ (see Sect.~\ref{sec:res}). For $a$ as small as $500$~au, Fig.~\ref{fig:a500} shows that such resonances allow quite large variations of the perihelion distance, but at the speed of only a few au per Gyr. When varying the fixed value of the Hamiltonian, we note that the resonances $\omega+\Omega$ and $2\omega+\Omega$ are by far the most prominent ones for prograde orbits (even if this is not immediately obvious with the parameters chosen to draw Fig.~\ref{fig:a500}). The same holds for the resonances $\omega-\Omega$ and $2\omega-\Omega$ for retrograde orbits. As expected, we also observe libration islands of $\omega$ at $I\approx 63^\text{o}$ and $117^\text{o}$, and libration islands of $\Omega$ at $I\approx 90^\text{o}$. Hence, the $90^\text{o}$ limit is not a barrier anymore for the inclination when considering the perturbed problem. Narrow chaotic regions are present (bottom graph of Fig.~\ref{fig:a500}), as predicted analytically in Sect.~\ref{sec:res}, but this chaos acts on extremely large timescales. We also notice thin resonances that we did not mention in Sect.~\ref{sec:res}: using a perturbative approach, such resonances are only of order $\varepsilon_{\mathrm{G}_\mathrm{V}}^2$ or more.
   
   When we increase the semi-major axis, chaos spreads near the separatrices of the main resonances and libration islands of $\omega$ and $\Omega$. Figure~\ref{fig:a600} shows that chaotic flips between prograde and retrograde orbits are possible for $a>600$~au, but the very long timescales involved make these flips of little practical interest.
   
\begin{figure*}
      \centering
      \includegraphics[width=0.85\columnwidth]{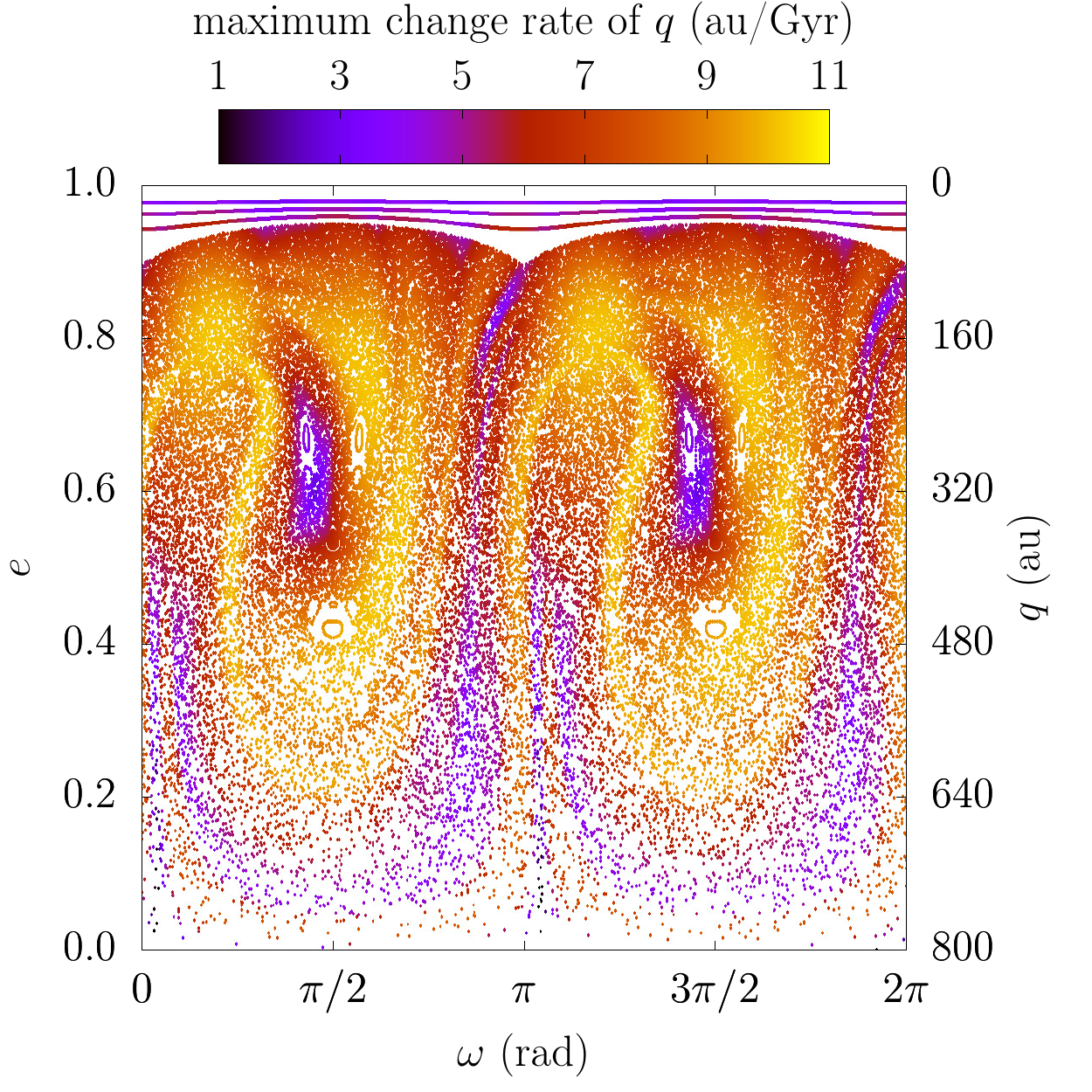}
      \hspace{1cm}
      \includegraphics[width=0.85\columnwidth]{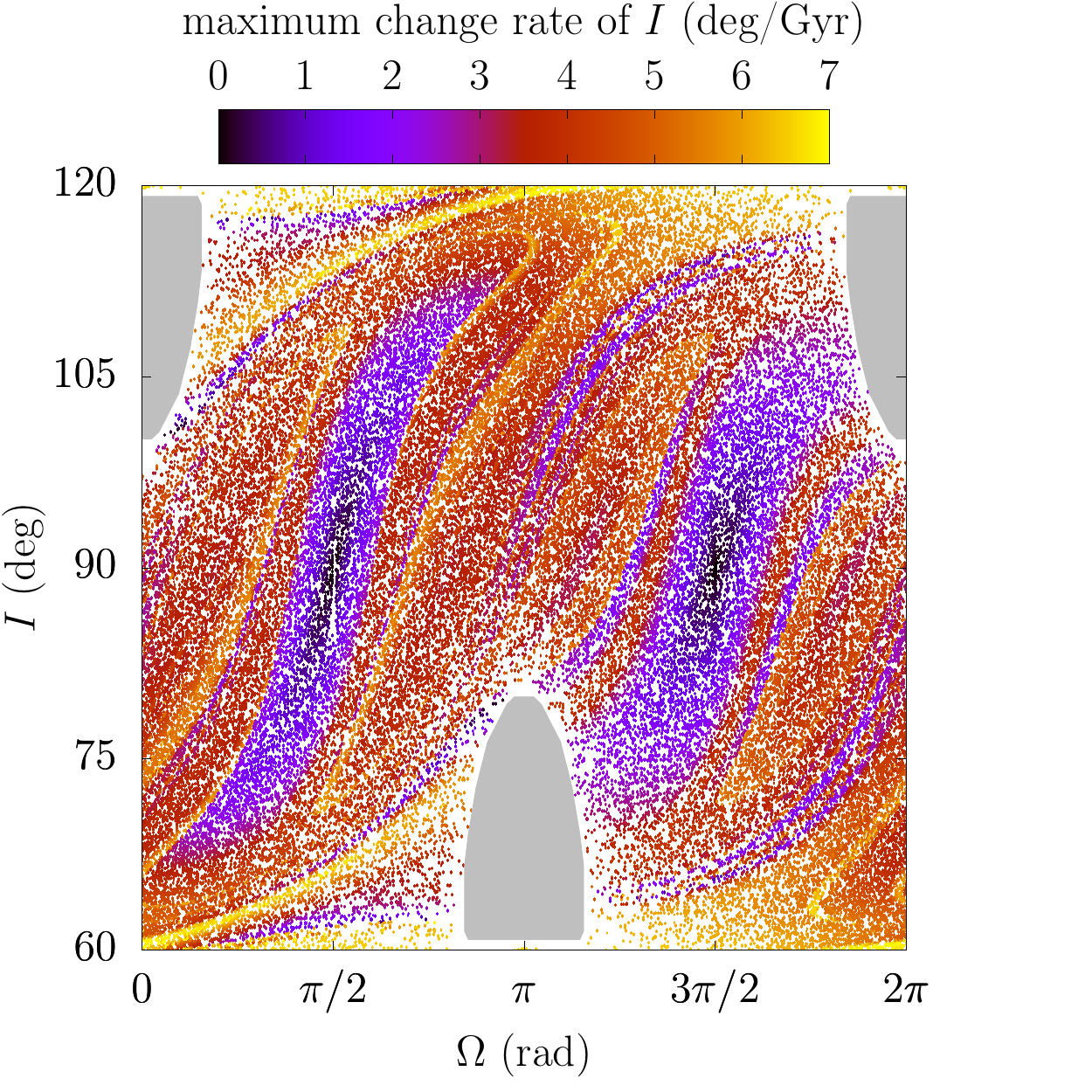}
      \caption{Same as Fig.~\ref{fig:a500} for $a=800$~au (left) and $a=1100$~au (right). \emph{Left:} section for $\mathcal{F}=1\times 10^{-9}$~au$^2$yr$^{-2}$ at $\Omega_\mathrm{G}=0$ decreasing. \emph{Right:} section for $\mathcal{F}=8\times 10^{-9}$~au$^2$yr$^{-2}$ at $\omega_\mathrm{G}=0$ decreasing.}
      \label{fig:a800}
   \end{figure*}

   For semi-major axes between about $800$ and $1100$, the phase space is almost completely filled with chaos (Fig.~\ref{fig:a800}). Stable trajectories only persist for very eccentric orbits, because they are governed almost entirely by the strong planetary perturbations. Moreover, the perihelion distance and the inclination evolve much faster than for $a<800$~au, making their variations substantial in a duration comparable to the age of the solar system. This confirms that eccentric orbits evolve much faster than circular orbits studied in Sect.~\ref{sec:preliminary}.
   
   \begin{figure*}
      \centering
      \includegraphics[width=0.85\columnwidth]{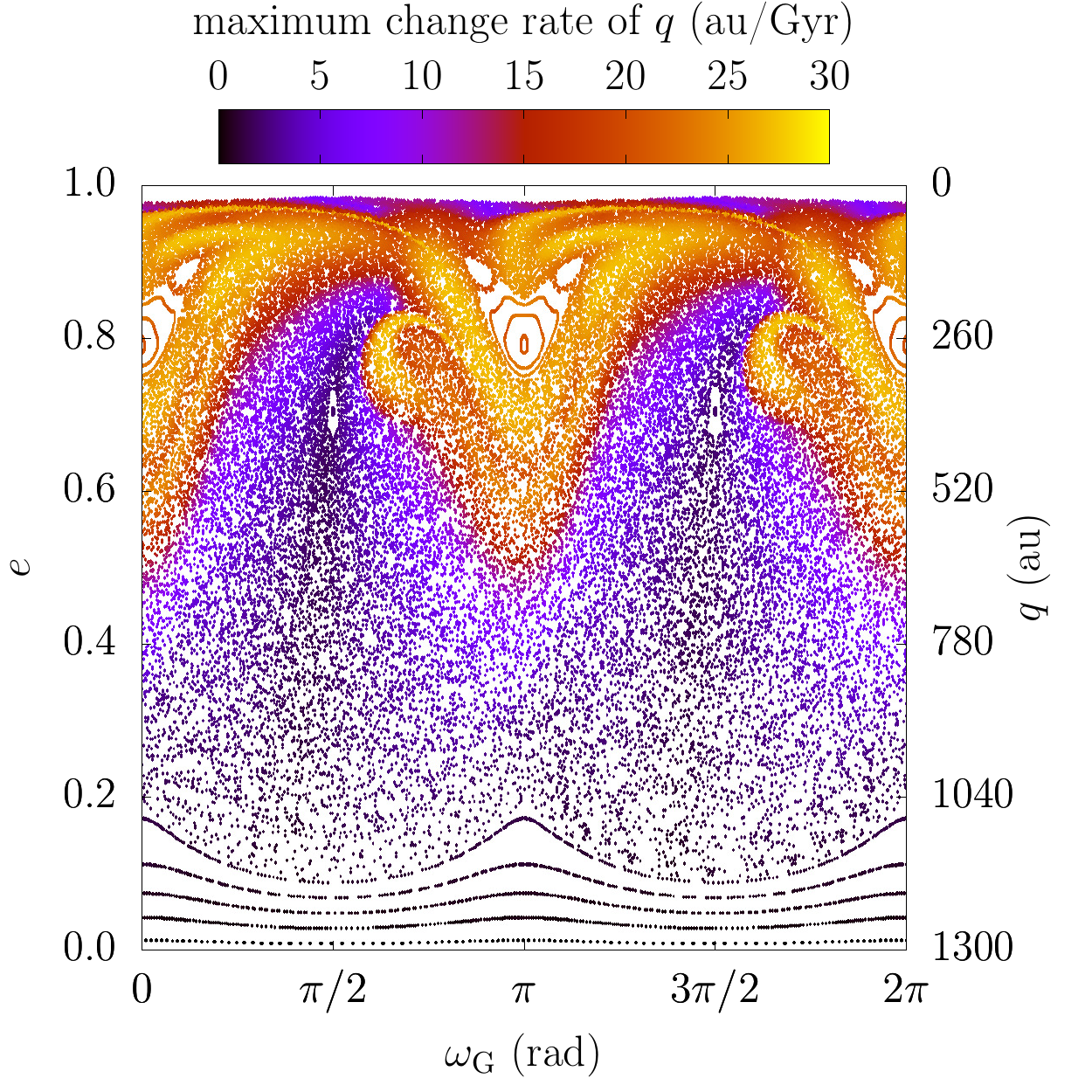}
      \hspace{1cm}
      \includegraphics[width=0.85\columnwidth]{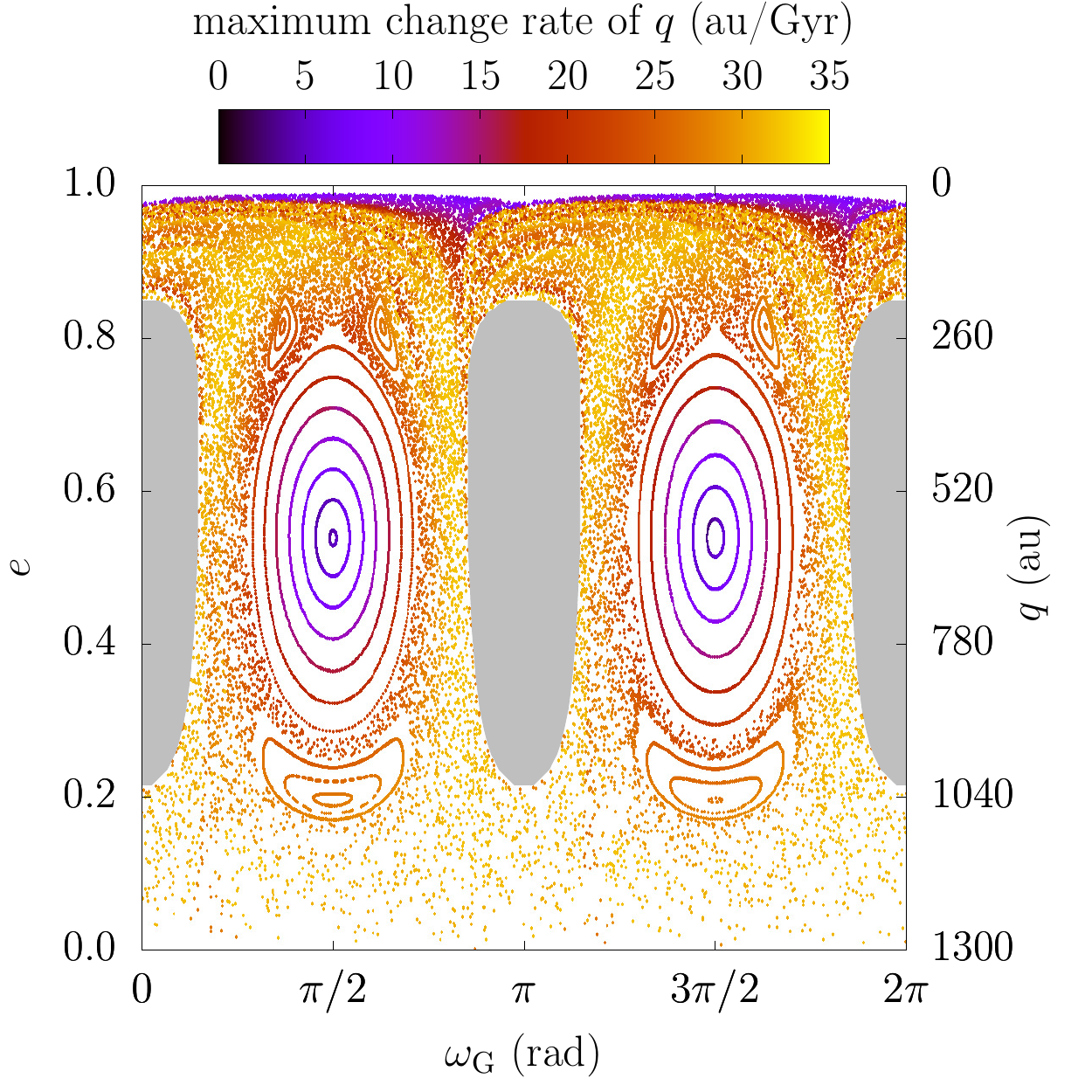}
      \caption{Same as Fig.~\ref{fig:a500} for $a=1300$~au. \emph{Left:} section for $\mathcal{F}=1\times 10^{-10}$~au$^2$yr$^{-2}$ at $\Omega_\mathrm{G}=0$ decreasing. The upper islands are due to librations of $\omega_\mathrm{G}$ with $e^2\approx 1-K^2$. \emph{Right:} section for $\mathcal{F}=2\times 10^{-9}$~au$^2$yr$^{-2}$ at $\Omega_\mathrm{G}=0$ decreasing. The islands are due to librations of $\omega_\mathrm{G}$ and oscillations of $e^2$ around $1-\sqrt{5K^2}/2$.}
      \label{fig:a1300}
   \end{figure*}
   
\begin{figure*}
      \centering
      \includegraphics[width=0.85\columnwidth]{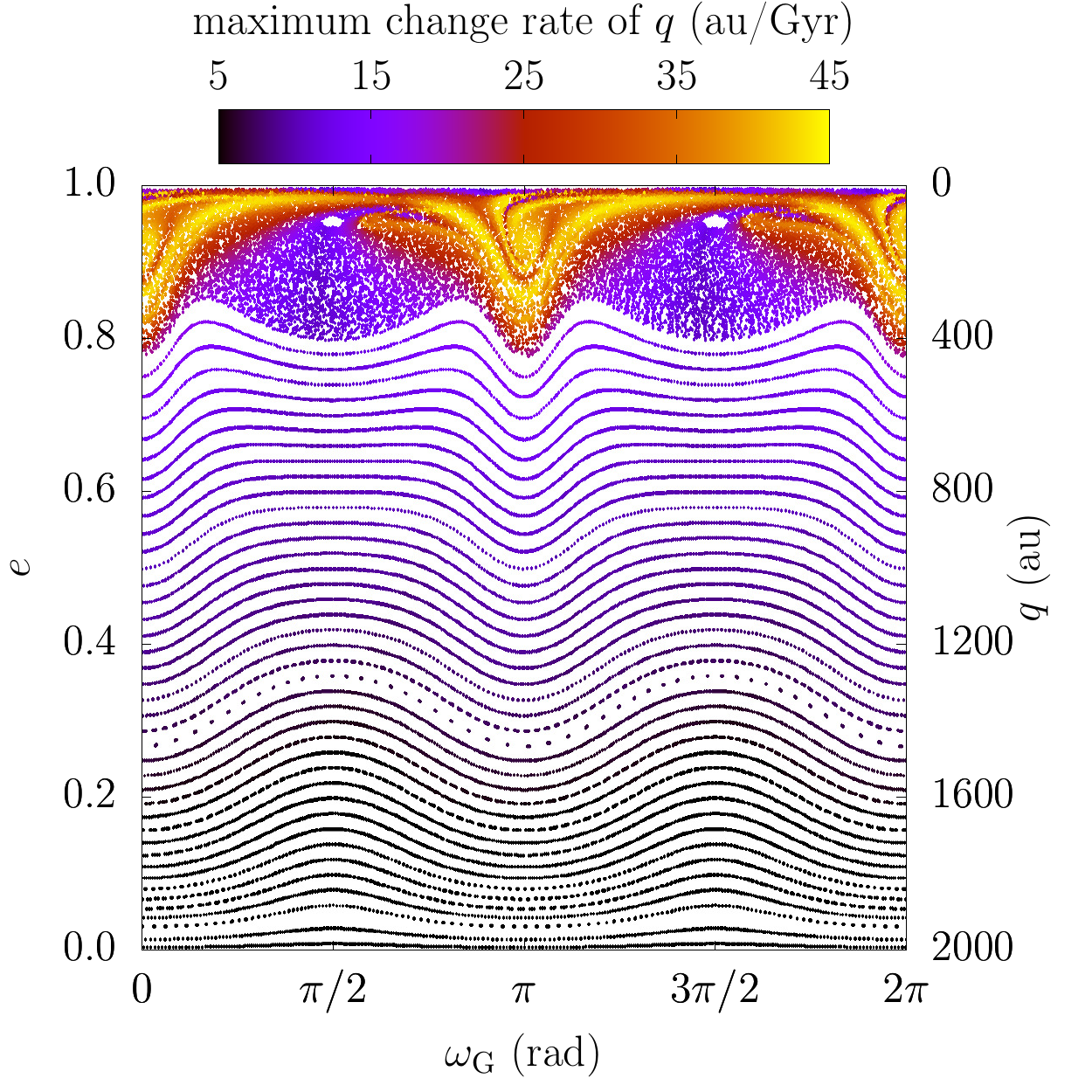}
      \hspace{1cm}
      \includegraphics[width=0.85\columnwidth]{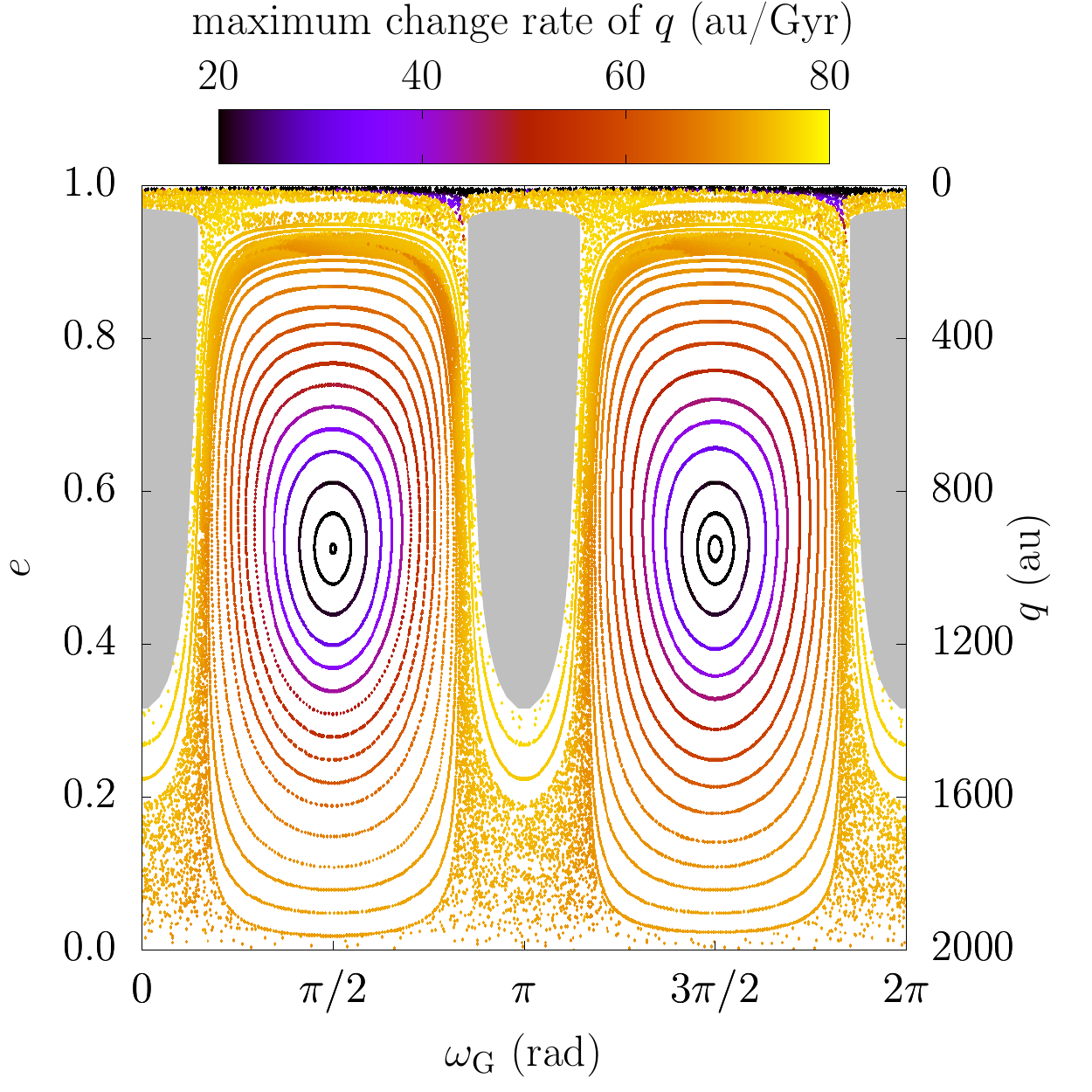}
      \caption{Same as Fig.~\ref{fig:a500} for $a=2000$~au. \emph{Left:} section for $\mathcal{F}=7\times 10^{-10}$~au$^2$yr$^{-2}$ at $\Omega_\mathrm{G}=0$ decreasing. \emph{Right:} section for $\mathcal{F}=5\times 10^{-9}$~au$^2$yr$^{-2}$ at $\Omega_\mathrm{G}=0$ decreasing. The big islands are due to librations of $\omega_\mathrm{G}$ and oscillations of $e^2$ around $1-\sqrt{5K^2}/2$.}
      \label{fig:a2000}
   \end{figure*}

   Finally, Figs.~\ref{fig:a1300} and \ref{fig:a2000} show that when the semi-major axis exceeds about $1300$~au, the galactic structure of the phase space emerges from the chaotic sea and progressively dominate. Resonances and libration islands of $\omega$ and $\Omega$ disappear, and the situation is now better characterised in galactic coordinates: we retrieve the structure described in Sect.~\ref{sec:galdyn} and illustrated in Fig.~\ref{fig:Hgalniv} for $K^2$ smaller or larger than $4/5$. This means that the ecliptic inclination $I$ oscillate with a very large amplitude while the ecliptic longitude of ascending node $\Omega$ stays around $0$ or $\pi$, as expected from the Laplace plane (Sect.~\ref{sec:preliminary}). It should be noted, however, that whatever the value of the semi-major axis, there is always a chaotic region for very eccentric orbits, and a stable region for even higher $e$, because the denominator of $\bar{\mathcal{H}}_{\mathrm{P}_2}$ diverges and makes the planetary perturbation dominate again (see Eq.~\ref{eq:Hp}).
   
\section{Implications for real objects}\label{sec:app}
   \subsection{Dynamics of known inert-Oort-cloud bodies}\label{sec:real}
\begin{figure*}
      \centering
      \includegraphics[width=0.85\columnwidth]{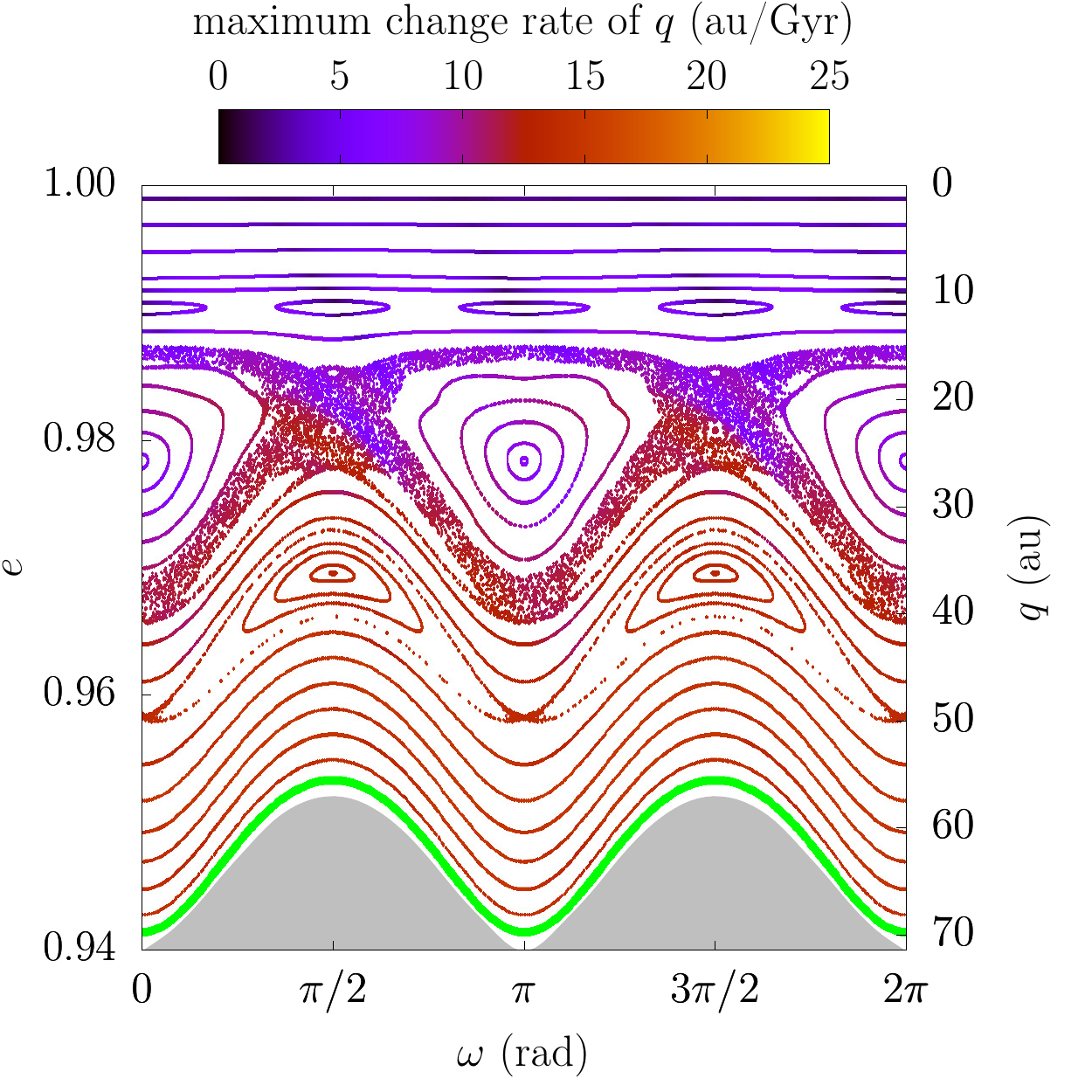}
      \hspace{1cm}
      \includegraphics[width=0.85\columnwidth]{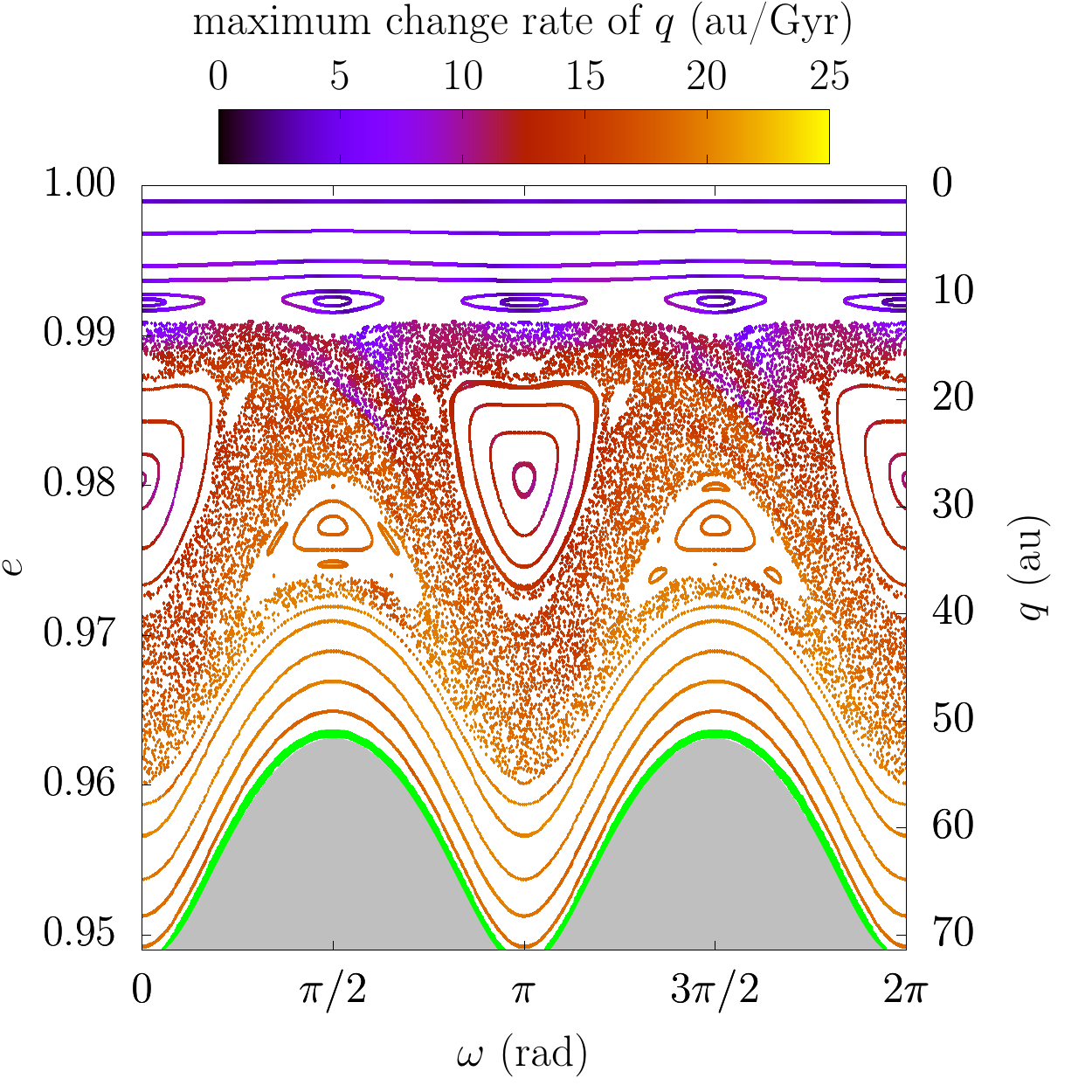}
      \caption{Same as Fig.~\ref{fig:a500} for the parameters of $2015$\,TG$_{387}$. The points of its trajectory that cross the section are represented in green. \emph{Left:} nominal orbital elements given by \cite{SHEetal:19}. \emph{Right:} semi-major axis increased by $3\sigma$. In both panels, the largest islands are due to the resonances $\omega+\Omega$ (above) and $2\omega+3\Omega$ (below).}
      \label{fig:2015TG387}
   \end{figure*}

   The dynamics of the three observed members of the inert Oort cloud (Sedna, $2012$\,VP$_{113}$, and $2015$\,TG$_{387}$) are known to be stable, even though \cite{SHEetal:19} mentioned that $2015$\,TG$_{387}$ is at the limit of destabilisation by galactic tides (one of its numerical clones was ejected from the solar system). As expected from previous results, Poincar{\'e} sections computed in the vicinity of the orbit of $2012$\,VP$_{113}$ ($a\approx270$~au) show that $e$ and $I$ are almost constant while $\omega$ and $\Omega$ circulate. This is also the case for Sedna ($a\approx 540$~au), despite its larger semi-major axis, because its orbital elements, and in particular its low inclination with respect to the ecliptic, make it unable to reach any of the features shown in Fig.~\ref{fig:a500}. The case of $2015$\,TG$_{387}$ is more interesting ($a\approx 1190$~au), because its high semi-major axis corresponds to the region where both the planetary and galactic perturbations have substantial effects. Figure~\ref{fig:2015TG387} shows that $2015$\,TG$_{387}$ is not far from a chaotic zone surrounding the  $\omega+\Omega$ resonance. Its trajectory, however, is strictly quasi-periodic in our simplified model. Increasing the value of its semi-major axis in the uncertainty range leads to faster oscillations of $q$ with a larger amplitude (see Fig.~\ref{fig:2015TG387}, right panel), but $2015$\,TG$_{387}$ is unable to reach the nearest chaotic zone. The unlikely ejection path found by \cite{SHEetal:19} involves the scattering effects of the giant planets, triggered when the perihelion distance reaches its minimum (see Fig.~\ref{fig:inertOC}).
   
\begin{figure*}
      \centering
      \includegraphics[width=0.85\columnwidth]{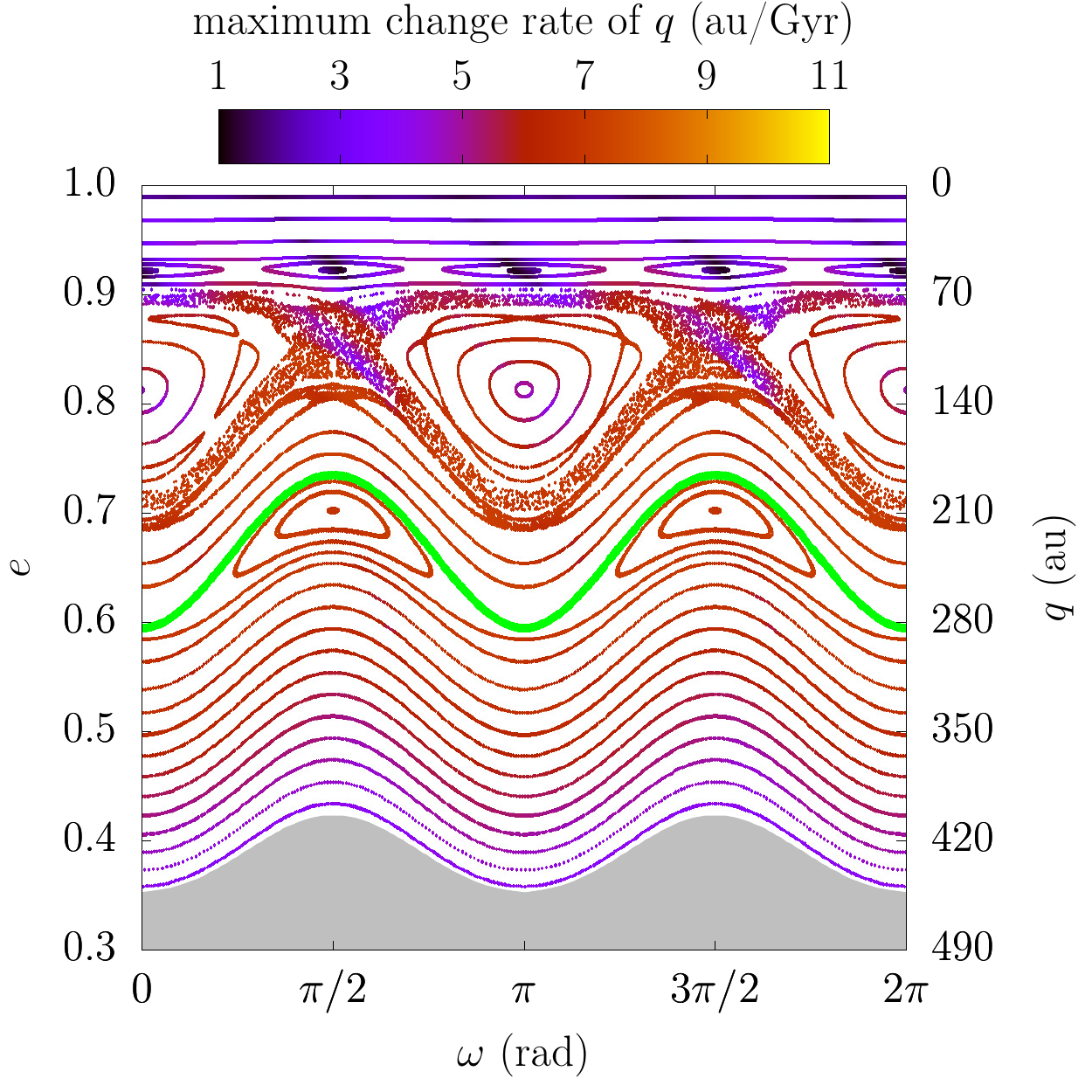}
      \hspace{1cm}
      \includegraphics[width=0.85\columnwidth]{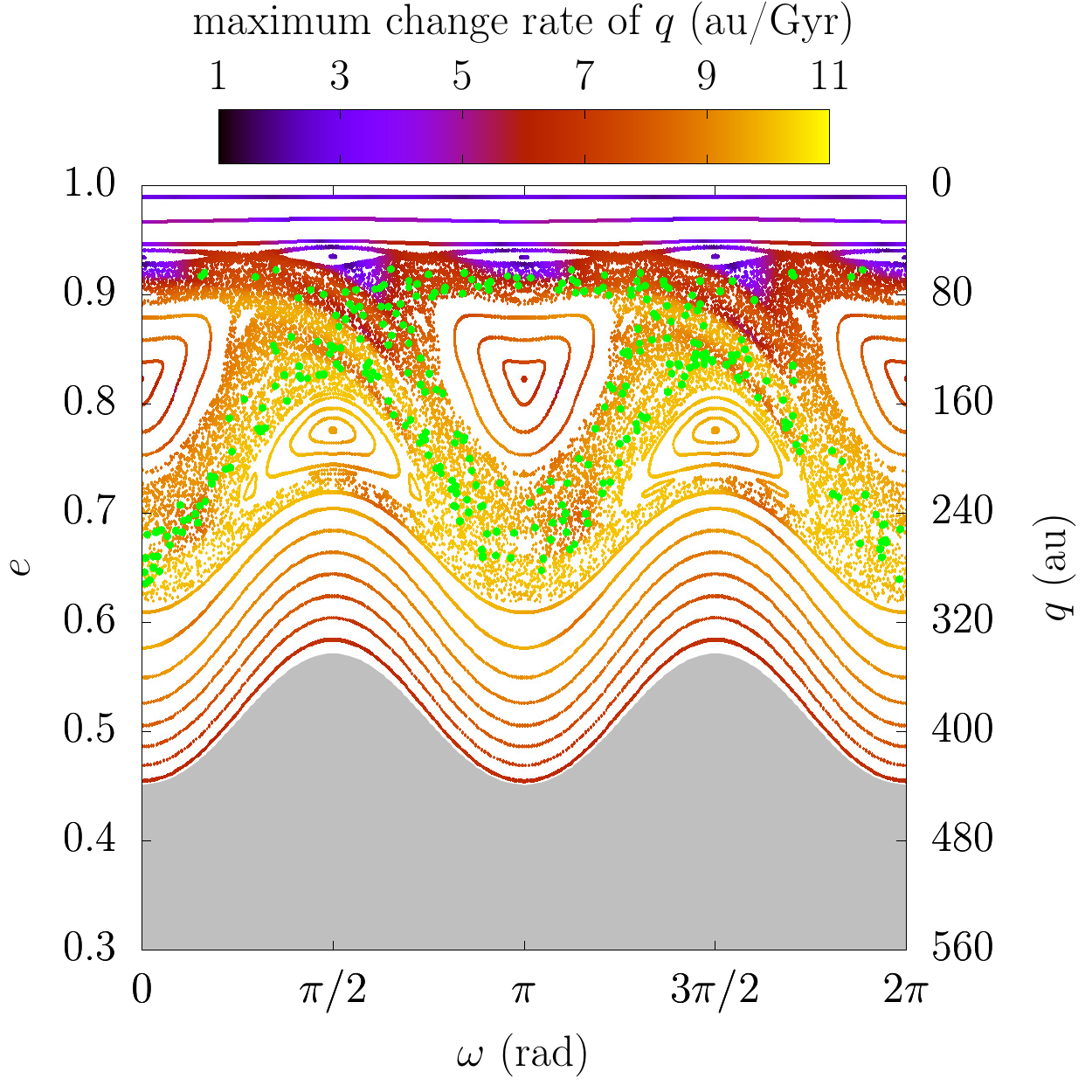}
      \caption{Same as Fig.~\ref{fig:2015TG387} for the parameters of the Planet 9 hypothesised by \cite{BAT.BRO:16}. \emph{Left:} nominal orbital elements (see text). \emph{Right:} semi-major axis increased by $100$~au. In both panels, the largest islands are due to the resonances $\omega+\Omega$ (above) and $2\omega+3\Omega$ (below).}
      \label{fig:P9}
   \end{figure*}

   In addition to its few observed members, the inert Oort cloud could also contain the hypothetical Planet 9 (`P9') proposed by \cite{BAT.BRO:16}. Fixing its initial conditions to the nominal orbital elements adopted for instance by \cite{FIEetal:16}, in particular $a=700$~au, $q=280$~au, and $I=30^\text{o}$, we obtain the left panel of Fig.~\ref{fig:P9}. The situation is similar to that of $2015$\,TG$_{387}$, meaning that the trajectory of P9 is regular but close to a chaotic zone emerging from the $\omega+\Omega$ resonance. Since P9's orbit is still quite undetermined (if P9 ever exists), we cannot rule out the possibility that it actually lies inside the chaotic zone. We did not investigate in detail the structure of the chaos in the vicinity of P9's orbit, but we can mention that the chaos reaches P9 if we increase its semi-major axis by only $50$~au. For $a=800$~au, the chaotic region extents down to $q=45$~au (right graph of Fig.~\ref{fig:P9}), and for $a=900$~au it extents down to Neptune-crossing orbits. The long timescale involved would prevent P9 to actually encounter Neptune in $4.5$~Gyrs, but its perihelion distance could anyway vary quite substantially, possibly modifying over time the characteristics of its shepherding effect on distant trans-Neptunian objects. This remains true for the updated P9 orbit obtained by \cite{BATetal:19}, even though it has a somewhat smaller semi-major axis.
   
   \subsection{Evolution of a sample of objects over 4.5~Gyrs}\label{sec:samples}
   The exploration of the inert-Oort-cloud dynamics conducted in Sect.~\ref{sec:explor} allowed us to characterise the structure of the phase space in this region, including the location of the chaotic regions. This structure should appear as imprints in the orbital distribution of a large sample of small bodies. For instance, the existence of a Laplace plane (Sect.~\ref{sec:preliminary}) that is distinct from the ecliptic should naturally produce an accumulation of $\Omega$ near the ascending node of the galactic plane (here located at $\Omega=\pi$). Moreover, we found in Sect.~\ref{sec:intermed} that the combination $\varpi=\omega+\Omega$ is among the strongest resonances in the transitional regime, which should preferentially orient $\varpi$ around $\pm \pi/2$ minus the galactic node. However, this picture was drawn for an infinite timescale, and numerous trajectories flagged as chaotic in the Poincar{\'e} sections actually wander over the chaotic zones in a dramatically long timescale, even though no dynamical barrier prevents these trajectories from freely wandering around. In order to determine which of the dynamical structures could be discernible during a time span restricted to the age of the solar system, we monitor the evolution of a swarm of test particles over $4.5$~Gyrs. The simplicity of the system under study (Eq.~\ref{eq:F}) allows for the propagation of millions of trajectories in a reasonable amount of computation time.
   
   Since we aim to fully explore the parameter space, rather than modelling a realistic population of trans-Neptunian objects, we do not restrict our sample to a particular distribution. Our setup is organised as follows. At first, we use a uniform distribution of semi-major axis $a$ between $100$~au and $2000$~au. In order to manipulate easily understandable variables, we build our initial conditions in ecliptic coordinates. The angles $\omega$ and $\Omega$ are set uniformly between $0$ and $2\pi$, and we structure our exploration in slices of perihelion distance (e.g. $q\in[40,60]$~au, $[60,80]$~au, etc.) and slices of ecliptic inclination cosine (e.g. $\cos I\in[0.9,1]$, $[0.8,0.9]$, etc.). Each slice is uniformly populated by a sample of $10^5$ test particles, which are integrated numerically for $4.5$~Gyrs using Hamilton's equations of motion applied to the Hamiltonian function from Eq.~\eqref{eq:F}. These propagations still do not contain the planetary scattering, active for low perihelion distances (see Fig.~\ref{fig:inertOC}), that would add fuzziness in our distributions.
   
   \begin{figure}
      \centering
      \includegraphics[width=\columnwidth]{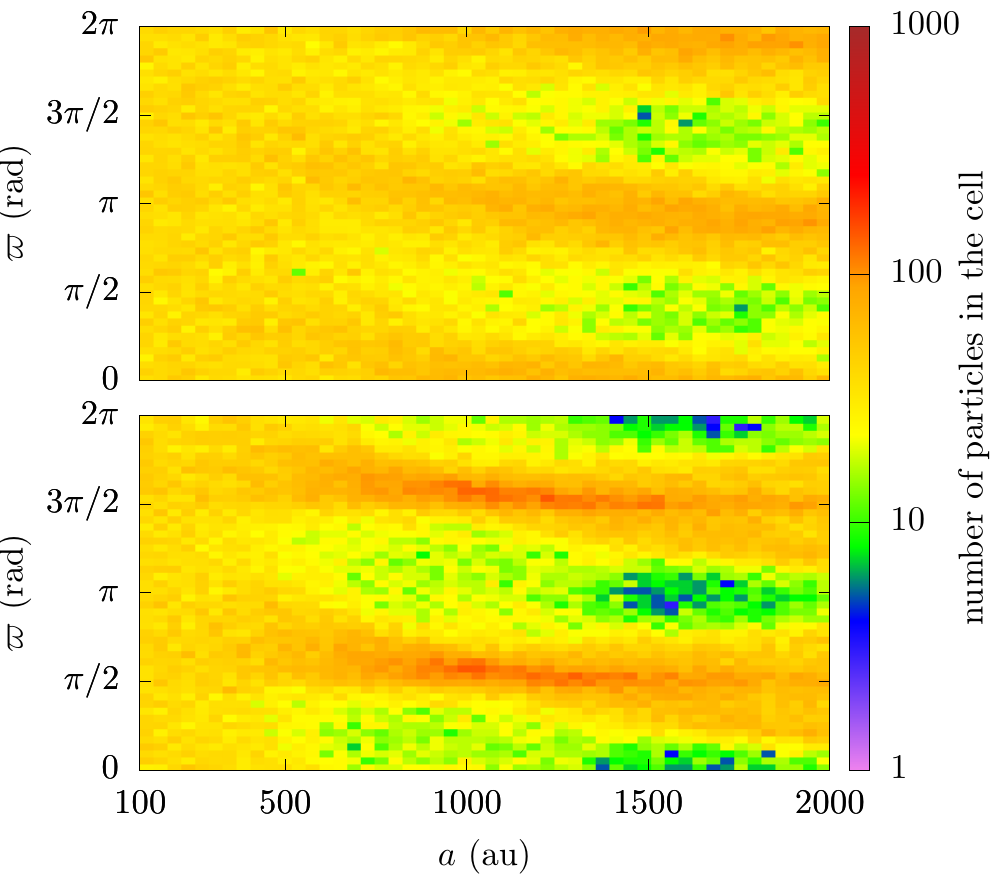}
      \caption{Density of particles in the plane $(a,\varpi=\omega+\Omega)$ after $4.5$~Gyrs for two slices of initial conditions taken as examples. \emph{Top:} initial values $q\in[40,60]$~au, and $\cos I\in[0.7,0.8]$. \emph{Bottom:} initial values $q\in[40,60]$~au, and $\cos I\in[-0.1,0]$. See text for the remaining initial orbital elements.}
      \label{fig:varpiE}
   \end{figure}
   
\begin{figure}
      \centering
      \includegraphics[width=0.95\columnwidth]{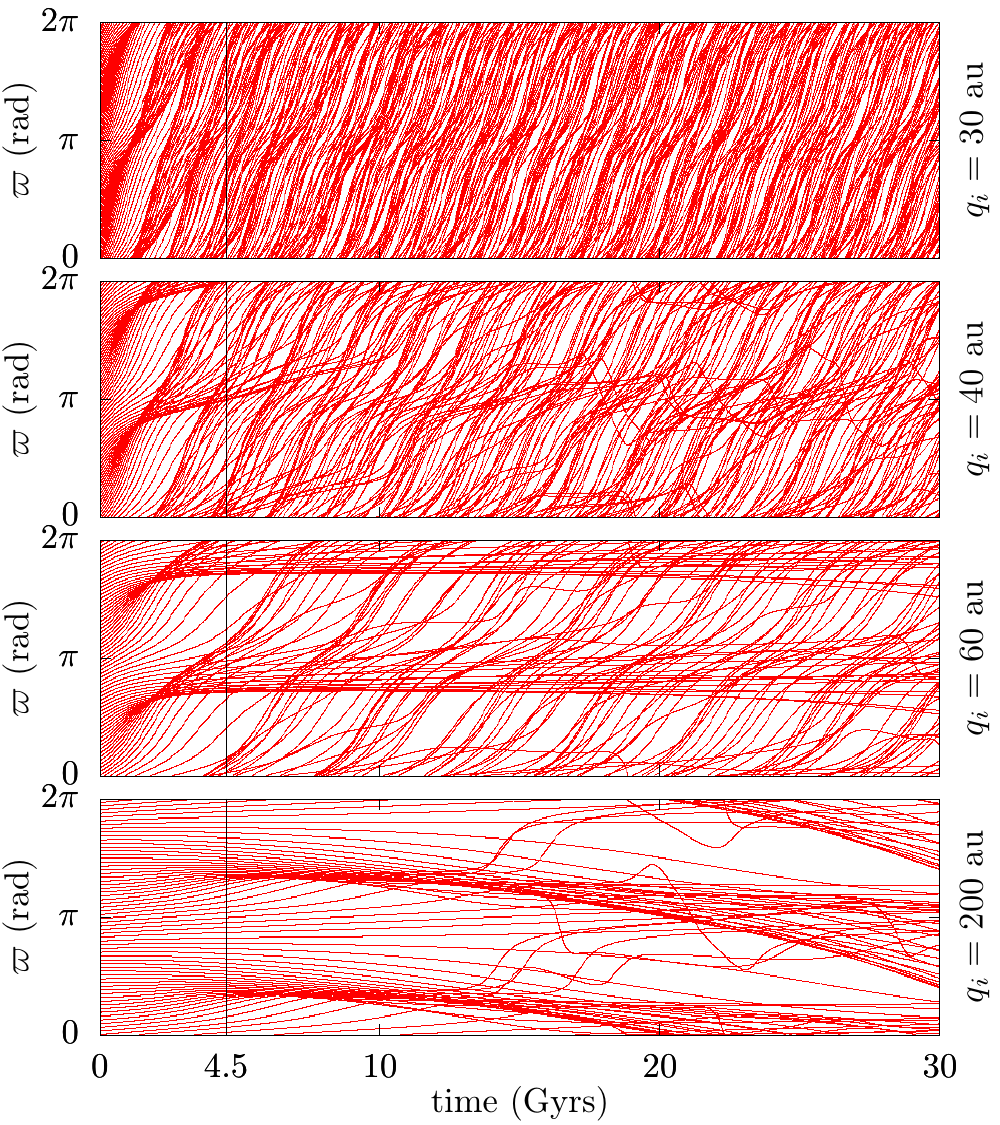}
      \caption{Temporal evolution of orbits with $a=1800$~au. The initial conditions are $I=0^\text{o}$, and $\varpi=\omega+\Omega$ equally distributed in $[0,2\pi]$; the initial value $q_i$ of the perihelion distance is written on the right of each graph. The vertical black line marks the $4.5$~Gyrs time, at which our density maps (e.g. Fig.~\ref{fig:varpiE}) are computed.}
      \label{fig:overdst}
   \end{figure}

   After $4.5$~Gyrs, all our samples feature overdensity regions for the ecliptic angles $\omega$ and $\Omega$ at large semi-major axes. As illustrated in Fig.~\ref{fig:varpiE}, the extension and shape of these regions are different according to the sample considered, and in some cases, overdensities are noticeable even below $a=500$~au. The planetary perturbations alone cannot modify the angular distributions in our samples, because they induce precession velocities that are independent of the angles (see Eq.~\ref{eq:plregime}). Consequently, Fig.~\ref{fig:varpiE} demonstrates that the galactic tides have a noticeable effect in $4.5$~Gyrs even for moderate values of the semi-major axis. It happens, however, that these overdensity regions have little to do with the dynamical mechanisms (libration zones, resonances) revealed in Sect.~\ref{sec:intermed}. In fact, as shown by Fig.~\ref{fig:overdst}, most of the particles follow only a small portion of the dynamical cycles involved, due to the large timescales at play. The galactic tides induce a gradient of precession velocities with respect to $\omega$ and $\Omega$; therefore, orbits that initially precess faster catch up with orbits that precess slower, before all orbits go away on their respective dynamical paths (which can be totally different). This `phase effect' produces temporary overdensity regions, like the ones shown in Fig.~\ref{fig:varpiE}. The gradient of precession velocities is different for each of our slices of inclination and perihelion distance, producing differing patterns. As shown in Fig.~\ref{fig:overdst}, the sharp patterns disappear as time goes by, replaced by actual dynamical features like resonances and libration zones. In other words, the patterns that we observe after $4.5$~Gyrs are a direct relic of our initial distribution of particles. At this point, it would be tempting to conclude that the orbits of distant trans-Neptunian objects still keep a clear memory of their primordial distribution, and that all that is needed to extract the relevant dynamical patterns is to find a sufficient number of them. However, other dynamical mechanisms, like the randomisation by passing stars or the presence of distant unseen perturbers could erase the signature that we are looking for. We also stress that the patterns mentioned above (e.g. the ones appearing in Fig.~\ref{fig:varpiE}) cannot be linked to the clustering of objects that motivates the Planet 9 hypothesis \citep{BATetal:19}, mostly because they form at too large semi-major axes values. Still, we find it a surprising coincidence that the ecliptic plane, the galactic plane, and the proposed P9 orbital plane intersect almost along the same line\footnote{Measured on the ecliptic, the longitude of ascending node of the proposed P9 is close to the longitude of descending node of the galactic plane. The galactic Laplace plane mentioned in Sect.~\ref{sec:preliminary} is therefore rotated by $180^\text{o}$ with respect to the Laplace plane raised by P9, and it has nothing to do with the clustering of orbital planes mentioned for instance by \cite{BATetal:19}.}.
   
   \begin{figure*}
      \centering
      \includegraphics[width=\textwidth]{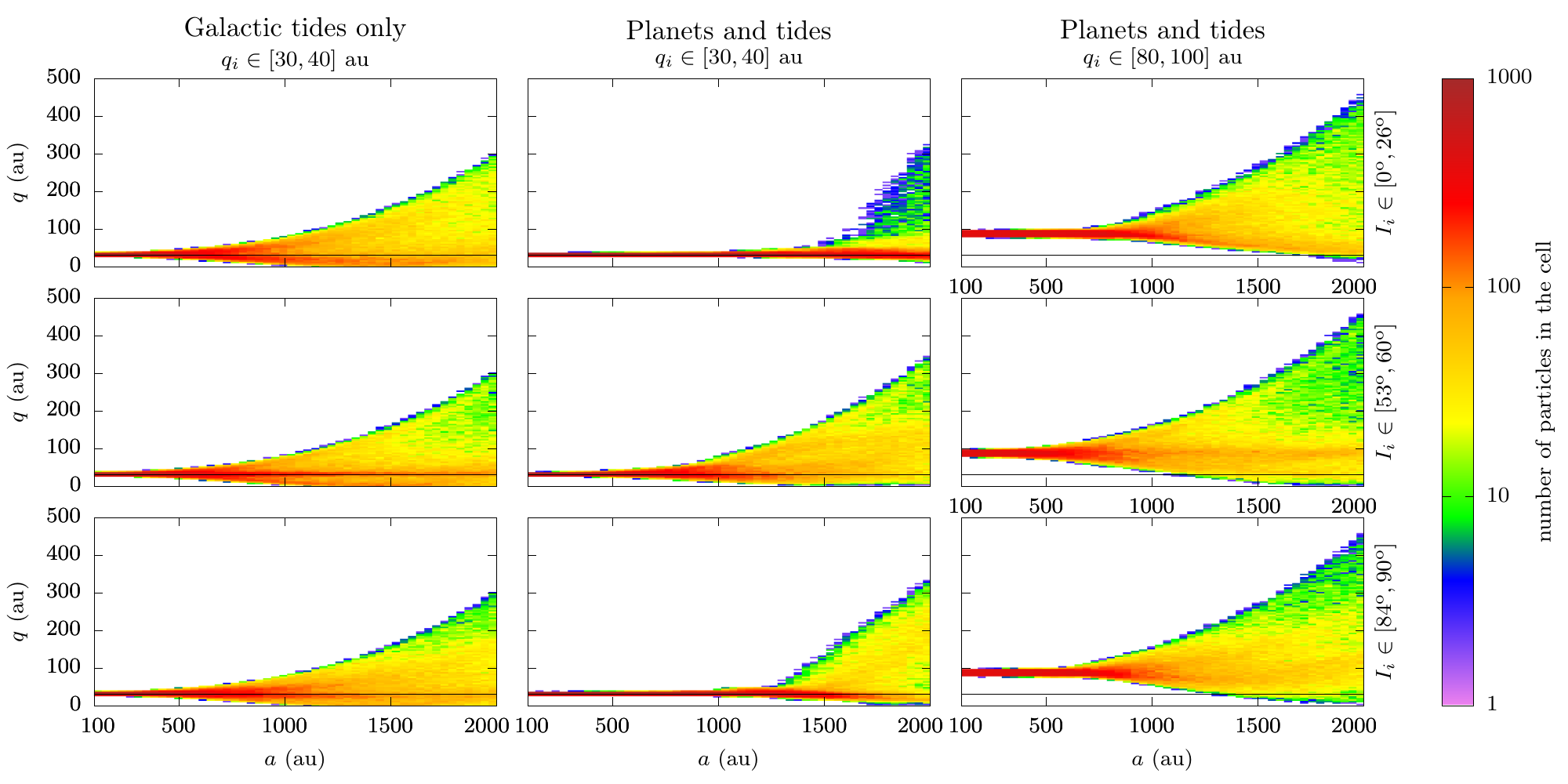}
      \caption{Distribution of a few samples of particles after $4.5$~Gyrs in the plane $(a,q)$. As indicated in the titles, the first column is for an evolution with only the galactic tides (Hamiltonian $\varepsilon_{\mathrm{G}_\mathrm{V}}\bar{\mathcal{H}}_{\mathrm{G}_\mathrm{V}}$, Eq.~\ref{eq:Hg}), and the second and third columns are for an evolution with both planets and galactic tides (Hamiltonian $\mathcal{F}$, Eq.~\ref{eq:F}). Six slices of initial conditions are shown, as written in the top and right side of the graphs: the first two columns are for initial perihelion distance $q_i$ in $[30,40]$~au, and the third one in $[80,100]$~au, while the cosine of the initial ecliptic inclination $I_i$ is distributed in $[0.9,1]$ for the top row, $[0.5,0.6]$ for the middle row, and $[0,0.1]$ for the bottom row. On each graph, the horizontal black line shows the semi-major axis of Neptune ($\sim 30$~au) for reference.}
      \label{fig:qsamples}
   \end{figure*}
   
   \begin{figure}
      \centering
      \includegraphics[width=\columnwidth]{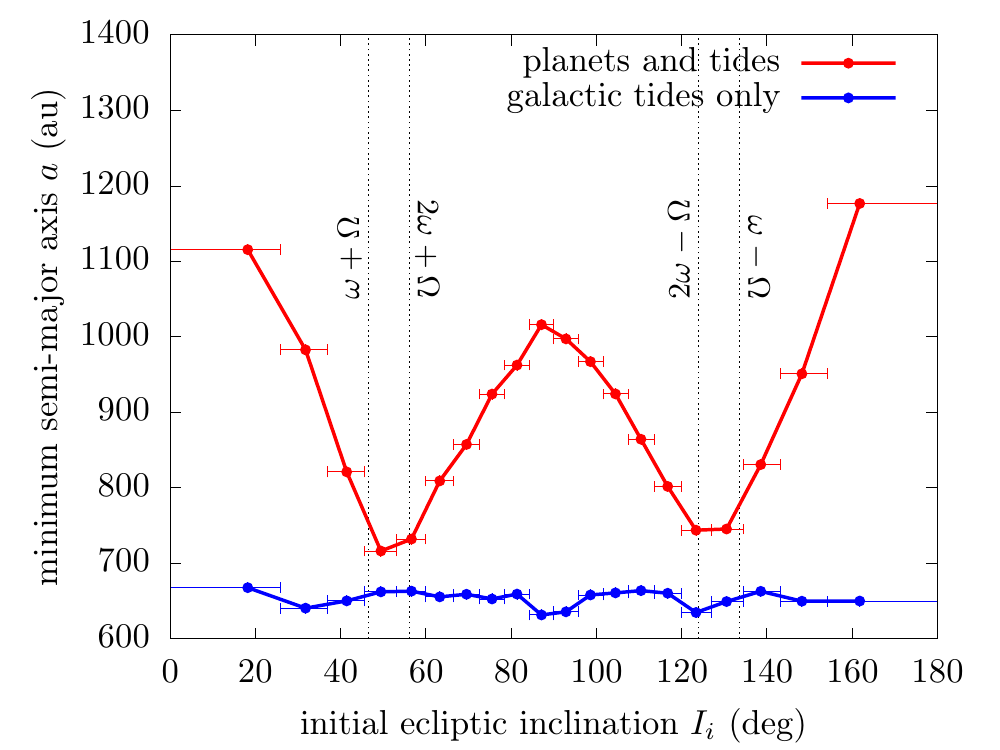}
      \caption{Minimum value of the semi-major axis above which particles initially sampled with $q\in[40,60]$~au spread beyond $q=80$~au in $4.5$~Gyrs, with respect to their initial ecliptic inclination. The horizontal bars show our twenty $0.1$-width slices of $\cos I$, connected in their centre by a full curve. The location of the four major resonances for the weakly perturbed planetary Hamiltonian are indicated by dotted black lines.}
      \label{fig:amin}
   \end{figure}
   
   We now focus on the excursion in perihelion distance $q$ of our samples after $4.5$~Gyrs. We recall that the galactic tides produce large cycles of eccentricity and inclination (Sect.~\ref{sec:galdyn}), at a rate that increases with the semi-major axis value. The planets, on the contrary, do not change the eccentricity and ecliptic inclination, but induce a precession of $\omega$ and $\Omega$ that is faster for smaller semi-major axes and smaller perihelion distances (Sect.~\ref{sec:pladyn}). If this precession is fast with respect to the galactic cycles, it has the consequence of averaging to zero the galactic contribution. Hence, as a rule of thumb, the planetary perturbations block the cycles raised by the galactic tides, with an efficiency that decreases for growing semi-major axis and perihelion distance. This is indeed what we observe in Fig.~\ref{fig:qsamples}, by comparing the behaviour of the samples with and without the planetary perturbations. This `blocking' effect is very efficient at $a=100$~au (and even up to about $1500$~au in the top middle panel), but almost null at $a=2000$~au. It is also less efficient for high perihelion distances (right column of Fig.~\ref{fig:qsamples}). The spreading of the distribution is damped most for small initial ecliptic inclinations (top row of Fig.~\ref{fig:qsamples}), because this corresponds to the maximum of the planetary-induced precession velocities (see Fig.~\ref{fig:res}). Moreover, we note that two ranges of initial ecliptic inclination are much more prone to orbital variations than other ranges, as exemplified by the middle row of Fig.~\ref{fig:qsamples}. This is particularly visible in Fig.~\ref{fig:amin}, showing the value of the semi-major axis above which the perihelion distance of small bodies, starting from $[40,60]$~au, can get beyond $80$~au in $4.5$~Gyrs. The two favoured inclination ranges are approximatively $I\in[45^\text{o},55^\text{o}]$ and $[125^\text{o},135^\text{o}]$, which corresponds to the location of the two pairs of resonances $(\omega+\Omega,2\omega+\Omega)$ and $(\omega-\Omega,2\omega-\Omega)$, and the regions where they overlap (see Fig.~\ref{fig:res}). These resonances being by far the strongest ones in terms of their widths in perihelion distance (see Sects.~\ref{sec:res} and \ref{sec:intermed}), they favour variations of~$q$.
   
\begin{figure*}
      \centering
      \includegraphics[width=\textwidth]{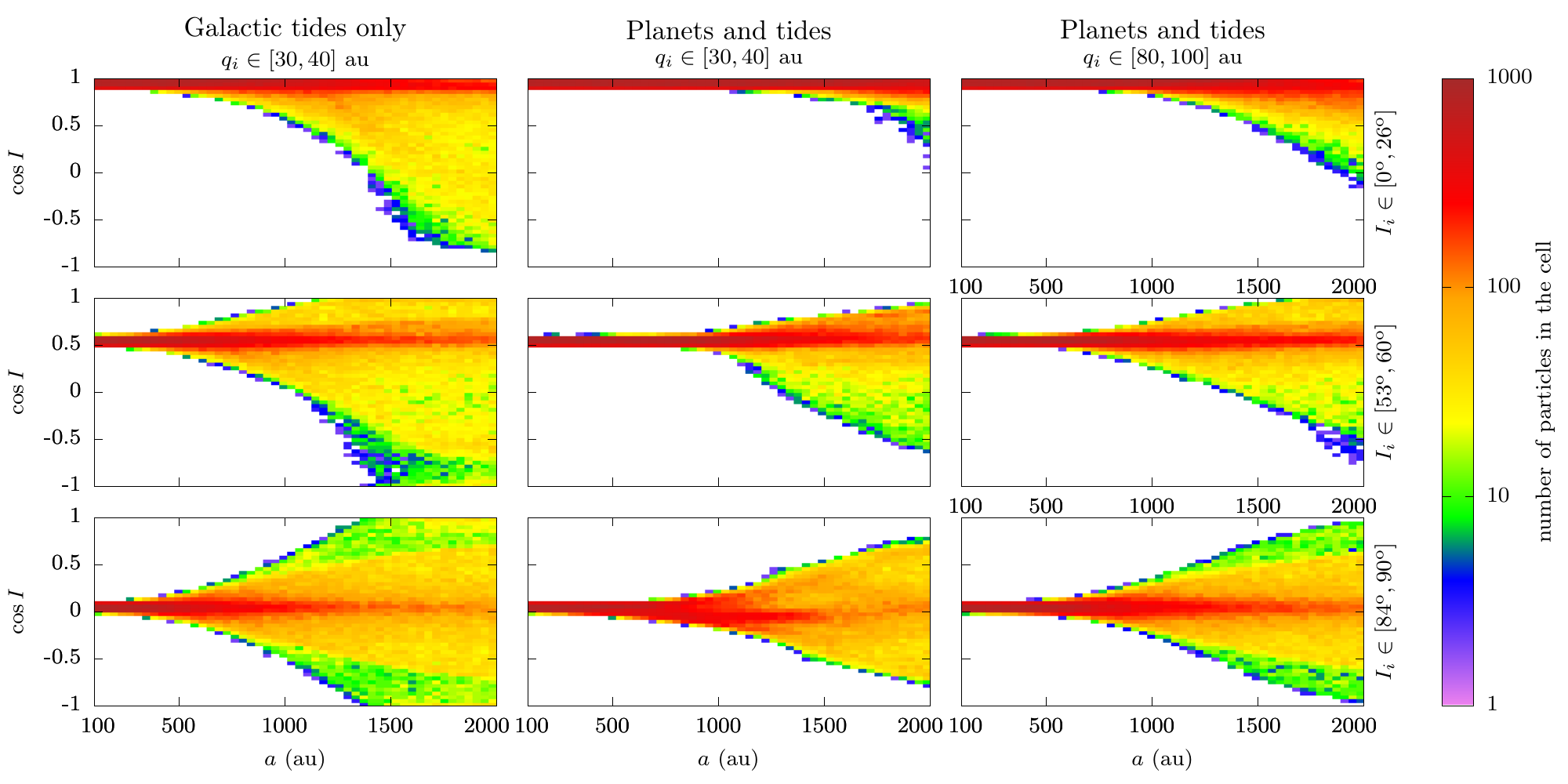}
      \caption{Same as Fig.~\ref{fig:qsamples}, but showing the distribution of the ecliptic inclination.}
      \label{fig:isamples}
   \end{figure*}

   We finally focus on the excursion in ecliptic inclination $I$ of our samples after $4.5$~Gyrs. Figure~\ref{fig:isamples} shows that for small perihelion distances, the spreading of the inclination distribution is strongly damped by the planetary perturbations for initial inclinations near $I=0^\text{o}$ (and $180^\text{o}$). This is similar to what we observed for the perihelion distance (Fig.~\ref{fig:qsamples}). However, this time, the smallest value of the semi-major axis at which the spreading is substantial is reached for initial ecliptic inclinations $I\sim 90^\text{o}$ (see the middle column of Fig.~\ref{fig:isamples}), and we observe no marked enhancement of the inclination excursions for other specific ranges of initial inclination. As before, these results are a direct consequence of the form of the galactic potential (see Sect.~\ref{sec:res}). Actually, the particles naturally spread in inclination as they precess about an inclined axis, corresponding to the tilt of the galactic Laplace plane (Sect.~\ref{sec:preliminary}). For very eccentric orbits, the classic Laplace plane is severely bent towards the ecliptic (Appendix~\ref{asec:eccLap}), but the orthogonal equilibrium at $I\sim 90^\text{o}$ produces large oscillations of the inclination (Sect.~\ref{sec:res}).
   
   \subsection{Limits of the inert region}\label{sec:limits}
   Looking at how samples of particles initially distributed in localised regions of the orbital elements space spread under the secular action of the planets and galactic tides, we are now able to answer one of the main questions: what are the limits of the inert Oort cloud? Where in the $(a,q,I)$ space can small bodies like Sedna and $2012$\,VP$_{113}$ remain efficiently fossilised since the early stages of the solar system evolution?
   
   For each value of $(a,q,I)$, we now look in the plane $(\omega,\Omega)$ for the initial condition producing the maximum variation of $q$ or $I$ in $4.5$~Gyrs. To this end, we save the extrema $q_\mathrm{min}$ and $q_\mathrm{max}$ reached by $q$ (resp. $I$) in the course of each numerical integration, and we apply an optimisation algorithm to maximise their difference $\Delta q=q_\mathrm{max}-q_\mathrm{min}$ (resp. $\Delta I$). We opted for the Particle Swarm Optimisation method \citep{POLetal:07} in order to limit the cases of convergence towards local maxima. The initial condition $(\omega,\Omega)$ that maximises $\Delta q$ is generally different from the one that maximises $\Delta I$, so two separate optimisation procedures are needed.
   
\begin{figure*}
      \centering
      \includegraphics[width=\textwidth]{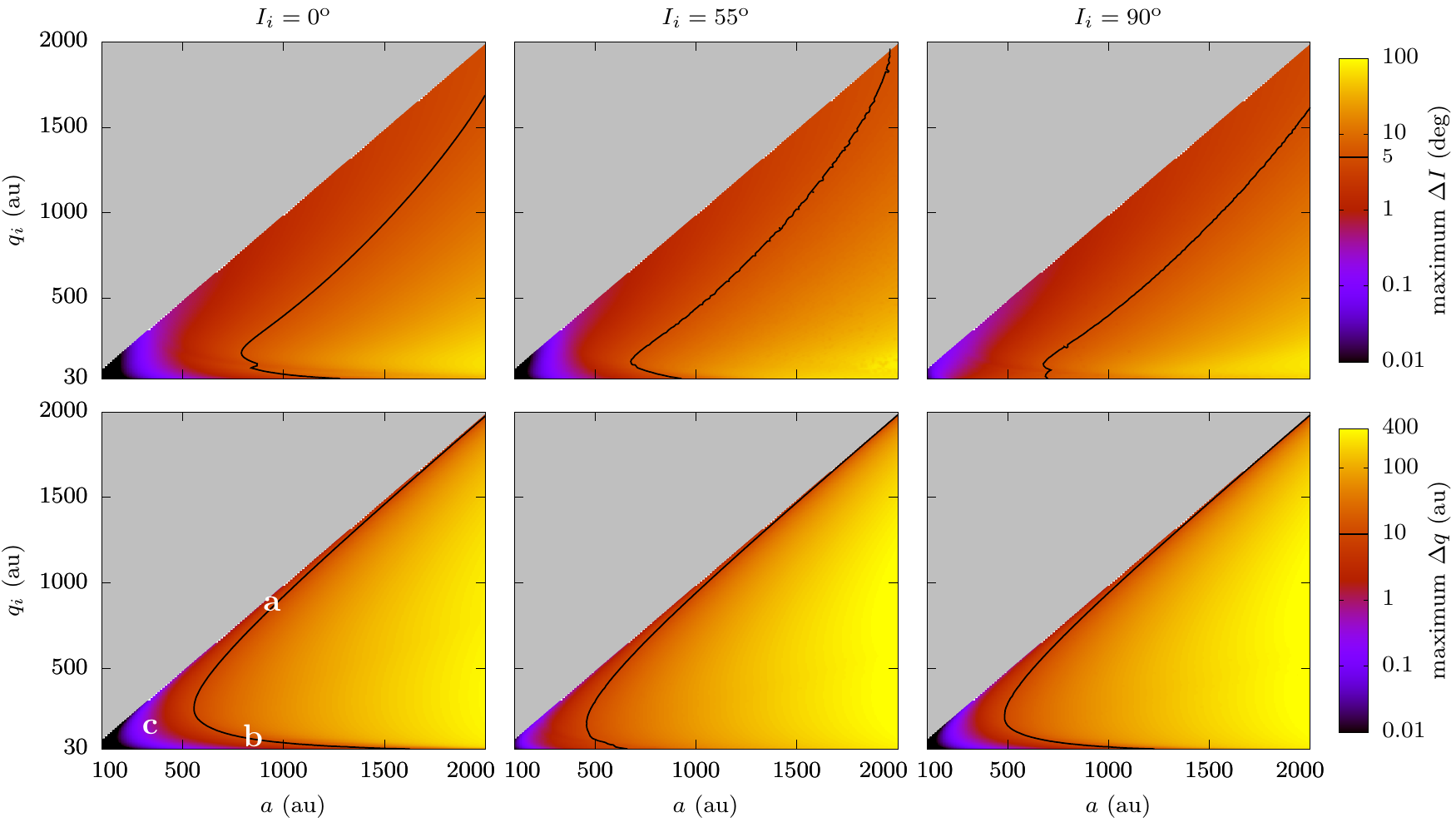}
      \caption{Limits of the inert region in the $(a,q)$ plane. Each column corresponds to a different value of the initial ecliptic inclination (see titles). The colour scale represents the maximum possible orbital variations in $4.5$~Gyrs: the top row shows the variation of ecliptic inclination, and the bottom row shows the variations of perihelion distance (see labels on the right). The black level corresponds to a variation of $5^\text{o}$ in inclination (top row) or $10$~au in perihelion distance (bottom row). Below the black level, the region can be considered inert. See text for the white symbols.}
      \label{fig:aqmap}
   \end{figure*}

\begin{figure*}
      \centering
      \includegraphics[width=\textwidth]{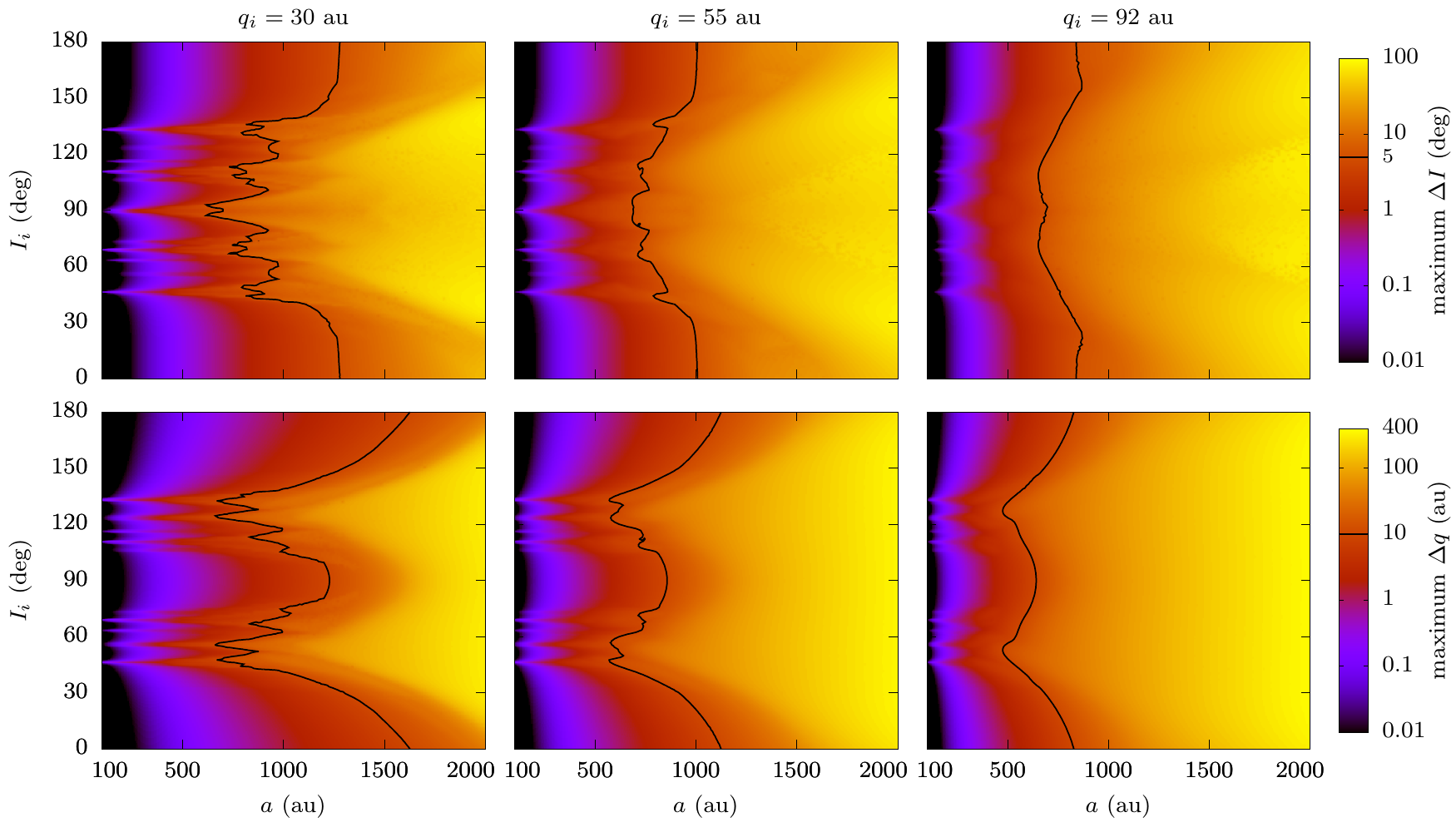}
      \caption{Same as Fig.~\ref{fig:aqmap} but in the $(a,I)$ plane. Each column corresponds to a different value of the initial perihelion distance (see titles).}
      \label{fig:aimap}
   \end{figure*}

   Using this method, we obtain the full three dimensional structure in the $(a,q,I)$ space of the largest orbital changes produced within our simplified model. Figures~\ref{fig:aqmap} and \ref{fig:aimap} show representative sections of this space in the $(a,q)$ and $(a,I)$ directions. As expected, the highest orbital variations are reached for orbits with largest semi-major axes, in the regime where the galactic tides strongly dominate over the planetary perturbations (see Sect.~\ref{sec:explor}). The black curve in Figs.~\ref{fig:aqmap}-\ref{fig:aimap} delimits the `inert' portion of the space, defined arbitrarily as $\Delta q<10$~au or $\Delta I<5^\text{o}$. We see that the naive picture depicted in Introduction from previous works largely overestimates the inert region. Fig.~\ref{fig:aqmap} shows that orbits are truly inert only if:
   \begin{enumerate}[a)]
      \item $a\gtrsim 500$~au and the orbit is nearly circular.
      \item $500\lesssim a\lesssim 1500$~au and $q$ is close to the planetary region.
      \item $a\lesssim500$~au.
   \end{enumerate}
   These three inert regions are labelled on the bottom left panel of Fig.~\ref{fig:aqmap}. They are dynamically distinct:
   \begin{enumerate}[a)]
      \item For nearly circular orbits, we know from Sect.~\ref{sec:preliminary} that very large orbital variations are actually allowed by the dynamics, but that the timescale is dramatically long; this means that such orbits hardly even precess in $4.5$~Gyrs (unless the semi-major axis is extremely large, see Fig.~\ref{fig:LaplaceFreq}).
      \item For small perihelion distances, the planetary perturbations produce a fast precession of the orbits, which averages out the galactic contribution; this means that such orbits have frozen $q$ and $I$ even when considering an infinite timescale. However, a very small perihelion distance implies that the planetary scattering, not taken into account here, is triggered (see Introduction). In case of scattering, of course, the orbit cannot be considered inert.
      \item For $a\lesssim 500$~au, the inert regions a and b merge. These orbits precess substantially, similarly to region b, even for large perihelion distances. This is where Sedna ($a\approx 540$~au) and $2012$\,VP$_{113}$ ($a\approx 270$~au) are located. 
   \end{enumerate}
   In regions b and c, Fig.~\ref{fig:aimap} confirms that the border of the inert region has a complex structure that is directly linked to the main resonances between $\omega$ and $\Omega$ (see Sect.~\ref{sec:res}). As discussed in Sect.~\ref{sec:samples}, this complex structure produces a very marked differential spreading of small bodies in the space of orbital element. This structure disappears for large perihelion distances, as all resonances overlap. Sedna and $2012$\,VP$_{113}$ have large perihelion distances ($76$~au and $81$~au), but not large enough to completely suppress the effect of resonances. However, due to their relatively small ecliptic inclinations ($12^\text{o}$ and $24^\text{o}$), both of them are out of any of the main resonances. This makes them true `inert' objects with precessing orbits. As expected from Sect.~\ref{sec:real}, on the contrary, $2015$\,TG$_{387}$ is out of the inert region depicted in Fig.~\ref{fig:aqmap}-\ref{fig:aimap} ($a\approx 1190$~au, $q\approx 65$~au, $I\approx 12^\text{o}$). It is however close to its border.
   
   For completeness, Appendix~\ref{asec:maps} gives sections of the $(a,q,I)$ space in the $(q,I)$ direction: we retrieve the resonance structure described analytically in Sect.~\ref{sec:res}.

\section{Summary and conclusions}\label{sec:disconc}
   We studied the long-term orbital dynamics of small bodies in the intermediate regime between the Kuiper belt and the Oort cloud, that is, where the planetary perturbations and the galactic tides have the same order of magnitude. The two kinds of perturbations are weak in this region, and we call it the `inert Oort cloud' in reference to the few observed detached Kuiper belt objects, which have extremely stable orbits.
   
   The problem is formally close to the case of a satellite perturbed by the $J_2$ flattening of its host planet and the averaged attraction from the star. As such, it possesses a tilted Laplace plane (the `galactic Laplace plane'), with a crossover located at about $1000$~au. This means that for semi-major axes much smaller than this value (say $500$~au), circular orbits precess about the ecliptic pole, whereas for semi-major axes much larger than this value (say $1500$~au) they precess about the galactic pole. In between, they precess about an intermediately tilted pole. In this regime, however, the precession period for circular orbits counts in hundreds of Gyrs, meaning that these orbits hardly change at all in practice.
   
   These dramatically long timescales are greatly reduced for eccentric orbits. The dynamics is integrable in the small and large semi-major axis regimes, when one kind of perturbation strongly dominates the other one. Between about $800$ and $1100$~au, however, the phase space is almost completely filled with chaos, from very eccentric down to nearly circular orbits. The chaotic diffusion timescales are quite large, but they decrease with the semi-major axis value. For semi-major axes as small as $800$~au, the joint action of planets and galactic tides can produce a chaotic diffusion of perihelion distance $q$ over tens of astronomical units in a few billion years. Even though frozen orbital regions do exist (this is the case for Sedna and $2012$\,VP$_{113}$), we conclude that this region is far from being inert, contrary to what one could expect from the weakness of the perturbations.
   
   In $4.5$~Gyrs, the galactic tides have noticeable effects down to semi-major axes of about $500$~au. At $2000$~au, the orbital excursions induced can exceed $400$~au in perihelion distance and $80^\text{o}$ in inclination. Interestingly, the largest changes of perihelion distance are reached for ecliptic inclinations $I$ in the ranges $[45^\text{o},55^\text{o}]$ and $[125^\text{o},135^\text{o}]$. These ranges are delimited by the two pairs of strong resonances $(\omega+\Omega,2\omega+\Omega)$ and $(2\omega-\Omega,\omega-\Omega)$, which ease perihelion variations. When monitoring swarms of particles over $4.5$~Gyrs, we also observe accumulations of orbital angles in localised zones, for semi-major axes larger than about $500$~au. Indeed, due to the long timescales at play, particles do not have the time to go away on their respective dynamical paths, but they rather spread in a non uniform manner, creating (temporary) overdensities. Such accumulations are a relic of the initial distribution of small bodies, but they have little observational consequences at this stage, considering the very distant objects involved.
   
   In conclusion, when mapping the truly `inert' region ($\Delta q<10$~au and $\Delta I<5^\text{o}$ over $4.5$~Gyrs), we find that it is remarkably small. The precise limits of the inert region can be found in Fig.~\ref{fig:aqmap}-\ref{fig:aimap}. It is either composed of: a) nearly circular orbits with $a\gtrsim 500$~au, b) orbits with $500\lesssim a\lesssim 1500$~au and perihelion close to the planetary region, or c) orbits with $a\lesssim 500$~au (as long as they are unaffected by mean-motion resonances). Moreover, orbits are truly inert only if their perihelion distance is high enough to avoid planetary scattering; in region b, this only leaves a thin inert zone. Out of the inert region, the excursions mentioned above in perihelion distance and inclination, as well as the angular accumulations, are direct effects of the galactic tides. They are quite noticeable after $4.5$~Gyrs, and can therefore be decisive when classifying observed bodies as `detached' or not, or when monitoring samples of them, as is done for P9 simulations (see e.g. \citealp{BATetal:19} and references therein). Hence, we advocate including the galactic tides in numerical simulations of trans-Neptunian object with semi-major axis larger than $500$~au.
   
\begin{acknowledgements}
   We thank Vacheslav Emel'yanenko for inspiring discussions about this manuscript, as well as for his independent verifications of our results. We also thank Auriane {\'E}gal for her help about the Particle Swarm Optimisation method. This work was supported by the Programme National de Plan{\'e}tologie (PNP) of CNRS/INSU, co-funded by CNES.
\end{acknowledgements}

\bibliographystyle{aa}
\bibliography{biblio}

\appendix
\section{Conversion formulas between the ecliptic and the galactic reference frames}\label{asec:conv}
   The galactic reference frame used here is defined with the third axis perpendicular to the galactic plane and the first axis directed towards the ascending node of the ecliptic\footnote{This is the same reference frame as used by \cite{HIGetal:07}; the minus sign in their Eq.~(22) is a typographical error.}. We write $(I_\mathrm{G},\omega_\mathrm{G},\Omega_\mathrm{G})$ the Keplerian elements of the small body measured in this frame. The ecliptic reference frame is defined with the third axis perpendicular to the ecliptic and the same first axis as the galactic reference frame. We write $(I,\omega,\Omega)$ the Keplerian elements of the small body measured in this reference frame. Passing from one frame to another corresponds to a rotation of $\pm\psi$ around the first axis, $\psi$ being the inclination of the ecliptic measured in the galactic reference frame. We obtain
   \begin{equation}
      \cos I_\mathrm{G} = \cos\psi\cos I - \sin\psi\cos\Omega\sin I \,,
   \end{equation}
   \begin{equation}
      \left\{
      \begin{aligned}
         \cos\Omega_\mathrm{G}\sin I_\mathrm{G} &= \cos\psi\cos\Omega\sin I + \sin\psi\cos I \,, \\
         \sin\Omega_\mathrm{G}\sin I_\mathrm{G} &= \sin\Omega\sin I \,,
      \end{aligned}
      \right.
   \end{equation}
   and
   \begin{equation}
      \left\{
      \begin{aligned}
         \cos\omega_\mathrm{G}\sin I_\mathrm{G} &= \cos\psi\cos\omega\sin I \\
         &- \sin\psi(\sin\omega\sin\Omega - \cos\omega\cos\Omega\cos I) \,,\\
         \sin\omega_\mathrm{G}\sin I_\mathrm{G} &=\cos\psi\sin\omega\sin I \\
         &+ \sin\psi(\cos\omega\sin\Omega+\sin\omega\cos\Omega\cos I) \,.
      \end{aligned}
      \right.
   \end{equation}
   The inverse formulas are obtained by replacing $\psi$ by $-\psi$. Using these expressions, the Hamiltonian $\mathcal{F}$ in Eq.~\eqref{eq:F} can be written both in ecliptic or in galactic coordinates. More precisely, we have
   \begin{equation}\label{eq:HpG}
      \begin{aligned}
         \bar{\mathcal{H}}_{\mathrm{P}_2} &= \frac{1-3\cos^2I}{8(1-e^2)^{3/2}} \\
         &= \frac{-1}{16(1-e^2)^{3/2}}&&\Bigg[ \big(3C^2 - 1\big)\big(3\cos^2I_\mathrm{G} - 1\big) \\
         &&&+ 12 \, CS  \cos I_\mathrm{G}\sin I_\mathrm{G} \, \cos(\Omega_\mathrm{G}) \\
         &&&+ 3  \, S^2 \sin^2I_\mathrm{G}      \, \cos(2\Omega_\mathrm{G})
         \Bigg] \,,
      \end{aligned}
   \end{equation}
   and
   \begin{equation}\label{eq:HgE}
      \begin{aligned}
         \bar{\mathcal{H}}_{\mathrm{G}_\mathrm{V}} &= \frac{\sin^2I_\mathrm{G}}{4}\left(1+\frac{3}{2}e^2-\frac{5}{2}e^2\cos(2\omega_\mathrm{G})\right) \\
         = \frac{-1}{32}&\Bigg[ 2 \, (3e^2 + 2)(2C^2\cos^2I + S^2\sin^2I - 2) \\
         &  - 8  \, CS         (3e^2 + 2) \cos I\sin I       \, \cos(\Omega) \\
         &  + 5 \, S^2        e^2        (\cos I + 1)^2                \, \cos(2\omega + 2\Omega) \\
         &  + 20  \, CS         e^2        (\cos I + 1)\sin I \, \cos(2\omega + \Omega) \\
         &  + 10 \, (3C^2 - 1) e^2       \sin^2I                        \, \cos(2\omega) \\
         &  + 20  \, CS         e^2        (\cos I - 1)\sin I \, \cos(2\omega - \Omega) \\
         &  + 5 \, S^2        e^2        (\cos I - 1)^2                \, \cos(2\omega - 2\Omega) \\
         &  + 2 \, S^2        (3e^2 + 2) \sin^2I                       \, \cos(2\Omega)
         \Bigg] \,,
      \end{aligned}
   \end{equation}
   where $C\equiv\cos\psi$ and $S\equiv\sin\psi$.

\section{Hamiltonian function for the satellite case}\label{asec:satellite}
   For comparison purpose, we present here the Hamiltonian function describing the secular evolution of a satellite perturbed by the Sun and the oblateness of its host planet in the quadrupolar approximation. Such a Hamiltonian can be written
   \begin{equation}\label{eq:Hsat}
      \bar{\mathcal{K}} = \varepsilon_\mathrm{J}\bar{\mathcal{K}}_\mathrm{J} + \varepsilon_\odot\bar{\mathcal{K}}_\odot\,,
   \end{equation}
   where
   \begin{equation}
      \varepsilon_\mathrm{J} = \frac{1}{a^3}\big(2\mu_\mathrm{P}J_2R_\mathrm{P}^2\big)
      \hspace{0.5cm},\hspace{0.5cm}
      \varepsilon_\odot = a^2\left(\frac{3\mu_\odot}{2a_\odot^3(1-e_\odot^2)^{3/2}}\right)
      \,,
   \end{equation}
   and
   \begin{equation}
      \left\{
      \begin{aligned}
         \bar{\mathcal{K}}_\mathrm{J} &= \frac{1-3\cos^2I}{8(1-e^2)^{3/2}} \,,\\
         \bar{\mathcal{K}}_\odot &= \frac{\sin^2I_\mathrm{G}}{4}\left(1 + \frac{3}{2}e^2 - \frac{5}{2}e^2\cos(2\omega_\mathrm{G})\right) - \frac{1}{4}e^2 \,.
      \end{aligned}
      \right.
   \end{equation}
   In these expressions, $\mu_\mathrm{P}$, $J_2$, and $R_\mathrm{P}$ are the gravitational parameter, the flattening coefficient, and the equatorial radius of the host planet, respectively, whereas $\mu_\odot$, $a_\odot$, and $e_\odot$ are the gravitational parameter, the semi-major axis, and the eccentricity of the Sun, respectively. We keep the same notations as in the rest of the article (see e.g. Appendix~\ref{asec:conv}) in order to emphasize the similarities with the distant trans-Neptunian case. This time, however, the indexless orbital elements are measured with respect to the equatorial plane of the host planet, and the $\mathrm{G}$ index refers to the orbital plane of the Sun.
   
   The overall Hamiltonian function in Eq.~\eqref{eq:Hsat} should be compared with Eq.~\eqref{eq:F}. It is well known that the averaged quadrupolar effect of inner bodies has the same form as a $J_2$ flattening of the central body \citep[see e.g.][]{TREetal:09}. As such, $\varepsilon_\mathrm{J}$ is strictly equivalent to the parameter $\varepsilon_\mathrm{P_4}$ used above (see Eq.~\ref{eq:eps}), and $\bar{\mathcal{K}}_\mathrm{J}$ is identical to $\bar{\mathcal{H}}_\mathrm{P_4}$ (see Eq.~\ref{eq:Hp}). Furthermore, we see here that $\varepsilon_\odot$ has the same $a^2$ multiplier as $\varepsilon_{\mathrm{G}_\mathrm{V}}$ (see Eq.~\ref{eq:eps}), and that $\bar{\mathcal{K}}_\odot$ has nearly the same form as $\bar{\mathcal{H}}_{\mathrm{G}_\mathrm{V}}$ (see Eq.~\ref{eq:Hg}), apart from the additional term $-e^2/4$. The two problems are therefore not strictly equivalent, unless $e=0$. Hence, the results obtained by \cite{TREetal:09} for strictly circular orbits remain valid in the trans-Neptunian case, like the location of the equilibrium points (see Fig.~\ref{fig:exLap}), and their stability against inclination variation.
   
   \begin{figure}
      \includegraphics[width=\columnwidth]{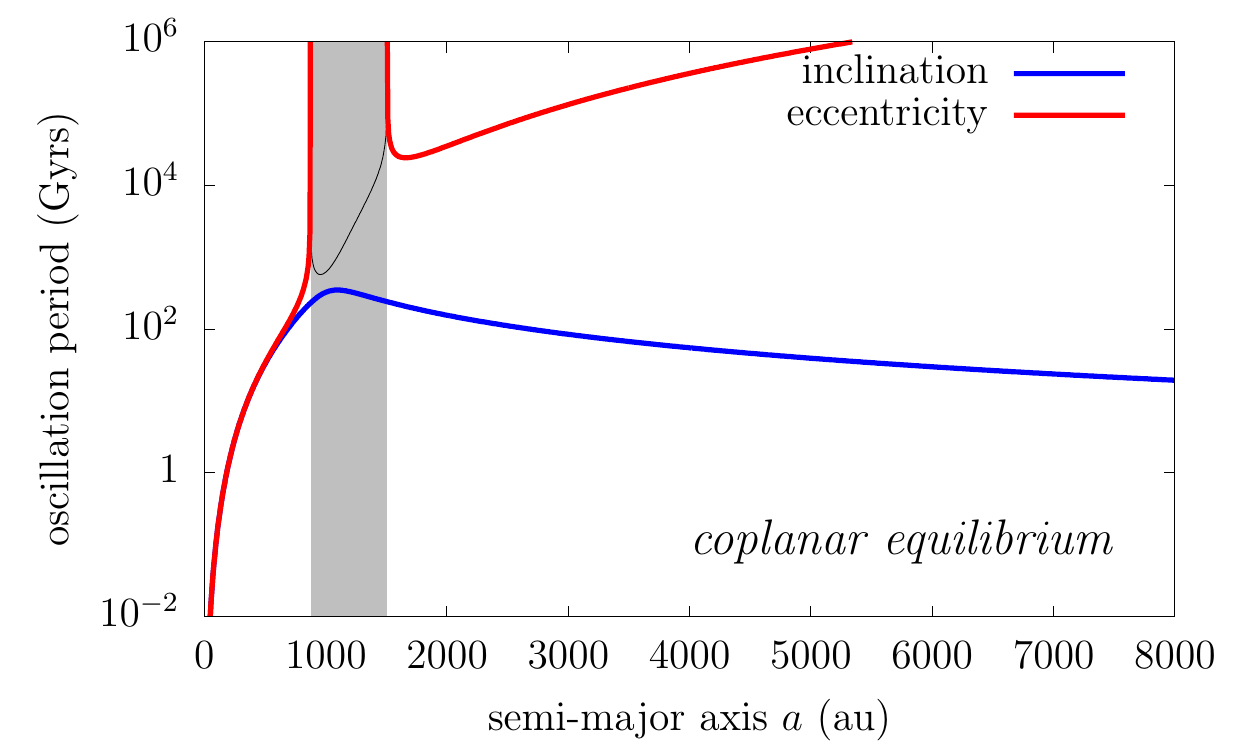}
      \includegraphics[width=\columnwidth]{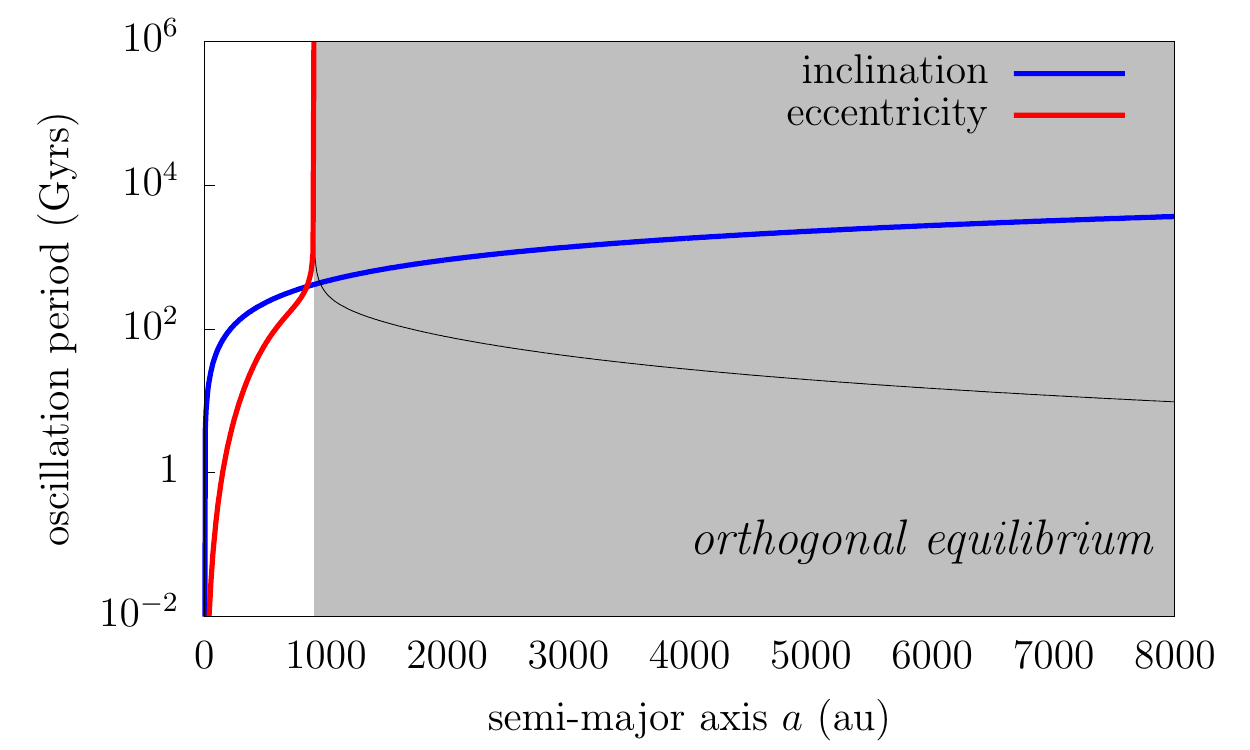}
      \caption{Period of small oscillations about the two kinds of circular Laplace equilibrium. \emph{Top:} coplanar equilibrium; \emph{Bottom:} orthogonal equilibrium. In the grey zone, the equilibrium point is unstable against eccentricity growth; accordingly, the oscillation period of eccentricity is replaced by the period $T$ after which the eccentricity is multiplied by $\exp(2\pi)\approx 535$ (black curve).}
      \label{fig:LaplaceFreqE}
   \end{figure}
   
   However, the stability of the circular equilibrium points (reusing the nomenclature of \citealp{TREetal:09}) against eccentricity growth are different. This can be shown using linearised equations around the $e=0$ equilibrium (and a set of variables that are not singular for circular orbits). In the satellite case, the circular coplanar equilibrium is stable for any $a$ as long as the obliquity of the host planet is smaller than $68.875^\text{o}$ (see \citealp{TREetal:09}). In the trans-Neptunian case, on the contrary, using the inclination of the ecliptic from Table~\ref{tab:physconst} (which is smaller than $68.875^\text{o}$), we find that the circular coplanar equilibrium is unstable against eccentricity growth for $a$ between $875$ and $1509$~au. This instability, however, is absolutely unable to affect real bodies because it acts on a dramatically long timescale (see Fig.~\ref{fig:LaplaceFreqE}). Finally, the stability of the circular orthogonal equilibrium against eccentricity growth is also different from the satellite case. Indeed, in the trans-Neptunian case, we obtain the stability condition
   \begin{equation}
      a^5 \leqslant \frac{3}{4}\frac{\sum_{i=1}^N\mu_ia_i^2}{\mathcal{G}_3} \,,
   \end{equation}
   which is twice the analogous limit obtained in the satellite case. This corresponds to a semi-major axis of about $904$~au. As before, though, the timescales involved here have no physical relevance.
   
\section{Laplace plane for eccentric orbits}\label{asec:eccLap}
\begin{figure}
      \includegraphics[width=\columnwidth]{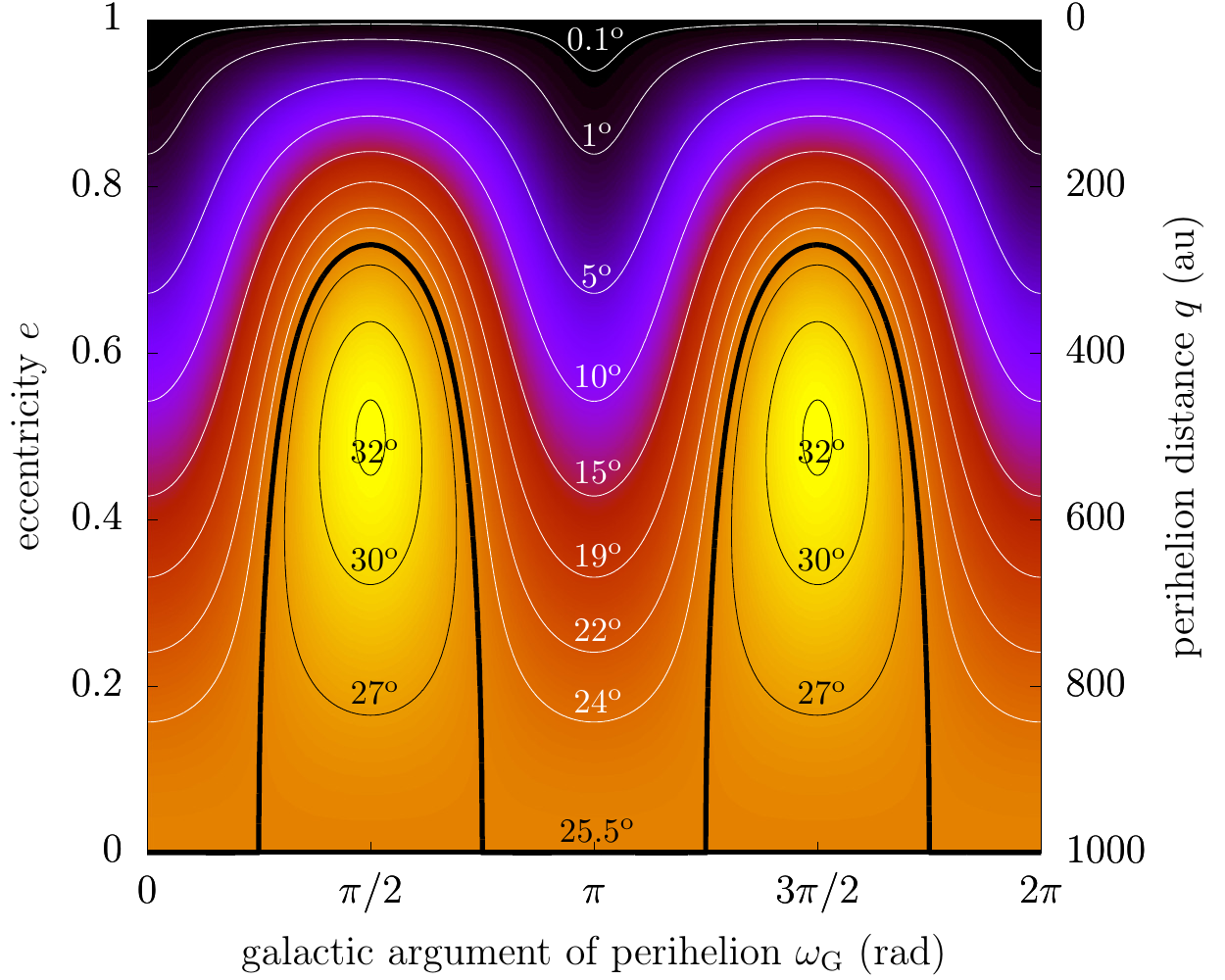}\\
      \vspace{0.1cm}\\
      \includegraphics[width=\columnwidth]{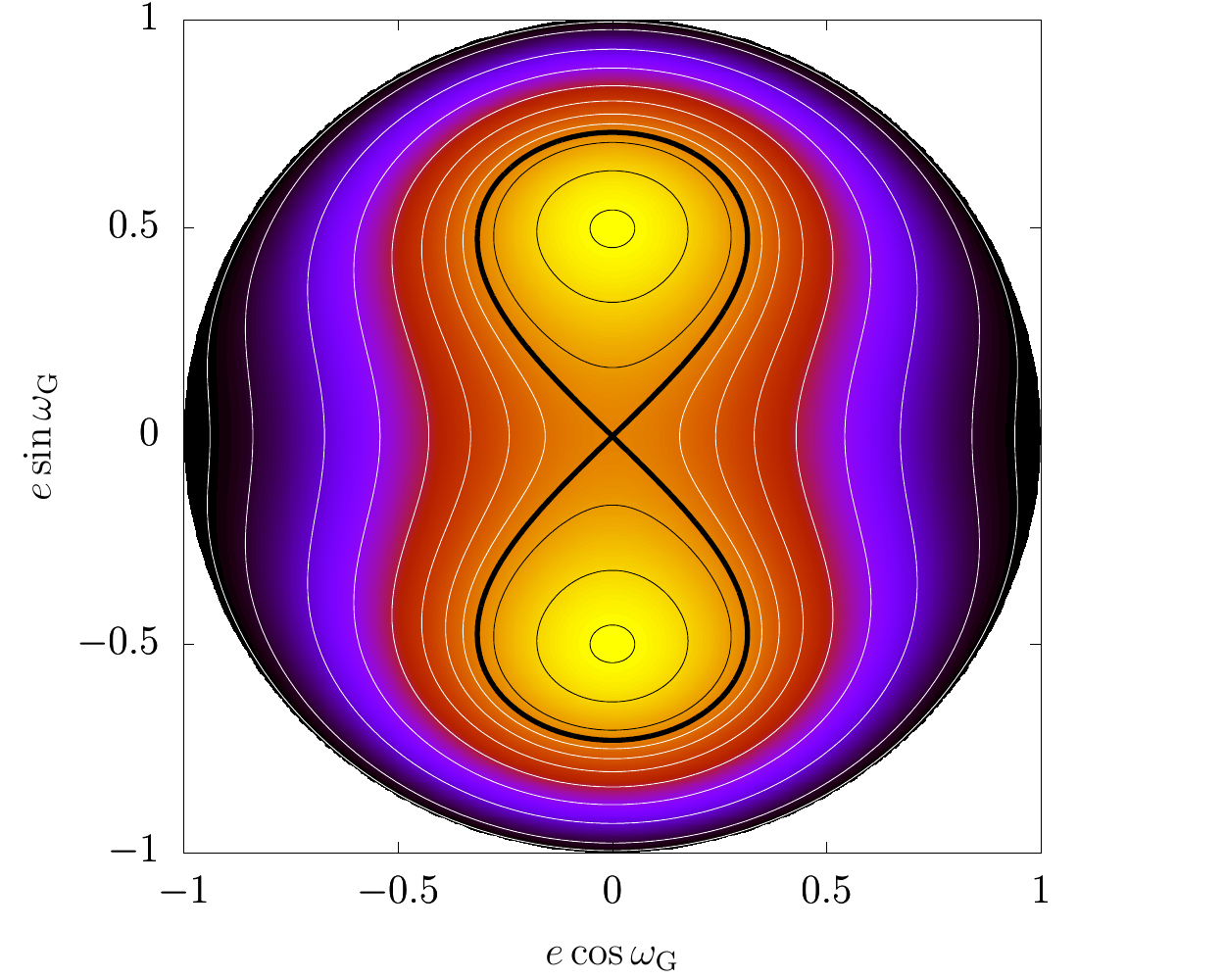}
      \caption{Inclination of the classic Laplace plane with respect to the ecliptic in the eccentric case. It is obtained by considering fixed values of $(e,\omega_\mathrm{G})$. The semi-major axis taken as parameter is $a=1000$~au. The two panels show the same level curves for two sets of variables. Except in the $e=0$ case, this inclination is only `instantaneous' because $e$ and $\omega_\mathrm{G}$ actually vary. Dark colours are low inclinations, and light colours are high inclinations, as shown by the labelled levels. The thick black level shows the inclination value that is equal to the one obtained in the circular case. The mean inclination, obtained by averaging over $\omega_\mathrm{G}$, corresponds to the vertical line $\omega_\mathrm{G}=\pi/4$ on the top panel, or the diagonal line on the bottom panel.}
      \label{fig:LaplaceEccOm}
   \end{figure}

\begin{figure}
      \includegraphics[width=\columnwidth]{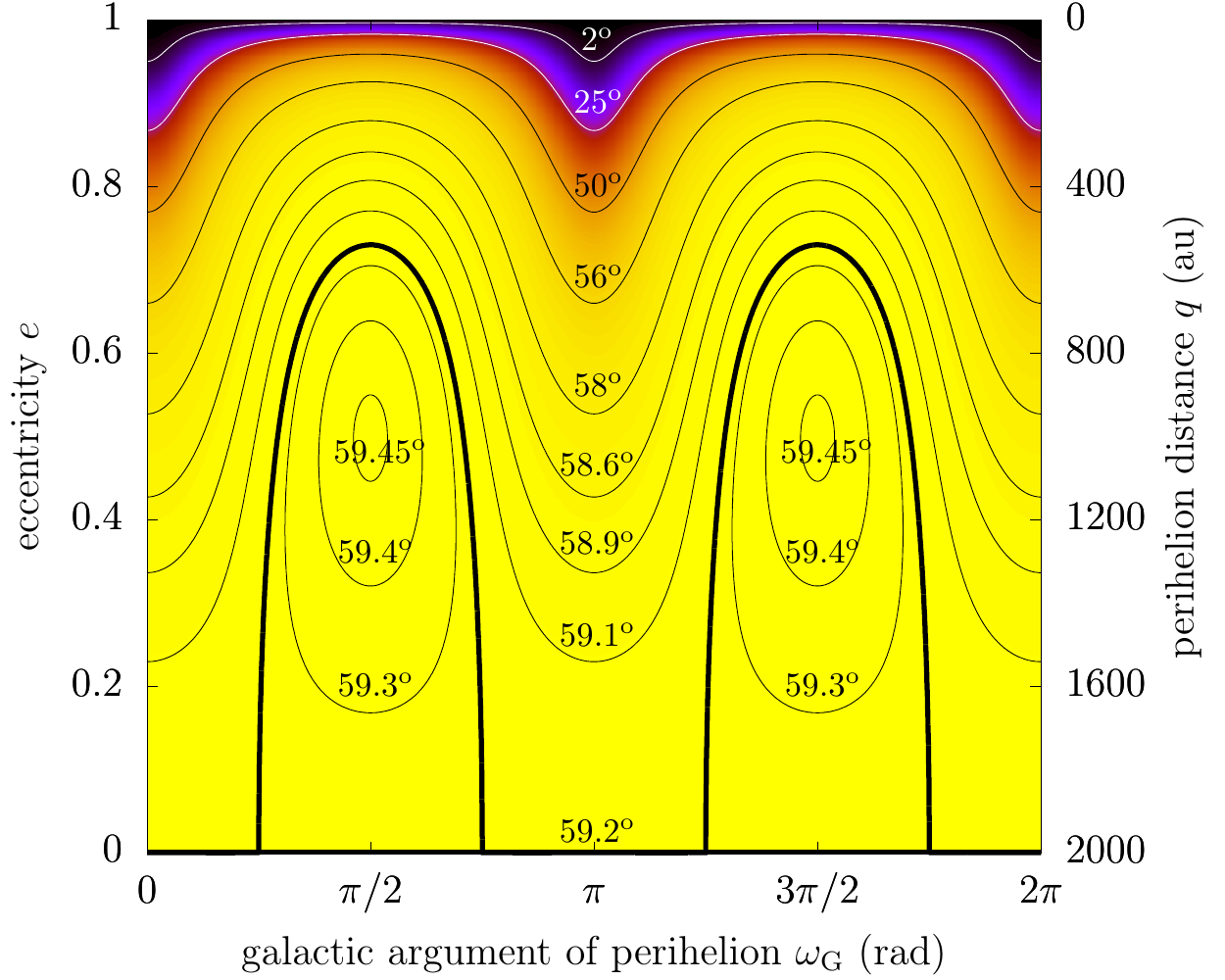}\\
      \vspace{0.1cm}\\
      \includegraphics[width=\columnwidth]{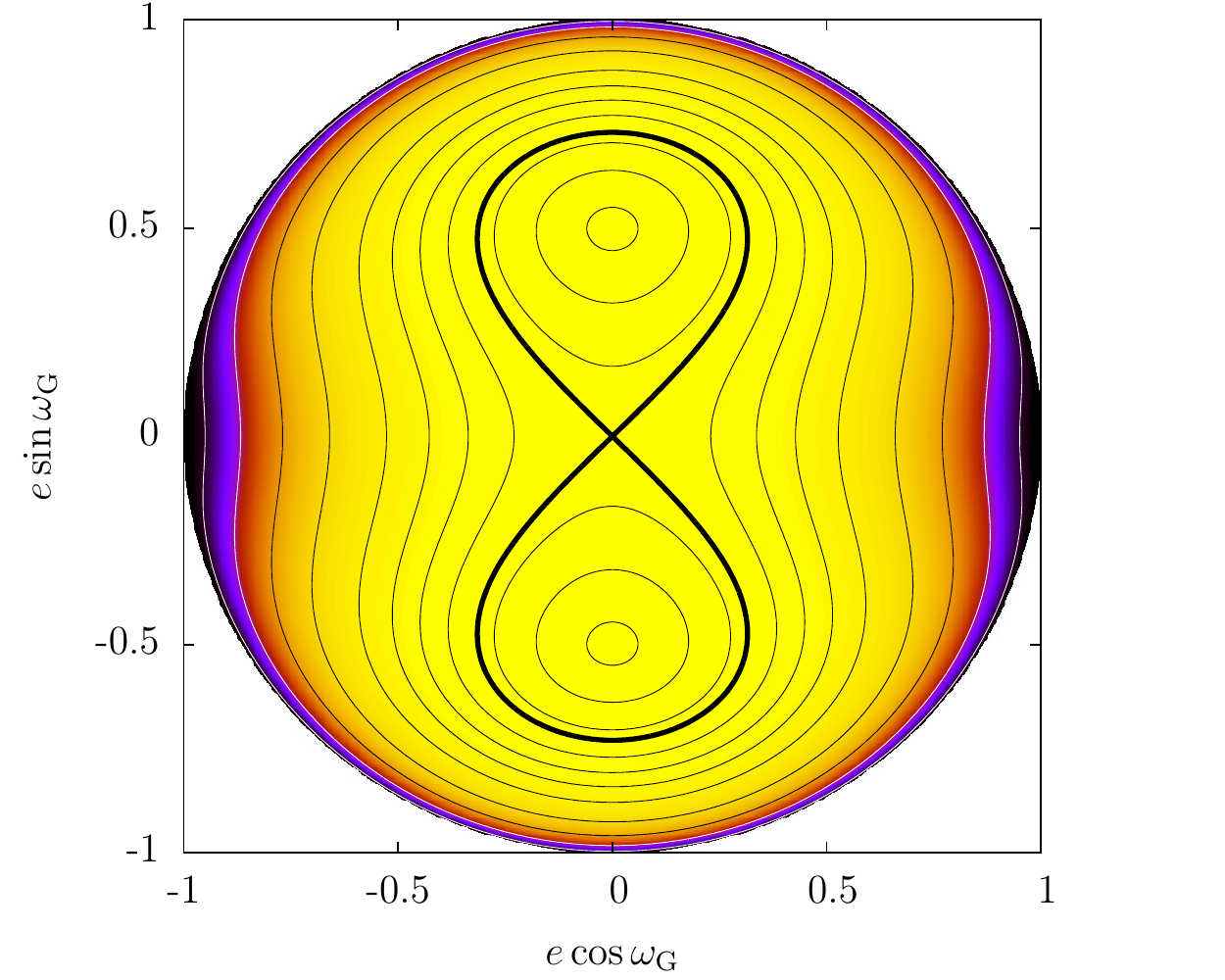}\\
      \caption{Same as Fig.~\ref{fig:LaplaceEccOm} for $a=2000$~au.}
      \label{fig:LaplaceEccOm2}
   \end{figure}

\begin{figure}
      \includegraphics[width=\columnwidth]{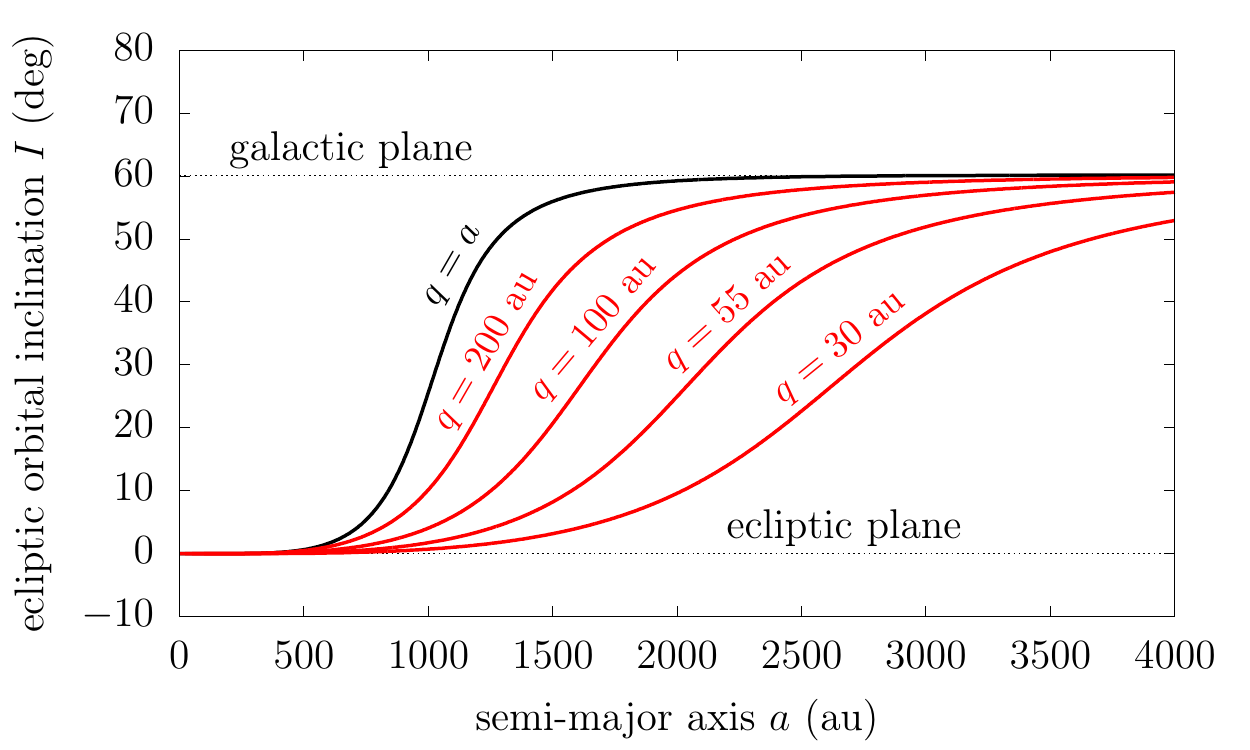}
      \caption{Inclination of the `eccentric Laplace plane' with respect to the ecliptic.  See text for a proper definition. The classic Laplace plane for a circular orbit is shown in black, while red curves show their eccentric counterparts, drawn for fixed values of the perihelion distance (see labels).}
      \label{fig:LaplaceEcc}
   \end{figure}

   Strictly speaking, the Laplace plane is defined for circular orbits (see Sect.~\ref{sec:preliminary} and \citealp{TREetal:09}). For eccentric orbits, the two degrees of freedom are fully coupled and the orbit does not precess around a fixed pole. However, one can still get an idea of the geometry of the phase space with $e>0$ by plotting the level curves of the Hamiltonian in the $(I_\mathrm{G},\Omega_\mathrm{G})$ plane for different values of $(e,\omega_\mathrm{G})$. As can be guessed from the expression of $\mathcal{F}$ (see Eq.~\ref{eq:F}), we obtain strictly the same geometry as for $e=0$ (Fig.~\ref{fig:exLap}), but where the equilibrium points at $\Omega_\mathrm{G}=0$ and $\pi$, denoting the classic Laplace plane, are shifted in inclination. For example, Figs.~\ref{fig:LaplaceEccOm} and \ref{fig:LaplaceEccOm2} show the inclination of this `instantaneous Laplace plane' with respect to $(e,\omega_\mathrm{G})$ for two given values of the semi-major axis. We see that the instantaneous equilibrium plane is close to the classic Laplace plane, except for very eccentric orbits, where it is severely bent towards the ecliptic (that it reaches for $e\rightarrow 1$). This property can be seen in Fig.~\ref{fig:LaplaceEcc}, where we plot the inclination of this plane averaged over $\omega_\mathrm{G}$ (`eccentric Laplace plane') as a function of the semi-major axis. As shown in Appendix~\ref{asec:pend}, the eccentric Laplace plane is not only informative, but it has a real dynamical meaning in the weakly perturbed planetary regime.
   
   Since the known distant trans-Neptunian objects are very eccentric, we expect from Fig.~\ref{fig:LaplaceEcc} that they precess about an axis that is quite closer to the ecliptic pole than predicted by the circular case shown in Fig.~\ref{fig:Laplace}.
   
\section{Dynamics in the weakly perturbed planetary regime}\label{asec:pend}
   We consider the dynamical system with Hamiltonian $\mathcal{F} = \varepsilon_{\mathrm{P}_2}\bar{\mathcal{H}}_{\mathrm{P}_2} + \varepsilon_{\mathrm{G}_\mathrm{V}}\bar{\mathcal{H}}_{\mathrm{G}_\mathrm{V}}$, where the expressions for each part can be found in Eqs.~\eqref{eq:HpG} and \eqref{eq:HgE}. If the planetary perturbations dominate over the galactic tides, $\varepsilon_{\mathrm{P}_2}\bar{\mathcal{H}}_{\mathrm{P}_2}$ acts as the integrable dominant part, whereas $\varepsilon_{\mathrm{G}_\mathrm{V}}\bar{\mathcal{H}}_{\mathrm{G}_\mathrm{V}}$ acts as a small perturbation.
   
   Luckily, the unperturbed part is already expressed in action-angle coordinates. This means that, neglecting terms in $\mathcal{O}(\varepsilon_{\mathrm{G}_\mathrm{V}}^2)$, the long-term behaviour of the system is simply given by the average of $\mathcal{F}$ over the non-resonant angles. This allows us to investigate the effects of each term one by one, and to study the structure of the flow in the vicinity of the resonances:
   
   \emph{i)} The first term of $\bar{\mathcal{H}}_{\mathrm{G}_\mathrm{V}}$ does not include the angles; its acts therefore only as a small modulation of the precession velocities $\dot{\omega}$ and $\dot{\Omega}$ governed by $\varepsilon_{\mathrm{P}_2}\bar{\mathcal{H}}_{\mathrm{P}_2}$ (see Eq.~\ref{eq:plregime} and Fig.~\ref{fig:res}).

   \emph{ii)} The second term of $\bar{\mathcal{H}}_{\mathrm{G}_\mathrm{V}}$ is factored by $\cos\Omega$. Strictly speaking, this term cannot be called `resonant', because it features no separatrix. It actually corresponds to the emergence of the classic Laplace plane: the orbit does not precess exactly about the ecliptic pole, as it would for $\varepsilon_{\mathrm{G}_\mathrm{V}}=0$, but about a tilted pole. By averaging over $\omega$, the momentum $G$ becomes a constant of motion, and we retrieve exactly the `eccentric Laplace plane' introduced in Appendix.~\ref{asec:eccLap}, that rules the dynamics of the variables $(I,\Omega)$.
   
   \emph{iii)} All the remaining terms of $\bar{\mathcal{H}}_{\mathrm{G}_\mathrm{V}}$ correspond to resonances and libration zones for $\omega$ and $\Omega$. Assuming that there is a single resonance, the resonant angle can be taken as a new independent variable by a linear canonical change of coordinate (unimodular matrix). For instance, we consider the resonance $\omega+\Omega$. From the ecliptic Delaunay coordinates $(P_\omega,P_\Omega,\omega,\Omega)$, the corresponding change of coordinates is
   \begin{equation}
      \left\{
      \begin{aligned}
         \sigma &= \omega+\Omega \\
         \gamma &= \Omega
      \end{aligned}
      \right.
      \hspace{0.5cm}
      \left\{
      \begin{aligned}
         \Sigma &= P_\omega \\
         \Gamma &= P_\Omega-P_\omega \,.
      \end{aligned}
      \right.
   \end{equation}
   After averaging over the circulating angle $\gamma$, we end up with one constant of motion $\Gamma$, and one degree of freedom $(\Sigma,\sigma)$. Table~\ref{tab:const} gives the constants associated to all resonances that appear at first order in $\varepsilon_{\mathrm{G}_\mathrm{V}}$ (that is, the ones directly appearing in the expression of $\bar{\mathcal{H}}_{\mathrm{G}_\mathrm{V}}$ in Eq.~\ref{eq:HgE}). The resonance centre is mostly governed by the unperturbed Hamiltonian $\varepsilon_{\mathrm{P}_2}\bar{\mathcal{H}}_{\mathrm{P}_2}$: it is fixed in inclination (see Fig.~\ref{fig:res}), but goes from $e=0$ to $e=1$ when varying the value of the constant quantity given in Table~\ref{tab:const}.
   
   \begin{table}
      \caption{Resonance centres and constants of motion arising from the dynamics in the vicinity of each resonance appearing at first order in~$\varepsilon_{\mathrm{G}_\mathrm{V}}$.}
      \label{tab:const}
      \vspace{-0.7cm}
      \begin{equation*}
         \begin{array}{rrl}
            \hline
            \text{resonant angle} & \text{resonance centre} & \text{constant quantity} \\
            \hline
            \hline
            2(\omega+\Omega) & (\sqrt{6}+1)/5   & \sqrt{1-e^2}(\cos I-1)   \\
            2\omega+\Omega   & (\sqrt{21}+1)/10 & \sqrt{1-e^2}(2\cos I-1)  \\
            2\omega                     & \sqrt{5}/5       & \sqrt{1-e^2}\cos I       \\
            2\omega-\Omega   & (\sqrt{21}-1)/10 & \sqrt{1-e^2}(2\cos I+1)  \\
            2(\omega-\Omega) & (\sqrt{6}-1)/5   & \sqrt{1-e^2}(\cos I+1)   \\
            2\Omega                     & 0                & e                                   \\
            \hline
         \end{array}
      \end{equation*}
      \vspace{-0.3cm}
      \tablefoot{The resonance centre given here is the value of $\cos I$. The eccentricity at the resonance centre depends on the value of the constant quantity, taken as parameter (right column). The retrograde cases are obtained by changing the sign of $\cos I$ and of $\Omega$.}
   \end{table}
   
   Since the resonances are thin in the weakly perturbed problem, we can use the pendulum approximation. This amounts to using a Taylor expansion of the Hamiltonian around the resonance centre $\Sigma_0$, keeping terms up to degree $2$ for the unperturbed part, and up to degree $0$ for the perturbation. The resulting Hamiltonian has the form
   \begin{equation}
      \mathcal{F}_\mathrm{res} = \alpha (\Sigma-\Sigma_0)^2 + \beta\cos 2\sigma \,,
   \end{equation}
   which is a pendulum of centre $\Sigma_0$ and half width $\sqrt{2|\beta/\alpha|}$. Using the constants of motion in Table~\ref{tab:const}, the widths can then either be expressed in terms of the eccentricity of in terms of the inclination.
   
   Figures~\ref{fig:secres500} and \ref{fig:secres700} show the location and widths of all the resonances that appear at first order in $\varepsilon_{\mathrm{G}_\mathrm{V}}$, computed analytically using this perturbative approach. In these figures, instead of using the value of the constant quantities as parameters, we directly use the perihelion distance of the resonance centre (this allows us to draw all resonances on a single graph). In the pendulum approximation, the upper and lower half widths are equal when they are expressed in canonical coordinates (i.e.~$\Sigma$), but, as shown by Figs.~\ref{fig:secres500}-\ref{fig:secres700}, this is not necessarily the case in $I$ nor in $q$ (and we indeed observe asymmetric resonances in Sect.~\ref{sec:intermed}).
   
   The hexadecapolar planetary term (see Eq.~\ref{eq:Hp}) can be easily incorporated using this perturbative approach; it only slightly changes the widths of the $\omega$ libration island.
   
\section{Maximum orbital variations reachable in $4.5$~Gyrs from the secular action of the planets and galactic tides}\label{asec:maps}
\begin{figure*}
      \centering
      \includegraphics[width=\textwidth]{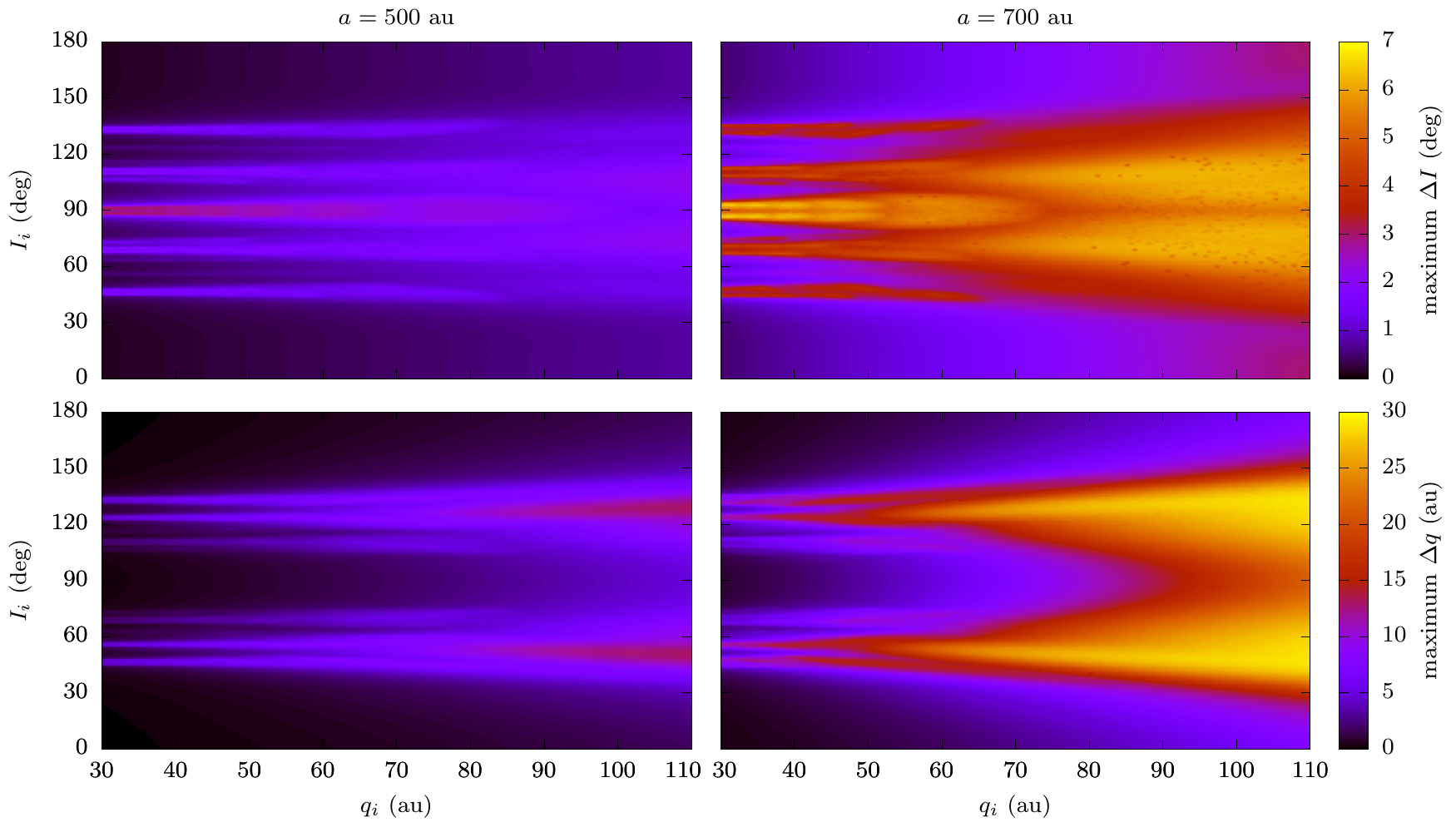}
      \caption{Maximum possible orbital variations produced in $4.5$~Gyrs in the $(q,I)$ plane. Each column corresponds to a different value of the semi-major axis (see titles). This figure is to be compared to Figs.~\ref{fig:secres500} and \ref{fig:secres700}, showing the location and widths of the main resonances obtained analytically.}
      \label{fig:qimap}
   \end{figure*}

   In Sect.~\ref{sec:limits}, we map the maximum possible variation for $q$ and~$I$ in $4.5$~Gyrs, according to the location in the $(a,q,I)$ space. Figures~\ref{fig:aqmap}-\ref{fig:aimap} show sections of this space in the $(a,q)$ and $(a,I)$ directions.  For completeness, Fig.~\ref{fig:qimap} shows here the maximum orbital variations in the $(q,I)$ direction. Only a restricted portion of the horizontal axis is shown in order to ease the comparison with Figs.~\ref{fig:secres500} and \ref{fig:secres700}. We retrieve the structure of the resonances studied in Sect.~\ref{sec:res}, except that Fig.~\ref{fig:qimap} also incorporates a timescale information. Indeed, for large initial perihelion distances, the precession due to the planets is so slow that the trajectories do not have enough time in $4.5$~Gyrs to explore the full width of the resonance (or the full extent of the chaotic region in case of resonance overlap). In Fig.~\ref{fig:qimap}, this extension of timescale produces a drop of the orbital variations inside the resonances. However, for even higher perihelion distances, this drop is compensated by the overall increase of the galactic potential, which tends to shorten the timescale (see the background colour gradient, for instance along the horizontal line $I_i=15^\text{o}$), while producing the generalised chaotic region studied in Sect.~\ref{sec:intermed}.
   
\end{document}